\documentclass[fleqn, usenatbib]{mnras}
\usepackage{amsmath}
\usepackage{amssymb}
\usepackage{mathtools}

\usepackage{wrapfig}
\usepackage{graphicx}
\usepackage{subcaption}
\usepackage{tcolorbox}
\DeclareMathOperator{\grad}{grad}

\title[Fragmentation in MHD disks]{Characterizing fragmentation and sub-Jovian clump properties in magnetized young protoplanetary disks}
\author[N. Kubli et al.]{Noah Kubli,$^{1}$
Lucio Mayer,$^{1}$
Hongping Deng$^{2}$
\\
$^{1}$ Institute for Computational Science, University of Zurich, Winterthurerstrasse 190, CH-8057 Zürich, Switzerland
\\
$^{2}$ Shanghai Astronomical Observatory, Chinese academy of Science, Nandan Rd 80th, 200030 Shanghai, China}

\pubyear{2023}

\begin{document}

\maketitle
\begin{abstract}

We study the initial development, structure and evolution
of protoplanetary clumps formed in 3D resistive MHD
simulations of self-gravitating disks.
The magnetic field grows by means of the recently
identified gravitational instability dynamo
\citep{riols_latter, deng1}.
Clumps are identified and their evolution
is tracked finely both backward and forward in time.
Their properties and evolutionary path is compared to clumps 
in companion simulations without magnetic fields.
We find that magnetic and rotational energy are important in the clumps' outer regions, while in the cores, despite appreciable magnetic field amplification,
thermal pressure is most important in counteracting gravity.
Turbulent kinetic energy is of a smaller scale
than magnetic energy in the clumps.
Compared to non-magnetized clumps,
rotation is less prominent, which results
in lower angular momentum in much better agreement
with observations.
In order to understand the very low sub-Jovian masses of clumps 
forming in MHD simulations, we revisit the perturbation theory
of magnetized sheets
finding support for a previously proposed magnetic
destabilization in low-shear regions.
This can help explaining why fragmentation ensues on a scale more than an order of magnitude smaller than that of the Toomre mass.
The smaller fragmentation scale and the high magnetic pressure in clumps' envelopes explain why clumps in magnetized disks are typically in the super-Earth to Neptune mass regime
rather than Super-Jupiters as in conventional disk instability.
Our findings put forward a viable alternative to
core accretion to explain widespread formation of intermediate-mass planets.

\end{abstract}

\section{Introduction}
\label{ch-introduction}
%\subsection{Fragmentation in magnetized disks}

With over 5000 confirmed detections of exoplanets\footnote{taken from \citet{nasa}}, the mass statistics of the population can now be inferred \citep{zhu-dong}.
It is known that the most common exoplanets lie in the intermediate-mass regime ranging from super-Earth to Neptune size \citep{schneider}.
Further, many gas giants have been detected.

Planet formation is addressed by two competing theories; core accretion and gravitational instability.
In core accretion \citep{safronov, pollack} a rocky planetary embryo grows through the accretion of planetesimals \citep{ndugu}. 
If it becomes massive enough, it might attract a gaseous envelope to become a gas giant \citep{helled}.
The process is slow compared to the disk's lifetime but can be significantly accelerated by pebble accretion \citep{ormel},
or even via a combination of pebble and planetesimal accretion \citep{alibert}.
On the other hand, with disk instability \citep{kuiper, boss-97, mayer-2002} the
timescale problem is circumvented by assuming that the formation process of 
a (massive) planet is driven by the self-gravity of fluid matter in the disk. If gas
is sufficiently cold and dense, direct collapse of a patch can occur despite the
counteracting action of shear,
where the Toomre theory \citep{toomre} provides a criterion 
for the occurrence of disk instability
in the
simple framework of linear perturbation theory, and widely verified by numerical
simulations across various domains of astrophysics \citep{durisen}.
In the context of planet formation, three-dimensional
numerical simulations of 
disk instability were first conducted by \citet{boss-97} 
in order to explain the formation of Jupiter and Saturn.
Gas collapse will  be eventually followed by accretion of solids
to form a rocky core and a metal-enriched envelope \citep{helled}.
A massive disk, of order 10\% of the mass of the star, can become gravitationally unstable on an orbital timescale. Disk instability could well explain massive planets (e.g. HR8799, see \citep{nero}).
It can also explain massive planets around low-mass stars (e.g. GJ3512b, see \citet{morales}) and wide-orbit 
gap-carving planets, (e.g. AS209, see \citet{bae}) both of which
cannot be explained by core accretion even when pebble accretion is considered.

On the population level the core accretion model predicts
a dip in the planet mass function around Neptune mass which
is due to the runaway gas accretion which is required to build gas
planets in this model.
This is however contrary to observations \citep{suzuki, schlecker}.
Traditional disk instability, neglecting magnetic fields,
is thought to be only relevant for gas giants
and thus cannot provide an explanation for intermediate-mass planets.
However, young gravitationally unstable disks exhibit spiral structures \citep{toomre, deng-ogilvie22}, 
such as observed in Elias 2-27 \citep{meru, veronesi, perez}, suggesting some role
of disk instability.

The spirals sustain a dynamo (even in poorly ionized disks, see \citet{riols-xu}) 
and lead to strong magnetic fields. 
This effect was described in 
\citet{riols_latter} and \citet{riols_latter2} and 
should not be confused with the magneto-rotational
instability (MRI).
The spiral-driven dynamo grows the magnetic field by
means of a feedback-loop amplification between field
stretching along the spirals and  field twisting across them
owing to vertical rolls triggered by shocks generated
by the spirals. In this way, an initial small toroidal field
is converted into a stronger poloidal field, and then
converted back into a proportionally stronger toroidal field.
Amplification in the vertical rolls is the key step, and makes
the dynamo inherently three-dimensional. Magnetic energy
grows at the expense of self-gravity, of which
the spirals are a manifestation, and rotational energy.
While the MRI breaks down at high values of resistivity, the
spiral-driven dynamo is resilient, and the resulting magnetic field
is much stronger than in the case of the MRI, as shown in \citet{deng1}.
They also demonstrated the global
nature of the dynamo, e.g. by measuring
the global toroidal field pattern and 
showing the importance of outflow 
boundary conditions at high altitude.
In \citep{riols-xu} the effect of ambipolar 
diffusion on the dynamo has been investigated  
showing that the dynamo is able to work 
on a large range of ambipolar Elsasser numbers.

Recent  simulations of \citep{deng} showed that magnetic fields may have
an important impact on the formation of planets through disk instability.
Protoplanetary clumps emerged with masses
one to two orders of magnitude smaller than one would expect
from conventional simulations and models of disk
instability \citep{durisen}. %what models?
Their masses clustered around Super-Earth to Neptune masses,
a mass range which is not prevalent in core accretion \citep{suzuki, mordasini} %, mordasini}
while conventional disk instability favors planets with masses from that
of Jupiter up to the brown dwarf regime \citep{helled}.
On the other hand, observations suggest that exoplanets, at least in our Galaxy,
are most abundant in this mass range \citep{schneider}.
In purely hydrodynamical simulations using identical 
disk models and cooling (without the magnetic field)
which they run for comparison, much fewer clumps resulted, none
survived till the end of the simulations, and their masses up to
the disruption were close within factors of a few from a Jupiter mass.

Besides the lower mass of the fragments,
the presence of the magnetic field also leads to differences
in the further evolution of the clumps.
The purely hydrodynamical simulations required a fast, physically unrealistic
cooling for
the clumps to survive, otherwise they would be disrupted by shear \citep{deng}.
On the other hand, even the lowest mass clumps
forming in the MHD simulations
could survive,  which was attributed to a shielding-effect by the magnetic field
which underwent amplification at their boundary, which also prevented
significant mass growth via the effect of magnetic pressure.

For magnetic fields to be present in protoplanetary disks, they
need to be ionized to a certain degree.
Although spiral shocks may heat the disk up to 
some hundred K \citep{podolak}, in general the temperature
in the simulations is too low to provide the necessary
ionization \citep{deng1}.
It has been discussed in \citep{deng1} that the ionization must stem
from other sources than temperature
which could be the central star or other close
stars
providing a source for ionizing radiation
or
cosmic rays.
The magnetic fields have shown to be 
dynamically important \citep{turner, masson}
in protoplanetary disks.

A physical understanding of the small masses of the clumps in
magnetized disks, from their very appearance
in the disk to their growth phase, is still lacking. In addition
in \citep{deng} many questions were left open concerning if and
how they differ from clumps in unmagnetized disks in other
ways than just their mass, and what is the relation between
their properties and the nature of the flow in the disk, which
is magnetized but also more turbulent than in conventional
disk instability \citep{deng1}.
In this paper, the properties of the magnetized clumps as well as 
their formation path from the disk material are studied
and characterized in great detail. In addition,
with the aid of the simulations, we propose a theoretical
framework that can provide an understanding of their low
masses.

The standard method to characterize disk instability
is the Toomre analysis \citep{toomre}.
Starting from the hydrodynamical fluid equations, and performing
a perturbative analysis, Toomre
derived a criterion for instability.
\begin{equation}
Q=\frac{c_s\kappa}{G\pi \Sigma} < 1
\end{equation}

The disk is destabilized through its self-gravity, encapsulated
in its surface mass density $\Sigma$,
and stabilized by rotation and gas pressure,
here expressed via
the epicyclic frequency $\kappa$ and the sound speed $c_s$, respectively.
The Toomre criterion is derived under the assumption of razor-thin
sheet with no pressure gradients, and is valid for local axisymmetric perturbations.

The case of disk fragmentation in the presence of magnetic fields
was investigated in \citet{gammie} and \citet{elmegreen} in the framework
of galactic disks.
Starting from the magnetized fluid equations (see section \ref{destabilization}) Gammie could
derive a relation similar to the Toomre analysis.
For axisymmetric perturbations and a toroidal orientation of
the magnetic field
he found that the magnetic field lead to an increasing stability
of the system.
A dispersion relation was derived for perturbations in 
such disks in which the magnetic field acts like 
the gas pressure:
\begin{equation}
\omega^2 - (\kappa^2 - 2\pi G\Sigma \lvert k\rvert + (c_s^2+V_a^2)k^2) =0
\end{equation}
where the magnetic field is expressed via the Alfvén velocity $V_a$.
Applying the same reasoning as in the Toomre theory \citep{toomre}
this allows to define a parameter for magnetized disks
$Q_B=\frac{\kappa\sqrt{c_s^2+V_a^2}}{G\pi\Sigma}$

Another approach, which was specialized for galactic
(hence non-Keplerian) disks was put forward by
\citep{elmegreen}. The latter author studied
the evolution of non-axisymmetric perturbations
in a differentially rotating  magnetized thin sheet
through numerical integration
of the perturbed fluid equations.
Since spiral structure is typically seen to develop
in fragmenting disks before fragmentation actually occurs,
the study of non-axisymmetric perturbations is most
relevant for our purpose.
In his study, \citep{elmegreen} found that the presence of a magnetic field can
lead to a destabilization of the system in certain
regions characterized by weak shear since
it can inhibit stabilization of a perturbation 
through the Coriolis force. As a result, 
perturbation with smaller wavelength can grow,
which would be otherwise stable.
This destabilisation mechanism, which is discussed in section
\ref{destabilization} and studied with our simulations, is appealing because it could provide a clue to  understand the different nature  of fragmentation in magnetized disks.

\section{Methods}
\label{ch-methods}
\subsection{Fragmenting MHD Simulations}
\label{ch-simulations}
In this section we briefly describe the simulations
that were analyzed in this work.
These simulations were already presented in 
\citet{deng} and are based on simulations from \citet{deng1}.

In the simulations, the self-gravitating MHD equations with resistivity
and cooling were solved:
\begin{equation}
\frac{\partial \rho}{\partial t} + \nabla(\rho v) = 0
\end{equation}
\begin{equation}
\frac{\partial v}{\partial t} + v \cdot \nabla v = -\frac{1}{\rho}\nabla(P + \frac{B^2}{8\pi})%
+ \frac{(B\cdot\nabla)B}{4\pi\rho}%
- \nabla\Phi
\end{equation}
\begin{equation}
\frac{\partial B}{\partial t} = \nabla \times (v\times B) + \eta\nabla^2B
\label{eq-ohm}
\end{equation}
\begin{equation}
\frac{\partial U}{\partial  t} + \nabla(U v) = -P\nabla v - \frac{U}{\tau_c}
\end{equation}
The cooling time was just assumed to be proportional to the orbital
time: $\tau_c = \beta / \Omega$
while the relation of pressure and internal energy
is determined via the ideal gas equation
$P=(\gamma - 1)U$ with $\gamma = 5/3$.
The simulations were conducted with GIZMO \citep{gizmo1, gizmo2, gizmo3}
which uses the MFM (meshless finite mass) method.
%Resolution
They simulated a disk of mass $0.07 M_{\text{sun}}$ in a radius
of $5-25\text{AU}$
with a central star of $1M_{\text{sun}}$ that is represented
by a sink particle.
The initialization of the simulations is described in \citet{deng1}:
They started with a surface density and a temperature profile 
of $\Sigma \propto r^{-1}$ and $T\propto r^{-1/2}$.
Also, a toroidal seed magnetic field was added in the MHD case.
The simulations were then run using a weak cooling rate ($\beta=8$)
until the disk's spiral structure was established.
Then the cooling was increased to $\beta = 6.28$ and the 
simulations were continued to saturate the magnetic field.
During this process, particle-splitting was applied to achieve 
the desired resolution.
The achieved resolution is very high: for the main  MHD
simulation more than 30 million particles were used to resolve MHD effects.
The same simulation was run in more than one variant (see below), such
as with or without Ohmic resistivity, and with a different cooling
prescription for the high density regions (see below).
Companion HD-simulations that did not include
a magnetic field
were also conducted; for those, lower resolutions were required (3 million particles, see also discussion).
Overall, these simulations took more than a year of computing time on the
Cray XC40 supercomputer "PizDaint" at the Swiss Supercomputing Center (CSCS).
This prevented us from running a large set of simulations with
different disk models so far.

These simulations were then taken and used as initial conditions for
the fragmenting simulation.
Fragmentation was then induced through an increase in cooling by changing to
$\beta = 3$ (see \citet{gammie-2001, deng-17}).
The results were then used as initial conditions
for subsequent runs that investigated the further evolution of the clumps
as described in \citep{deng}.
They also used a cooling-shutoff in the innermost regions of the clumps 
after they become gravitationally bound
noting that the high cooling rates there would be unrealistic
because the high density leads to highly optically thick conditions,
resulting in long photon diffusion times and nearly adiabatic evolution.
However, fragmentation and the early physical properties and initial evolution of clumps, the focus of this paper, are insensitive to the 
latter aspect, hence we will not use this variant of the simulations
for analysis here. Furthermore, the specific MHD simulation used for
the analysis of this paper includes Ohmic resistivity.
The companion HD-simulations are also used in this work for comparison.
The resistivity is set via the magnetic Reynolds number 
$R_m \equiv c_s H/\eta = 20$ with
$c_s$ the sound speed and $H$ the scale height of the disk \citep{deng}.

The analysis presented in this work is based on snapshots taken 
at equally-spaced time intervals of $10/2\pi$ years.
We describe the methods used to analyze the simulations in the next section.

\subsection{Identification of the clumps and backtracing}
\label{ch-identification}

\begin{figure}
    \centering
	\includegraphics[scale=0.55]{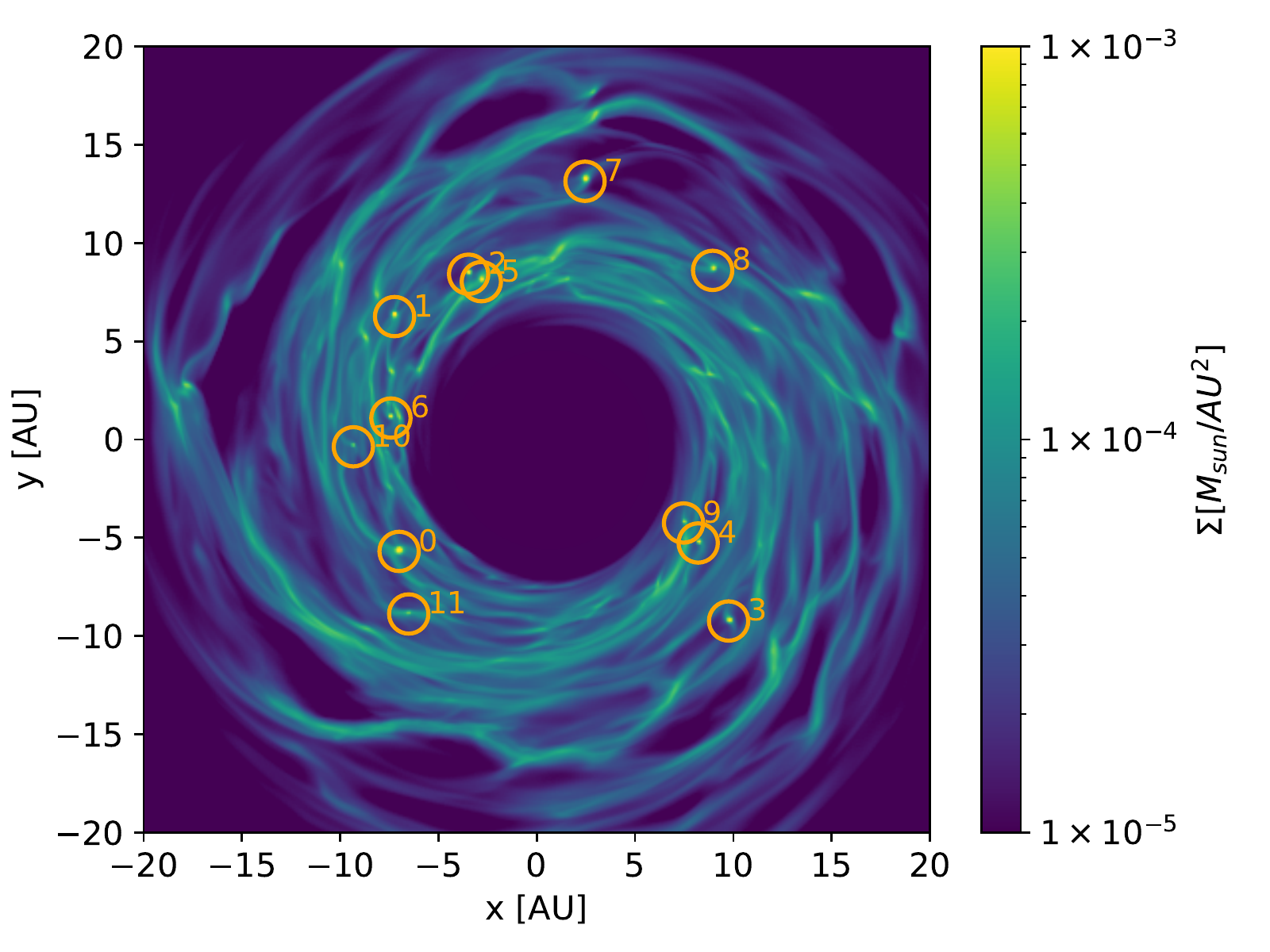}
\caption{Surface density plot of the disk towards the end of the considered time frame ($t = 156 \text{yr}$). 
The clumps are marked in orange circles.
The flocculent appearance of the disk can be seen as 
observed in \citet{deng1}.
}
\label{fig-disk-density}
\end{figure}

\begin{figure}
	\includegraphics[scale=0.5]{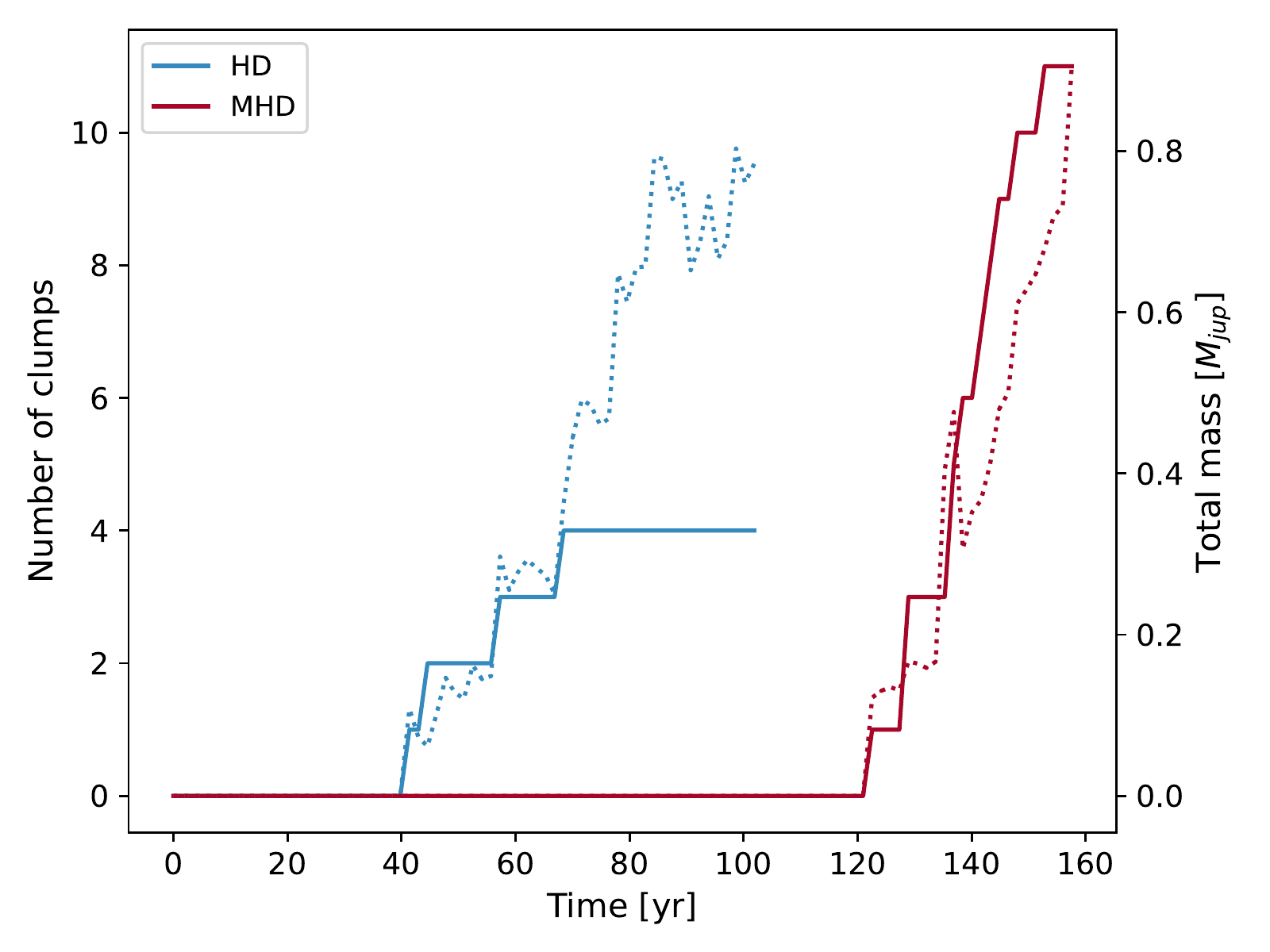}
\caption{Evolution of the clump formation:
Number of clumps (solid, left) and total mass contained in the clumps (dashed, right).
Many more clumps form in the MHD case.
}
\label{fig-clump_count}
\end{figure}
%\begin{wrapfigure}{R}{0.2\textwidth}
\begin{figure}
%\begin{subfigure}{0.24\textwidth}
    \centering
	\includegraphics[scale=0.6]{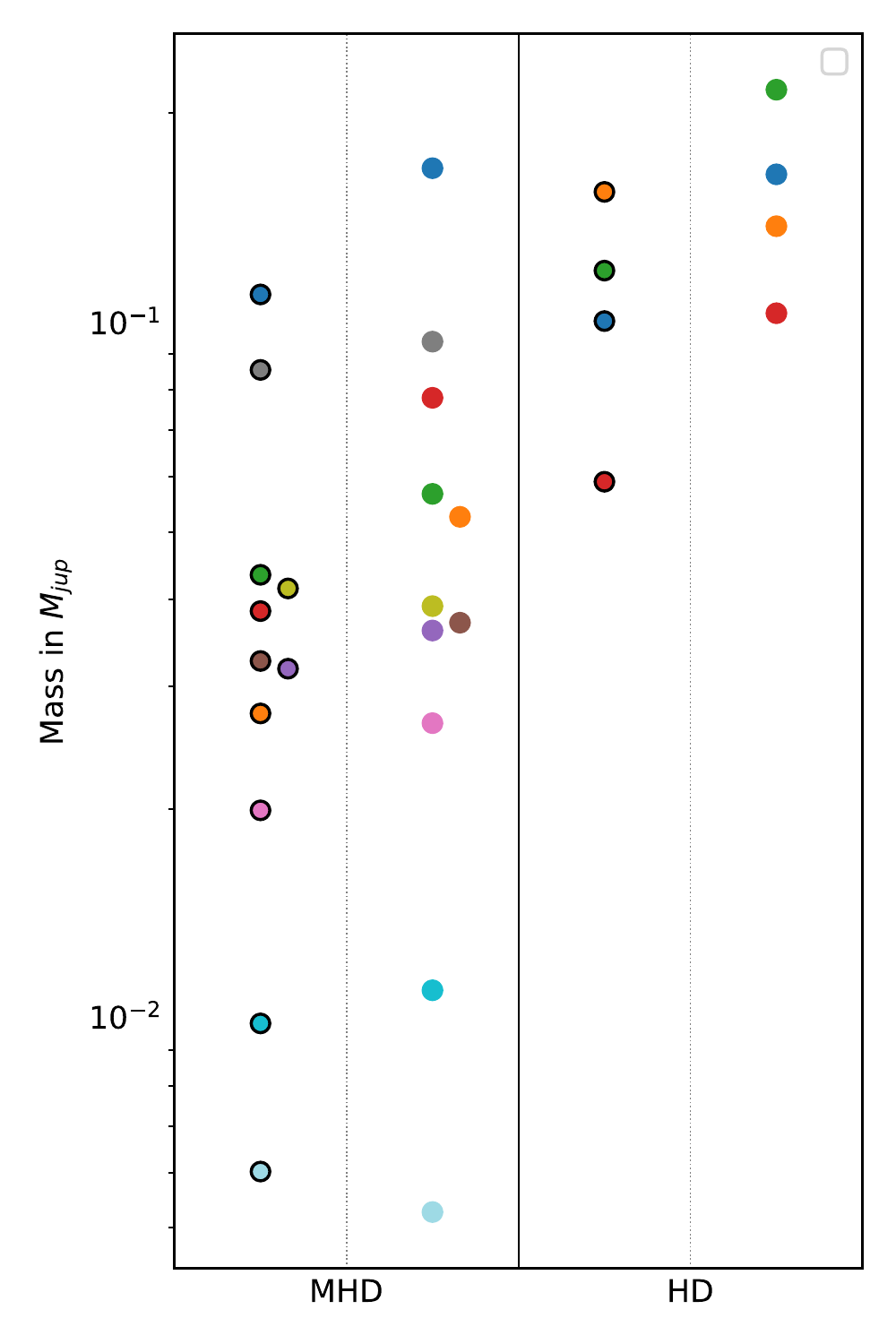}
 %\end{subfigure}
 %\hfill
 %\begin{subfigure}{0.24\textwidth}
 %	\includegraphics[width=\textwidth]{./images/masses_frag.pdf}
 %\end{subfigure}
\caption{Masses of the resulting clumps in the MHD and
the HD case.
The dots with the black edge colour represent their masses
at the time of fragmentation while the others show the mass
averaged over the lifetime.
In the MHD case, much smaller clumps can emerge, going below $10^{-2} M_\text{jup}$,
this is already true at the time of fragmentation, when the clumps emerge.}
\label{fig-masses}
%\end{wrapfigure}
\end{figure}

%\begin{figure}
%\centering
%	\includegraphics[width=0.25\textwidth]{./images/masses_frag.pdf}
%\caption{Masses of the clumps at their fragmentation time for the MHD
%and the HD case.
%The masses between the MHD and the HD case
%are already different at the time of fragmentation.}
%\label{fig-masses-initial}
%%\end{wrapfigure}
%\end{figure}

In this section we describe the procedure to find the clumps in the snapshots, and to analyze them.

Towards the end of the simulation,
at the last snapshot that we are considering,
we identify clumps as follows;
first, we find density peaks by selecting all particles
above a certain density threshold and assign them to a cell
on a grid that is superimposed on the particle distribution.
The cells that contain such particles are marked as dense cells.
We identify connected dense cells (clusters) on the grid and
define all particles (including those below the density threshold)
that lie in the corresponding
cells as belonging to that cluster.
Each of these clusters serves as the approximate location of a clump.
For the density threshold we chose a value of 
$10^{-8} \text{g}/\text{cm}^3$
which amounts to $\approx 100$ times the average density
in the simulation in the clump-forming radial extent of the disk.
The exact choice of this value does not make much of a difference
since the exact extension of the clumps are determined in the next step
by identifying their gravitational boundedness.

In the next step, we start by determining
the exact location (particle-wise) of a 
density peak within the cluster which we use
as a guess for the corresponding clump's centre.
Around this point we introduce concentric shells (radial bins).
Starting from the centre, we increase the radius by gradually
determining if a certain shell is bound to the clump that is defined
using all particles inside the corresponding radius of the shell.
The gravitational boundedness is determined by calculating
for each particle
the potential energy with respect to the clump, 
the kinetic and the internal energy.
The radius of the clump is finally defined 
using the inner end of the first unbound shell encountered.

Since we are interested in the clump's evolution and their origins,
we now trace them back to earlier snapshots.
This is done by determining all particles that are within the
clump's radius and then identifying the same particles at 
earlier snapshots.
Now there are two ways to proceed.

First, considering only this subset of particles, we find the position
of their density
maximum, i.e. where they congregate the closest.
This position is then taken as the clump's centre at the earlier snapshot.
The radius of the gravitationally bound region is then determined in
order to identify the clump.
We measure quantities such as their mass, angular momentum and energy.
The  method just described, however cannot be used indefinitely backward in time since 
there is no well defined density maximum in the very early stage of
clump formation, before a bound clump is present.
It can only be extended back to the time when the clump first becomes
bound (and somewhat before that).
We check that our clump is still well-defined by 
evaluating the fraction of particles originally
found in the clumps identified at a later stage 
but not included in the overdense region.
%XXX set threshold?
At the time when the clump is bound, this fraction is low ($\lesssim 20\%$),
meaning that there is no big change in the particle membership of the clumps after they become bound, namely their initial stage
reflects the "in-situ" flow.

Additionally, we are also interested in the regions of the disk at earlier times
where the clumps will emerge
%before their formation
in order to
calculate disk properties relevant for their formation such as  
the Toomre or Jeans mass.
For the latter we use a second method;
we identify the clump-forming regions by finding the ensemble of
particles that will be later incorporated in the clumps.
%In this process we do not identify a clump but rather we look at the whole
%region.

%surface density plot 2d and evolution plot
Fig. \ref{fig-disk-density} shows a density plot of the simulation
towards a time when all the
clumps that become gravitationally bound have formed.
The small scale structure that was observed in \citet{deng} can 
also be recognized in this plot.
The evolution of the clump population is presented in fig. \ref{fig-clump_count}
showing the number of clumps and their total mass over time and also
comparing to the HD-case.
The clumps are counted as soon as they are determined as bound using the
method described above.
One can see that in the HD-case, there are much fewer clumps than in
the MHD-case.

In Fig. \ref{fig-masses}, the masses of the clumps are shown.
They are calculated by determining the bound radius as described above
and then summing up the mass of all the particles inside.
The plot then shows their average mass from the time they become bound.
Further, also the masses at the time when they first become bound are shown.
As can be seen in the plot, in the MHD case, the clumps' masses
are generally lower and can have a much greater variation than in the HD case. Indeed, in the magnetized disks there are many low-mass clumps going below
the mass of Neptune 
\citep{deng}.
%Also, we show the respective masses at the time of fragmentation
%in fig. \ref{fig-masses-initial}.
We also note that the difference between the MHD and the HD case is already present at  the onset of fragmentation.

In summary, the difference between
the clumps in the MHD case and the HD case
is two-fold;
first, the clumps evolve differently when they
are embedded in a magnetized disk (e.g. they have lower
growth, and are protected from tidal disruption, see \citet{deng}).
Second, the initial fragmentation
stage is different as the MHD clumps
have smaller masses from the beginning and fragmentation is
more plentiful.

\section{Numerical Results}
\label{ch-results}
\subsection{Predicted mass scales}
\label{ch-scales}

\begin{figure}
\centering
\begin{subfigure}{\columnwidth}
	\includegraphics[width=\textwidth]{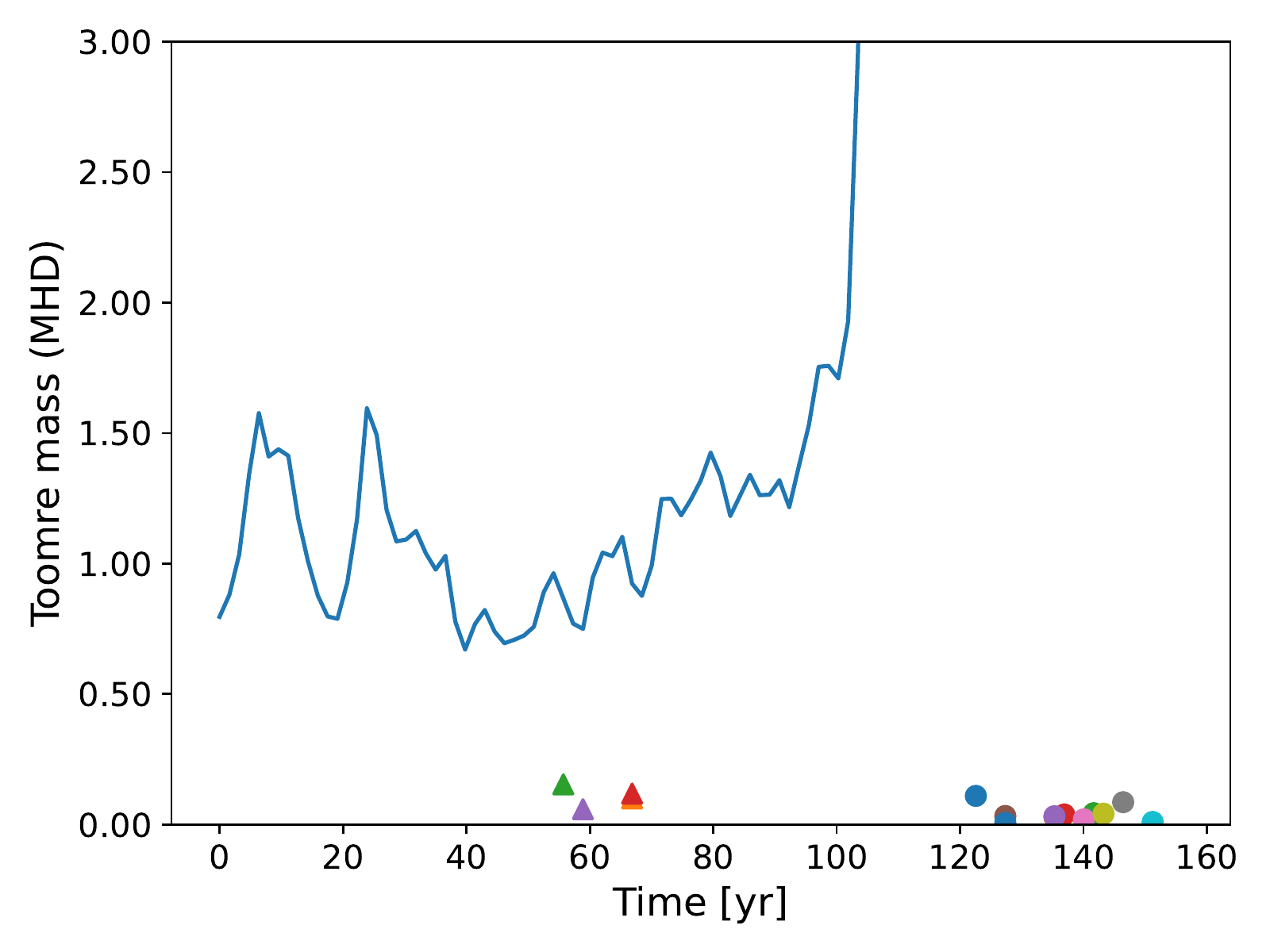}
 \caption{Toomre mass over time. 
 Generally, the masses of the clumps are greatly overestimated.}
 \label{fig-toomre-mass-toomre}
\end{subfigure}

\begin{subfigure}{\columnwidth}
	\includegraphics[width=\textwidth]{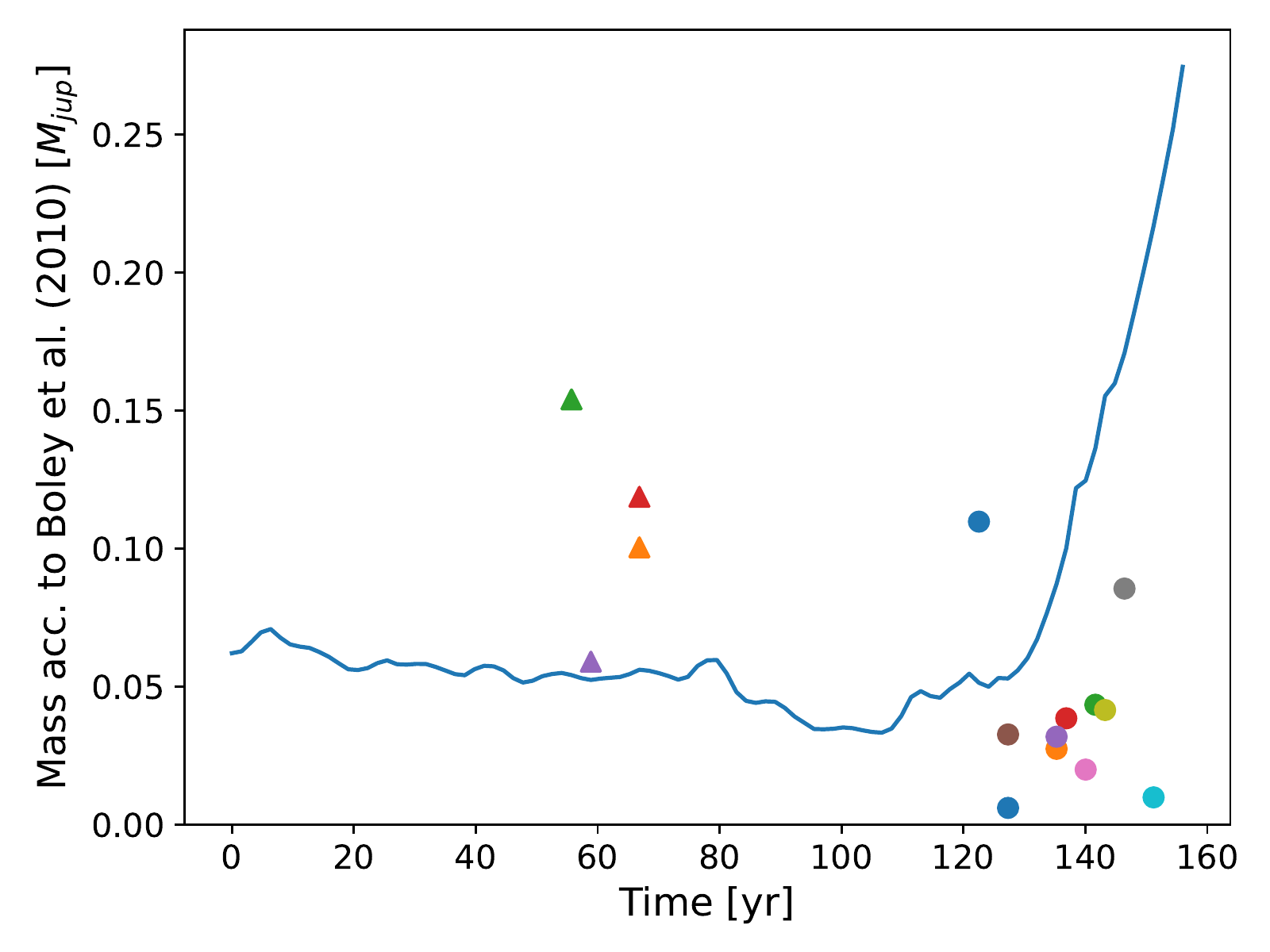}
 \caption{Predicted mass according
to \citet{boley-disruption} using the same procedure.
This theoretical prediction provides a better
estimation than the Toomre mass (fig. \ref{fig-toomre-mass}) 
but does not explain the difference between the MHD and the HD case.}
\label{fig-toomre-mass-boley}
\end{subfigure}
\caption{Toomre mass and mass prediction according to \citep{boley-disruption}
determined using the backtraced regions of particles.
At each snapshot in time, the particles that will later form the clump 
are identified and the quantities are measured.
Then we average over the different clumps so
the resulting values in the early stages
are an average prediction for the clump masses 
(see second method described in section \ref{ch-identification}).
At fragmentation, the Toomre mass and later also the Boley mass greatly increase since the background assumptions of the theories become invalid.
Shortly later, the clumps become bound and are marked with a corresponding dot.
As a comparison, the clumps arising in the HD simulation are shown with a triangle.}
\label{fig-toomre-mass}
\end{figure}

From Toomre's theory of disk instability one can derive an %Quelle
estimate for the mass of the clumps
by assuming that the collapsing region has a characteristic size 
of order the
Toomre-most-unstable wave length $\lambda_{T}$:
\hbox{$M_\text{Toomre} = \pi\left( \frac{\lambda_T}{2}\right)^2 \Sigma$}.
This estimate is shown in fig. \ref{fig-toomre-mass} where
we identified the fragmenting regions 
in the early snapshots of the MHD simulation
(using the second method described in section \ref{ch-identification})
and determined a representative Toomre mass at any given time by 
averaging over the Toomre mass values obtained from the back-traced
particles of the different clumps.
Only the early snapshots in fig. \ref{fig-toomre-mass} should be taken
into account since the Toomre theory assumes an equilibrated disk
which is better fulfilled before it fragments – when the clumps collapse
the measure becomes invalid.

It can be seen that the predicted mass is around $1 M_\text{jup}$.
The masses observed in the MHD simulation reach down to $0.006 M_\text{jup}$ –
(see fig. \ref{fig-masses})
all being much lower than the prediction.
Although higher, the masses of the clumps in the HD simulation are still 
below what would be expected from the Toomre mass.
This means that additional effects have to be considered
to study the nature of the collapse – both related and unrelated to the magnetic field.
These effects could be the vertical extension of the disk, 
turbulence which could be induced or altered through the magnetic field
or a direct effect of the magnetic field on fragmentation.
We investigate the latter effect in section \ref{section-instability}.

While Toomre's theory assumes a thin background axisymmetric disk with differential rotation, the minimum collapsing mass should be 
comparable to the Jeans mass
since that neglects rotation which affects the longer wavelength
branch in Toomre instability theory.
One can thus calculate the Jeans mass of the backtraced regions;
also, this simplistic estimate is definitely too high for explaining the 
observed clumps masses.

A different model to estimate clump masses has been presented
in \citet{boley-disruption}, which attempts to capture the
actual dynamics in a non-axisymmetric disk. The model was
verified against 3D radiative simulations of protoplanetary disks,
and more recently was also shown to match well the results
of fragmentation in high redshift galactic disks (Tamburello
et al. 2015). Fragmentation,
as seen from numerical simulations, does indeed occur
in spiral arms rather than directly from the axisymmetric
background flow (Mayer et al. 2004; Durisen et al. 2007).
Instead of a homogeneous, axially symmetric background flow,
the initial state is  a spiral density wave identified as
an overdensity whose strength is proportional to the local
Mach number of the flow, which leads to velocity gradients
that determine the region that can collapse. Considering
also finite thickness, namely that the spiral arm has
a vertical extent of order the disk scale height, they
obtain the following mass estimate:
\begin{equation}
M_f = 4\frac{c_s^3}{G \Omega f_g}
\end{equation}
with $c_s$ the sound speed, $\Omega$ the angular frequency
and $f_g$ a form factor to account for effects from self-gravity.
\citep{boley-disruption} estimated $f_g \approx 1.8 Q$.
This is compatible with the considerations in \citet{deng-ogilvie22} where they suggested 
a solitary ring structure as a transitory state
in which spiral density waves would emerge, and then
collapse would eventually ensue in the flow entrained
by them.
Measuring $Q$ in our simulations
by tracing back the clump's particles in time leads to
values of \hbox{$Q\approx 1.15$} in the early phases of the simulation.
We present the resulting estimate in fig. \ref{fig-toomre-mass}.
The resulting mass lies around $0.05 M_\text{jup}$, 
a bit higher at the beginning of the simulation.
While this estimate lies indeed much closer to the resulting 
clump masses, any eventual effect of the magnetic field
is not taken into account.
The lower-end of the clump masses is still well below the estimated value.
Also, when tracing back the particles of each clump individually
and determining the estimated mass separately, namely without averaging,
the mass estimates do differ significantly from the 
initial masses of the fragments.

%S: Höhe der Disk, Destabilisierung durch B, Turbulenz

%\subsection{Integrated model}

\subsection{Energetics of clump-forming sites before the collapse}
\label{ch-energy}
As we seek the reason for the higher fragmentation rate and the lower
mass fragments in the MHD case one should explore how the magnetic field
itself can influence the fragmentation process.
It could either directly change the physics during the collapse or indirectly, for example by stirring turbulence. Indeed, in their
study of the mean flow properties of self-gravitating disks with
and without magnetic fields, Deng et al. (2020) showed that 
magnetized disks are more turbulent relative to unmagnetized disks
because Maxwell and gravitational stresses concur to generate 
a larger overall stress, resulting in enhanced angular momentum
transport.
In either case, the magnetic field  is expected to affect the dynamics
of the material because at a minimum an additional force, namely
the Lorenz force, enters the equations of motion of fluid elements.
Therefore it is important to know the relative contribution of the
magnetic field, and of turbulence, to the energetics of those
regions of the disk, along spiral arms, that will turn into
clumps.
%It is certainly interesting to know the relative

\begin{figure}
\centering
\begin{subfigure}{\columnwidth}

	\includegraphics[width=\textwidth]{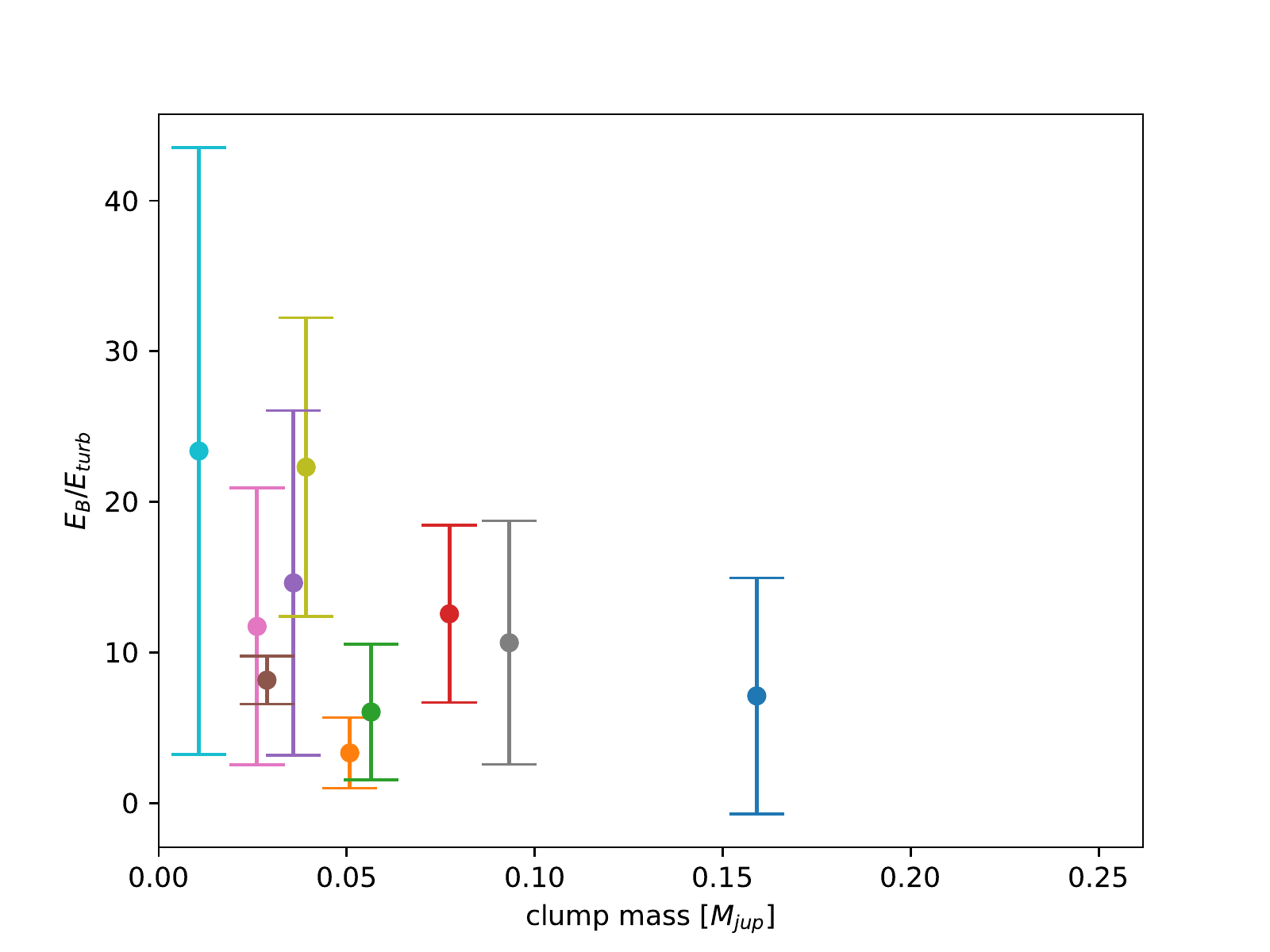}
\caption{Magnetic energy relative to turbulent 
energy}
\label{fig-bt_mT}
\end{subfigure}

\begin{subfigure}{\columnwidth}
	\includegraphics[width=\textwidth]{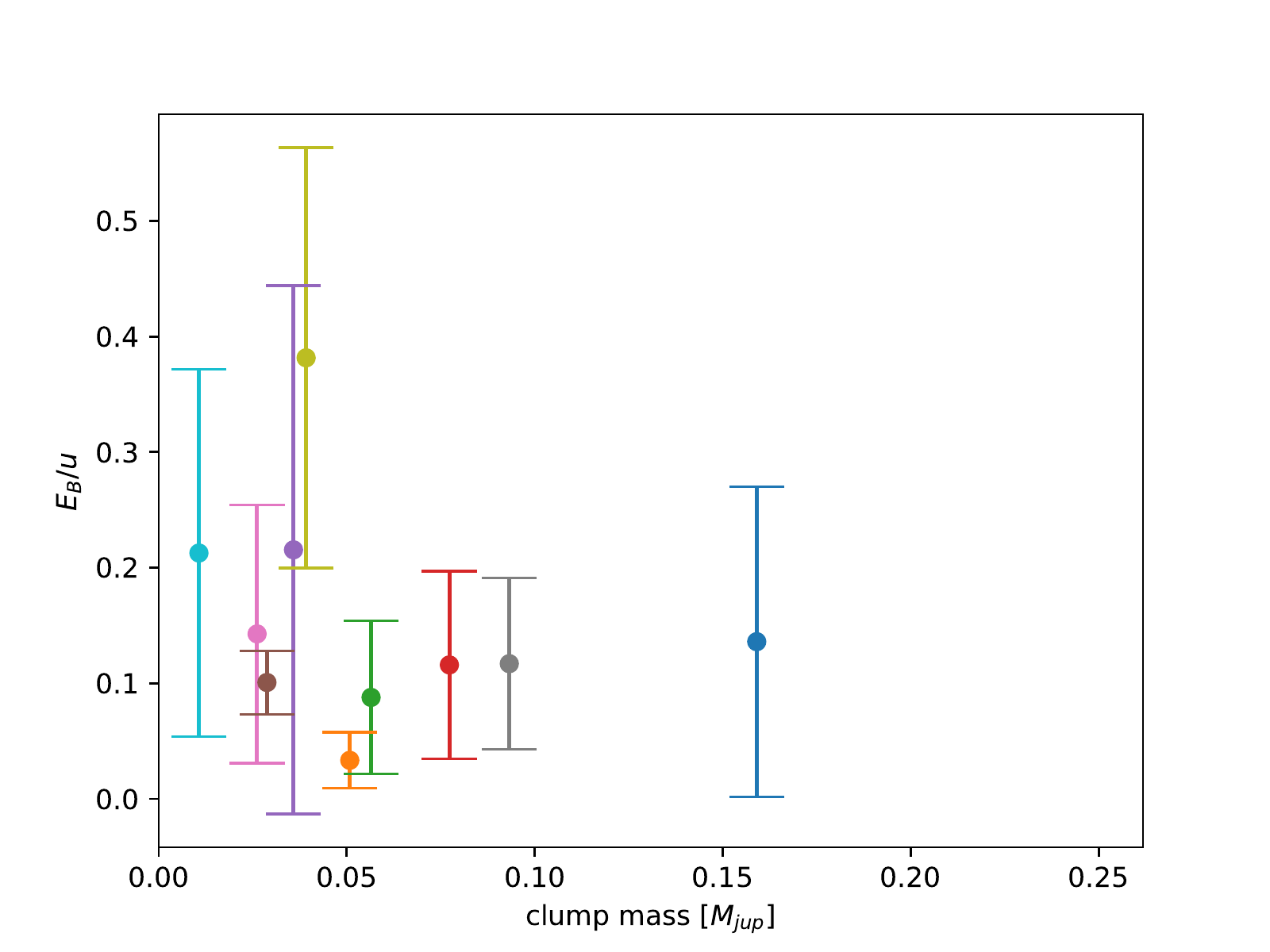}
\caption{Magnetic energy relative to internal energy  ($\beta_\text{plasma}^{-1}$)}.
\label{fig-bt_mU}
\end{subfigure}

\caption{
    Magnetic energy relative to turbulent 
energy and relative to internal energy, averaged 15 snapshots (corresponding to $\approx 24 \text{yr}$) before up until 
the existence of a bound structure.
Each dot represents a clump, the bars show the standard deviation
of observed values.
The magnetic energy is roughly an order of magnitude greater than the turbulent kinetic energy
compatible with a direct effect on fragmentation.
The internal energy is larger than the magnetic energy for all clumps.
}
\end{figure}

%First, we want to compare the magnetic field to the turbulent
%kinetic energy.
In fig. \ref{fig-bt_mT} the relation between the magnetic and the turbulent kinetic
energy is shown for the individual clumps.
The specific magnetic energy density is calculated via
$E_B = \frac{B^2}{4\pi \rho}$
where $B$ is the magnetic field and $\rho$ the density.
To quantify the turbulent kinetic energy, we first defined the velocity dispersion of a particle.
We used kernel-smoothing to calculate a mean velocity 
around a particle $i$ using a number of neighbours of $n_\text{smooth} = 32$,
\hbox{$\langle v_i\rangle = \sum_j v_j W\left(\frac{x_j - x_i}{h}\right) \frac{m_j}{\rho_j h^3}$}.
Here $W$ denotes the smoothing kernel
for which we used the cubic spline \citep{monaghan},
$h$ the smoothing length, $x_i$, $m_i$, $\rho_i$ the position, mass and density of particle $i$.
To arrive at the velocity dispersion we smoothed over the square deviation 
of the mean velocity $\langle v_i\rangle$:
\begin{equation}
\hbox{$\sigma_i = \sqrt{\sum_j \left(v_j - \langle v_i\rangle\right)^2 W\left(\frac{x_j - x_i}{h}\right) \frac{m_j}{\rho_j h^3}}$}.
\end{equation}
%We define the latter as the energy in the local deviation from a 
%mean flow and smooth it with a kernel.
%\begin{equation}
%\sigma_{\text{turb}, i}^2 = \Sigma_i \lvert v_i^2 - \langle v_i\rangle^2\rvert
%\end{equation}
%where $\langle v_i\rangle$ denotes the average velocity
%in the vicinity of a particle calculated with a kernel
%of $n_\text{smooth}$ = 32 and the sum goes over the coordinates.
%We take the absolute value coordinate-wise to prevent
%the value to become negative.
Then we calculated the turbulent kinetic energy via
\hbox{$E_{\text{turb}, i} = \frac{1}{2} m_i \sigma_{i}^2$}.

Similarly, in fig. \ref{fig-bt_mU} the relation between
the magnetic and the internal energy is shown.
The specific internal energy is available directly as a result of 
the simulation.
%In fig. \ref{fig-bt_mT} the relation between the magnetic and the turbulent kinetic
%energy is shown.
In both plots, we concentrate on the clumps 15 snapshots (corresponding to $\approx 24 \text{yr}$) of the simulation
before they become bound since we are interested in 
the influence of the magnetic field as a precondition of the fragmentation
process.
The results are then shown together with the mass of the respective clumps.

To trace the particles
back in time,
we used the first method described in section \ref{ch-identification}
namely we followed the density maxima as the centres of the distribution of particles
associated to a given clump backwards
in time. 
Despite of the clumps not yet being bound at earlier time, 
this method still traces the clump-forming regions  until a density maximum
can be defined and identified (see section \ref{ch-identification}).

It can be seen that the magnetic energy is larger than the
turbulent kinetic energy for all the clumps.
For most of them the difference is around an order of magnitude.
This means that the energy stored in the magnetic field
is much greater than in the turbulent motion of particles.
Therefore knowing the structure of the magnetic field 
is important to understand the gas motion.

%It can be seen that the magnetic energy is of a comparable order
%as the kinetic energy at this time,
However, when comparing with the internal energy of the gas, the 
magnetic field is significantly smaller.
Therefore the thermal gas pressure is still the most important of the quantities
considered so far.  We anticipate that, except in the inner regions
of the clumps, the total kinetic energy, including the non-turbulent
component of the velocity field, is the dominant contribution
counteracting gravitational potential energy, because clumps are rapidly
rotating. This will be studied in section \ref{ch-rotation}. In the interior of clumps,
instead, internal energy is always the main component establishing equilibrium,
as it will be further assessed in section \ref{ch-rotation}.
However, from \citet{deng1} we know that the presence of magnetic fields 
change the dynamics of the flow: first, the magnetic field ignites
turbulence in the disk and second, the presence of a magnetic field leads
to smaller-scale structures in the disk.
The question remains if the differences in fragmentation in the MHD case
is a direct consequence of the presence of the magnetic field during
this process or if it is a secondary effect arising from different
features of the environment.
The result from fig. \ref{fig-bt_mT} would be compatible with 
the magnetic field directly impacting fragmentation since it carries
much more energy in itself than the turbulent component of the kinetic
energy.
In section \ref{section-instability} we discuss a possible path
of how the magnetic field could affect fragmentation.
%Details of the clump formation observed in the simulations are shown 
%in the next two subsections where we also show more detailed radial
%profiles of the clumps.

%\todo[]{Change figure with new turbulence}

\subsection{Energetics of the clumps after collapse; the role of the magnetic field}
\label{ch-magnet}

In this section we want to give an overview 
of the evolution of the clumps during their early stages
right after they become bound
and to investigate the role of the
magnetic field during this time. %role of the magnetic field
For that, we focus on the magnetic, but also
%We focus on the magnetic but also on 
the turbulent kinetic energy and the internal energy of the gas.
%Further, we show the establishment of a magnetic shield around 
%a clump as it was observed in \cite{deng}.
It was already mentioned in \citet{deng} that
the magnetic field is amplified in and around the clumps
and may shield them from further growth
but may also prevent their disruption because
of its relative strength compared to the kinetic energy.
We also look at this amplification effect more closely in this section.

We now follow the evolution of a typical clump in the MHD simulations.
For that, we present two-dimensional cuts and radial profiles of the density, the magnetic field and the velocity field.
We choose clump nr. 5 from fig. \ref{fig-disk-density} as it belongs to the lower-mass end of the clumps distribution ($0.03 M_\text{jup}$),
our implicit assumption being that lower mass clumps should be most affected by the magnetic field as they are absent
in the non-magnetized case (HD simulations).
%XXX give reason for the choice of numbering -- while clump 5? Otherwise you can just skip calling it "5".
We start at a simulation time of 127 yr. %snapshot 80
At this time, some of the clumps are already bound while others are still forming.
Clump 5 is just becoming bound.
It is forming in a filament structure of increased density, along with three other clumps (0, 1 and 8).
In our subsequent analysis we are interested in the configuration of the magnetic field
and how it could affect the clump and its surroundings.

\begin{figure}
\centering
\begin{subfigure}{0.23\textwidth}
	\includegraphics[width=1.0\textwidth]{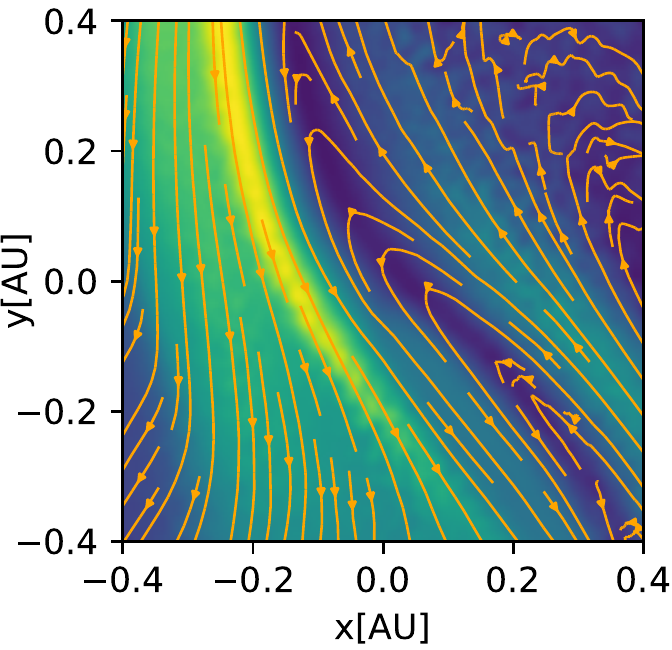}
	\caption{horizontal cut with magnetic field %
	lines, clump 5, 127 yr}
\end{subfigure}
\hfill
\begin{subfigure}{0.23\textwidth}
	\includegraphics[width=1.0\textwidth]{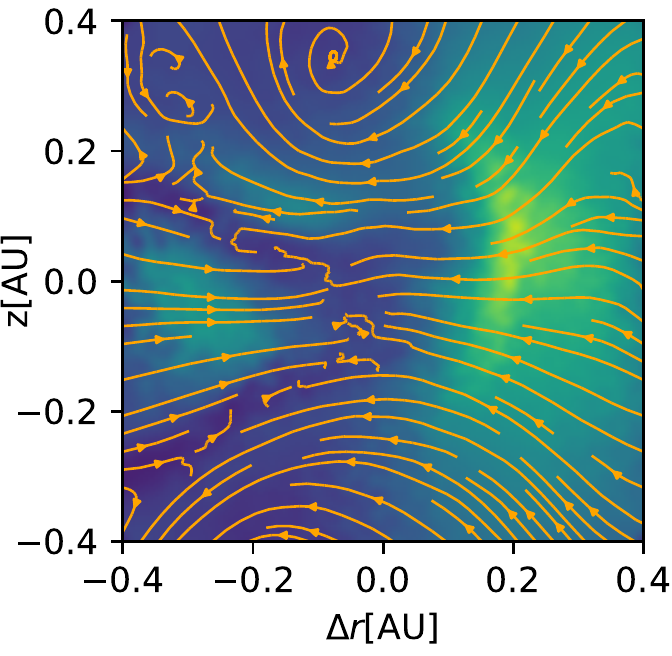}
	\caption{vertical cut with magnetic field % 
	lines, clump 5, 127 yr}
\end{subfigure}

\vspace{4mm}
\centering
\begin{subfigure}{0.23\textwidth}
	\includegraphics[width=1.0\textwidth]{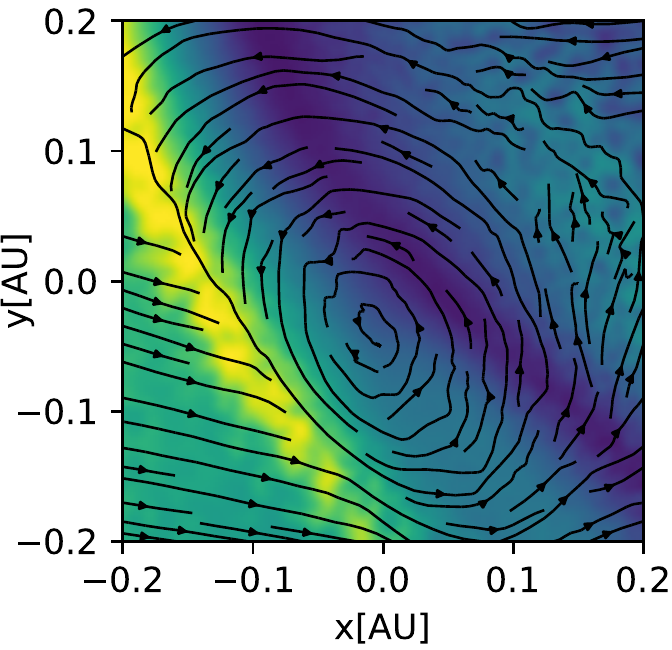}
	\caption{horizontal cut with velocity field 
	lines, clump 5, 127 yr}
\end{subfigure}%
\hfill
\begin{subfigure}{0.23\textwidth}
	\includegraphics[width=1.0\textwidth]{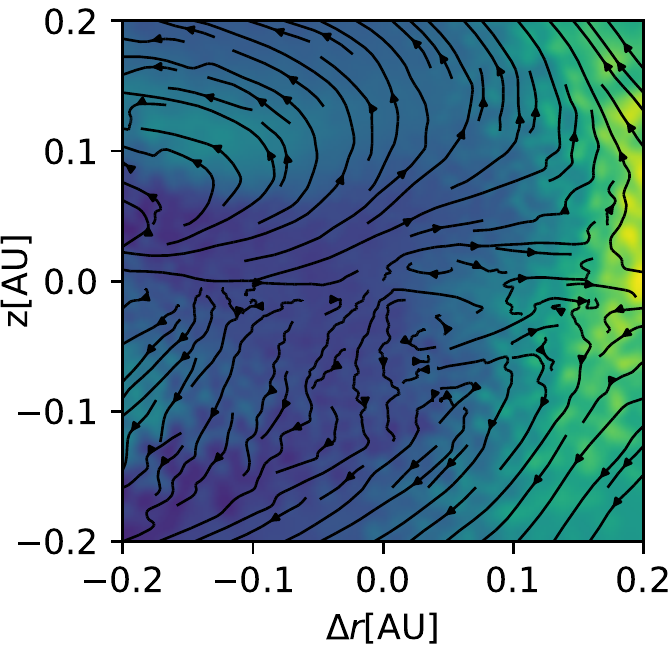}
	\caption{vertical cut with velocity field 
	lines, clump 5, 127 yr}
\end{subfigure}

\vspace{4mm}
\begin{subfigure}{0.46\textwidth}
\centering
	\includegraphics[width=0.8\textwidth]{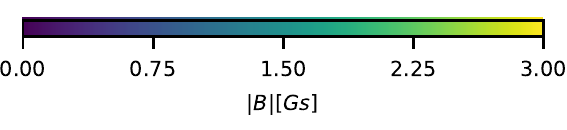}

\end{subfigure}
\caption{Evolution of clump 5: First bounded stage.
\textbf{Top:} magnetic field strength and magnetic field lines in a region
of $0.4 \text{AU}$ around the clump.
The horizontal cut on the left is aligned with the disks plane while
the vertical cut on the right shows the 
height $z$ over the disk's radial coordinate $\Delta r$.
\textbf{Bottom:} The same location but showing the velocity field lines 
(again magnetic field strength in the background) instead of the
magnetic field lines in a region of $0.2 \text{AU}$.
On both sides of the high-density filament, the magnetic field
is increased.
When comparing the magnetic field energy to the internal energy,
one arrives roughly at a value of $\beta_\text{plasma}^{-1}\approx 0.5$ for the outer regions in the plot.}
\label{fig-evolution-begin}
\end{figure}

\begin{figure}
\centering
\begin{subfigure}{0.23\textwidth}
	\includegraphics[width=1.0\textwidth]{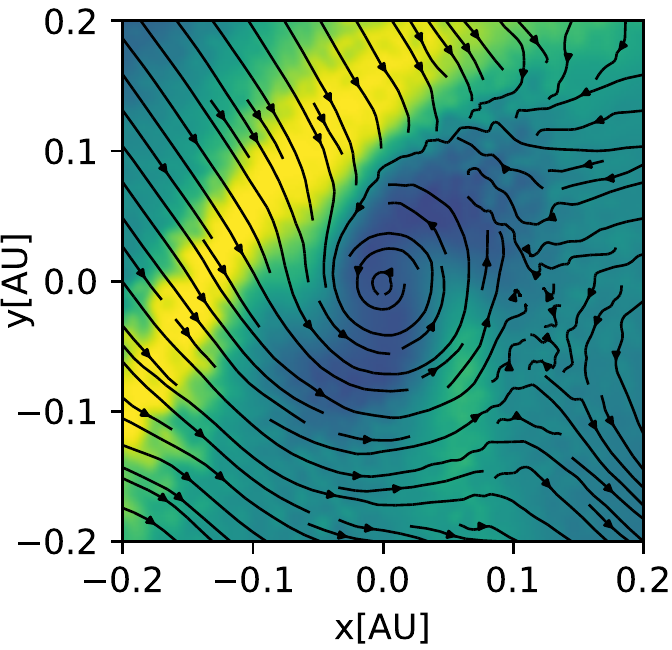}
	\caption{horizontal cut with velocity field lines,
	clump 5 at 141 years}
\end{subfigure}
\hfill
\begin{subfigure}{0.23\textwidth}
	\includegraphics[width=1.0\textwidth]{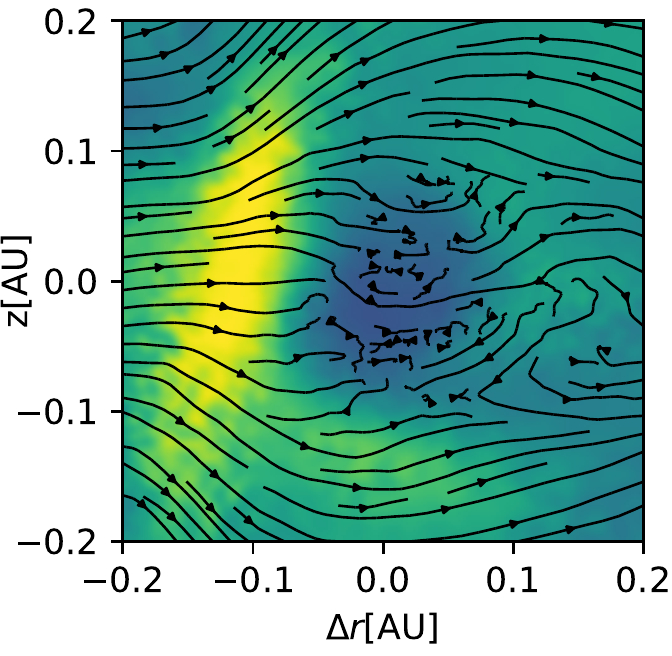}
	\caption{vertical cut with velocity field lines,
	clump 5 at 141 years}
\end{subfigure}

\vspace{4mm}
\begin{subfigure}{0.23\textwidth}
	\includegraphics[width=1.0\textwidth]{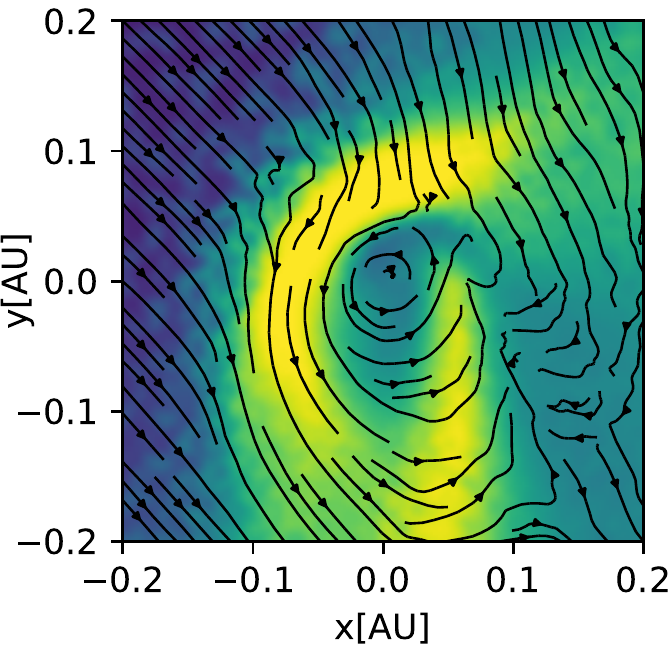}
	\caption{horizontal cut with velocity field lines,
clump 5 at 143 years}
\end{subfigure}
\hfill
\begin{subfigure}{0.23\textwidth}
	\includegraphics[width=1.0\textwidth]{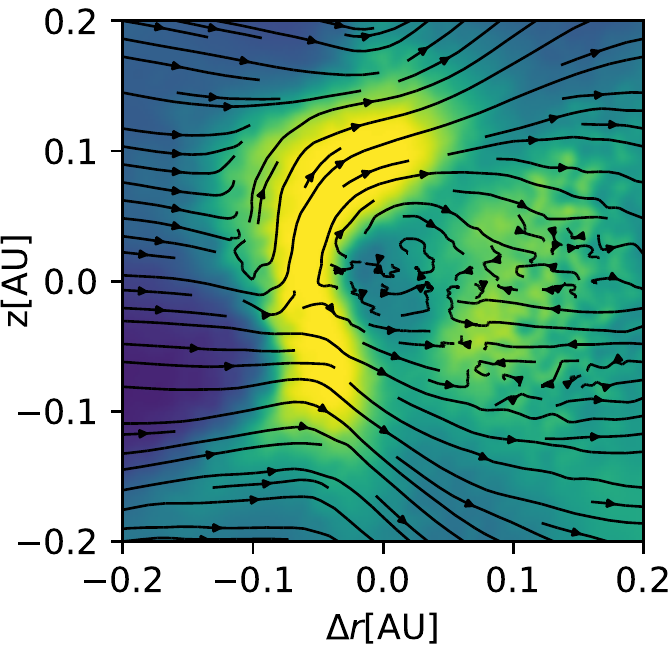}
	\caption{vertical cut with velocity field lines,
	clump 5 at 143 years}
\end{subfigure}

\vspace{4mm}
\begin{subfigure}{0.46\textwidth}
\centering
	\includegraphics[width=0.8\textwidth]{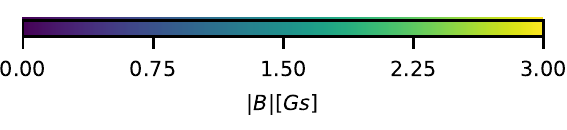}

\end{subfigure}

\caption{Evolution of clump 5: Build-up of a magnetic shield.
The plots on the left are horizontal cuts in the disk's plane, 
the plots on the right are vertical plotting the height $z$ and 
and the radial coordinate w.r.t. to the central star.
They show the magnetic field strength with the velocity
field lines at two consecutive snapshots (top: $142$ yr, bottom: $143$ yr).}
\label{fig-evolution-shield}
\end{figure}

Fig. \ref{fig-evolution-begin} shows the configuration of clump 5 at this time.
In the following two-dimensional profiles the clump is always centred using its
density maximum.
In the top of fig. \ref{fig-evolution-begin} the magnetic field strength is shown
around a region of $\pm 0.4\text{AU}$.
The clump will have a radial extension of $\approx 0.1\text{AU}$ (see e.g. Fig. \ref{fig-evolution-shield}).
The orange lines show the magnetic field lines.
In the horizontal cut on the left which is made parallel to the disk, 
it can be seen that the centre of the 
clump lies in an elongated region of low magnetic field strength
which is also of a higher density.
To the side of this region the magnetic field increases.
This effect of an increasing magnetic field along a thin elongated region
of higher density can also be observed around other clumps.

Also, contrary to the other clumps, here the magnetic field reverts its direction
when passing through the high-density filament.
%Also, in contrary to the other clumps, here the magnetic field
%changes its direction
%Notizartig
This can be explained by looking at the plots below
which show a smaller extract with the same centre.
Here, the velocity field (black lines) is drawn over the magnetic field strength:
The flow of the collapsing region moves inwards from two opposite sides,
thereby growing the high-density filament.
During this process magnetized material is transported close 
to the filament, enriching the magnetic field there.

\begin{figure}
\includegraphics[scale=0.5]{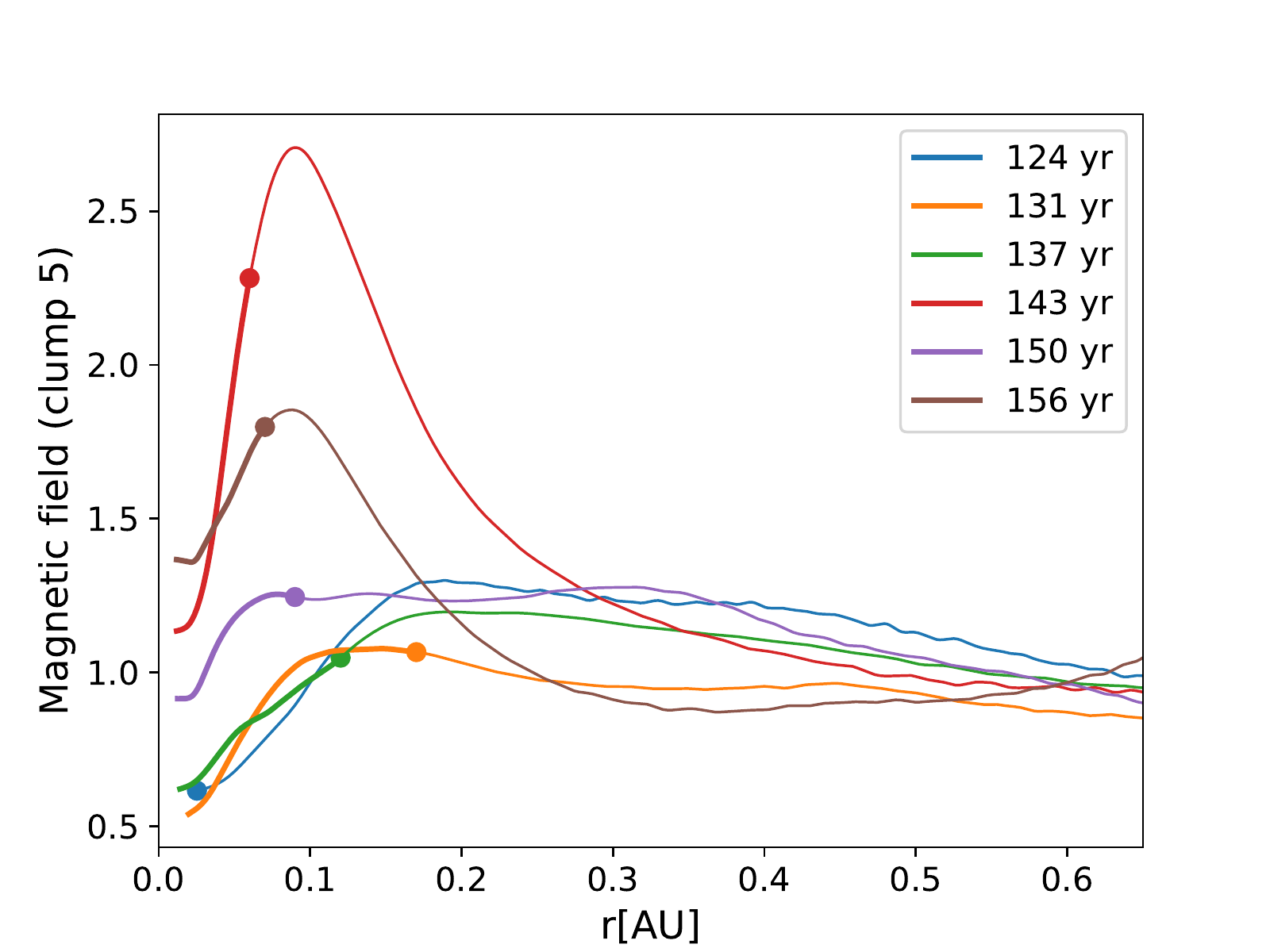}
\caption{Radial profile of the magnetic field strength (in Gs)
for clump 5 at various times.
The bound radius of the clump at each time is shown with a dot.
	At 143 yr, shortly after the clump
 became bound, a peak of the magnetic
	field develops in the region of the bound radius.
 Later, the magnetic field decays but begins
 to increase again at 156 yr.
 }
\label{fig-bpressure}
\end{figure}

%\begin{tcolorbox}
What magnetic field strength do we expect the filamentary structure
to have?
We assume that this structure emerged from a partial collapse in two directions
orthogonal to the direction of the filament.
Assuming also ideal MHD without resistivity, the flux through a surface
defined via any particles remains constant over time.
If we choose this surface to be orthogonal to the filament's elongated direction, 
then the area of the surface scales as $A\propto\rho^{-1}$ over time 
(since we assume no collapse in the elongated direction).
Since the conserved flux is defined as 
\begin{equation}
\Phi = \int B \cdot dS
\end{equation}
where the integral goes over the chosen surface,
the magnetic field $B$ has to scale as $B\propto\rho$.
The density at the boundary of the filament increases roughly four-fold.
So the magnetic field can be expected to also be four times as strong along the
filamentary structure.
This is about what is observed in fig. \ref{fig-evolution-density},
when assuming a background strength of the magnetic field
of $\approx 0.8 \text{Gs}$ (see also fig. \ref{fig-bpressure}).

%\end{tcolorbox}

While the magnetic field is increased at the boundary of the filament,
the low magnetic field region in the middle
probably arises because here the flow combines two regions of opposite
magnetic fields directions.
Further there is probably a higher gas pressure in this region because
of the higher density.
An effect of this can be seen at the bottom right of the figure where
%could arise because of the higher gas pressure in this high-density
%region. 
%This can be seen at the bottom right of the figure where
a vertical cut of the clump's region is shown.
In the vertical cut we show the height ($z$) and the change of the radial
component $\Delta r$ measured from the central star.
In this case, the prominent region of a strong magnetic field from
the horizontal cut on the left is now in the right side of the figure
at higher radii.
It can be seen that the flow escapes from the central region of higher density 
($z = 0 \text{AU}$) up- and downwards.
In the vertical cut on the right vortices of the magnetic field
can be seen above and below the clump.
Such vortices are also observed around other clumps.

Already $13$ years later, the development of a magnetic shield 
can be observed which is shown in fig. \ref{fig-evolution-shield}.
Again, horizontal cuts (left) and vertical cuts (right) are shown.
The magnetic shielding effect was already discovered in \citet{deng}
where it was found that the clumps were surrounded by regions of
high relative magnetic energy which control the flow around the clump
and thus shield it from disruption but also slow its further growth.
The top and bottom of fig. \ref{fig-evolution-shield} show the
situation at two consecutive snapshots, with $1.6$ years time difference
in between (compared to an orbital period of $23 $ years).
At this stage, the clump already developed a significant rotation and is quite
extended (see fig. \ref{fig-evolution-density} and section \ref{ch-rotation}).
The rotating flow drags the magnetic field around the clump thereby forming a
shield of increased magnetic field strength.
In the vertical cut on the bottom right it can also be seen that the
flow coming from the left does a sharp turn upwards when entering
the region of high-magnetic field strength.
This could reflect a deflection from the magnetic field.
In the later evolution, this magnetic field shield becomes weaker,
possibly due to Ohmic dissipation.
At later times it seems to regain strength.

In fig. \ref{fig-bpressure} we show the magnetic field strength for the same
clump as a radial profile at various times represented by the different
curves.
The radius $r$ measures the distance from the clump's centre – the bound
radius at each time is marked with a dot which is calculated as 
described in section \ref{ch-identification}:
At a time of 131 yr, the clump is first bound.
%Fig. 11 erklären und auch reinforcement erwähnen.
It can be seen that at a time of 143 yr, when we observed the magnetic shield,
a strong increase of the magnetic field is measured.
At $150$ yr, the magnetic field is again low but begins to increase
at $156$ yr.
Such an oscillating behaviour can also be seen for other clumps, however
mostly the peak is inside the bound radius.
%\begin{tcolorbox}
	What is the timescale for the Ohmic dissipation?
	Starting from eq. \ref{eq-ohm} and assuming zero velocity
	$v = 0$, one can estimate the time scale as:
	\begin{equation}
		\frac{\Delta B}{\Delta t_\text{diss}} \approx%
		\nu \frac{\Delta B}{\Delta r^2} \implies%
		\Delta t_\text{diss} \approx \frac{\Delta r^2 R_m\Omega}{c_s^2}
	\end{equation}
	where we substituted $\eta = c_s H/R_m$ with $H$ the scale height
	and $c_s$ the sound speed and used that $H\approx c_s/\Omega$.
	$\Omega$ is the rotational frequency around the central star.
	In this equation we take for $\Delta r$ the radial size of the 
	magnetic shield.
	When plugging in values that arise at the boundary of the filament
	one arrives at a dissipation time of roughly $t_B\approx 5 yr/2\pi$
	which means that changes in the magnetic field should be expected 
	between the snapshots which are taken every $10 yr/2\pi$.
%\end{tcolorbox}

We now
%It is now interesting to 
compare the magnetic energy to other
quantities to determine its importance. First, it is compared
to the turbulent kinetic energy.
The turbulent kinetic energy and the magnetic energy are measured as described in section
\ref{ch-energy}.

\begin{figure*}
\centering
\begin{subfigure}{0.33\textwidth}
	\includegraphics[scale=0.35]{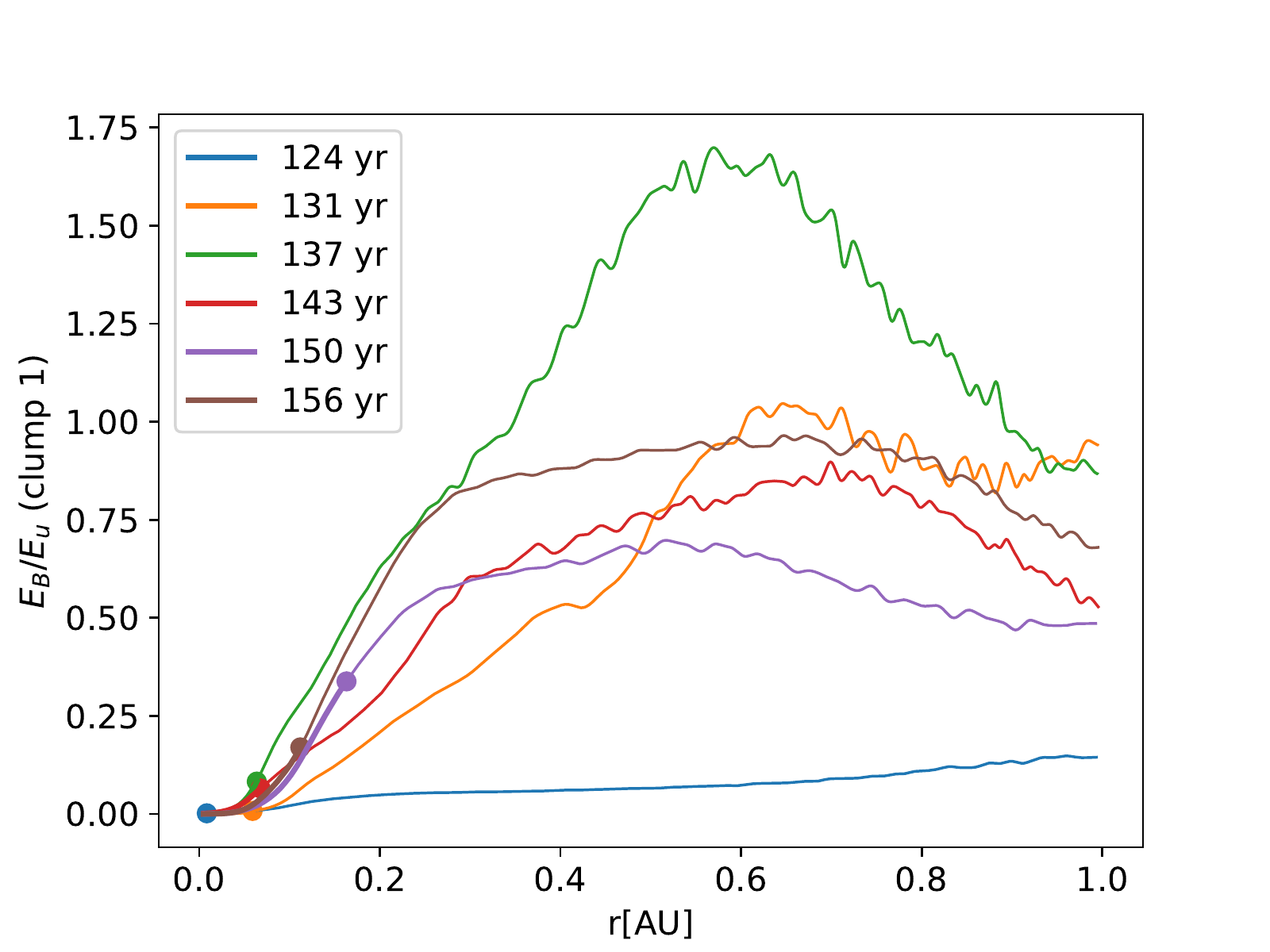}
	\caption{$E_B/E_u$ of clump 1}
\end{subfigure}%
\begin{subfigure}{0.33\textwidth}
	\includegraphics[scale=0.35]{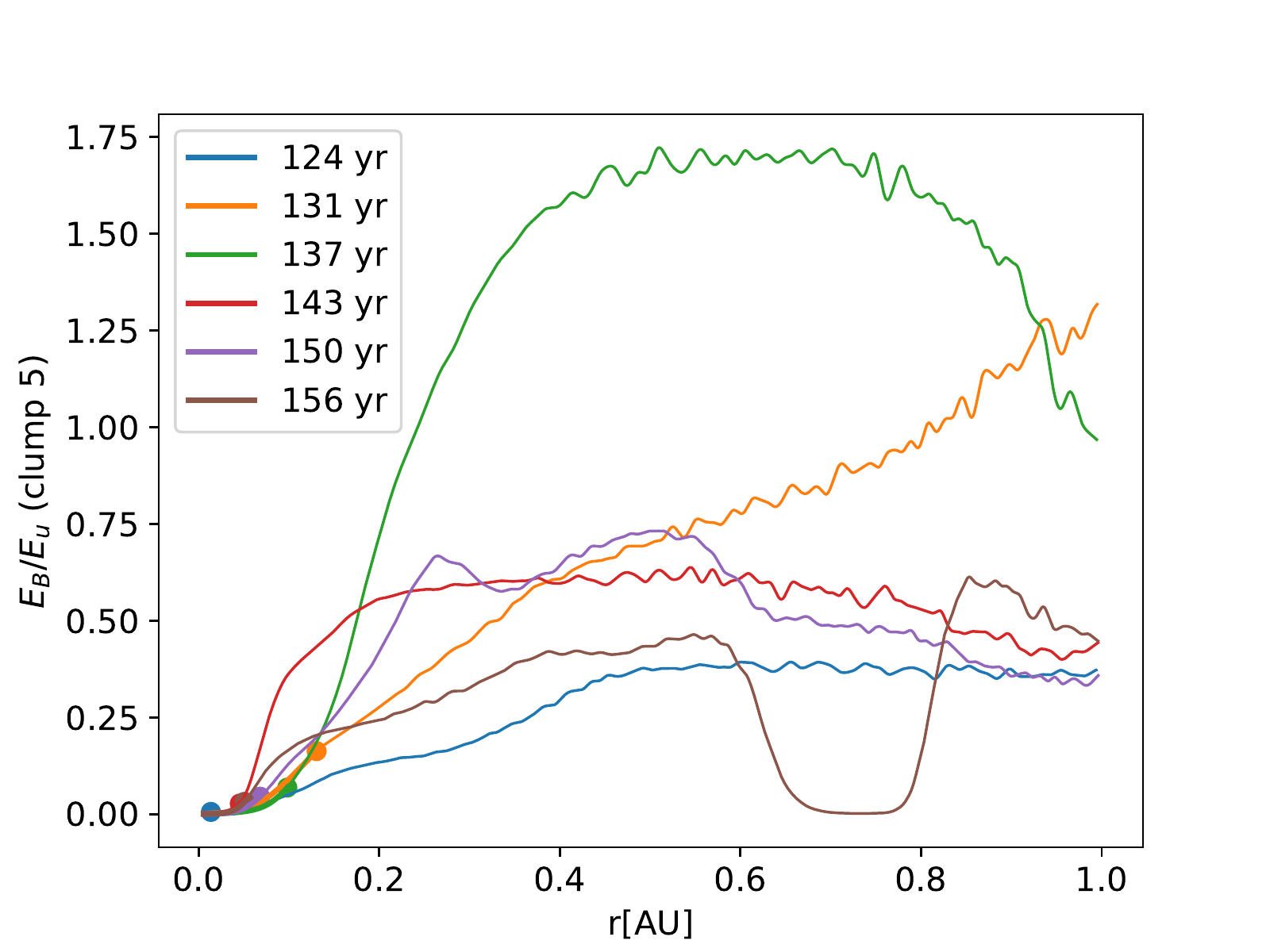}
	\caption{$E_B/E_u$ of clump 5}
\end{subfigure}%
\begin{subfigure}{0.33\textwidth}
	\includegraphics[scale=0.35]{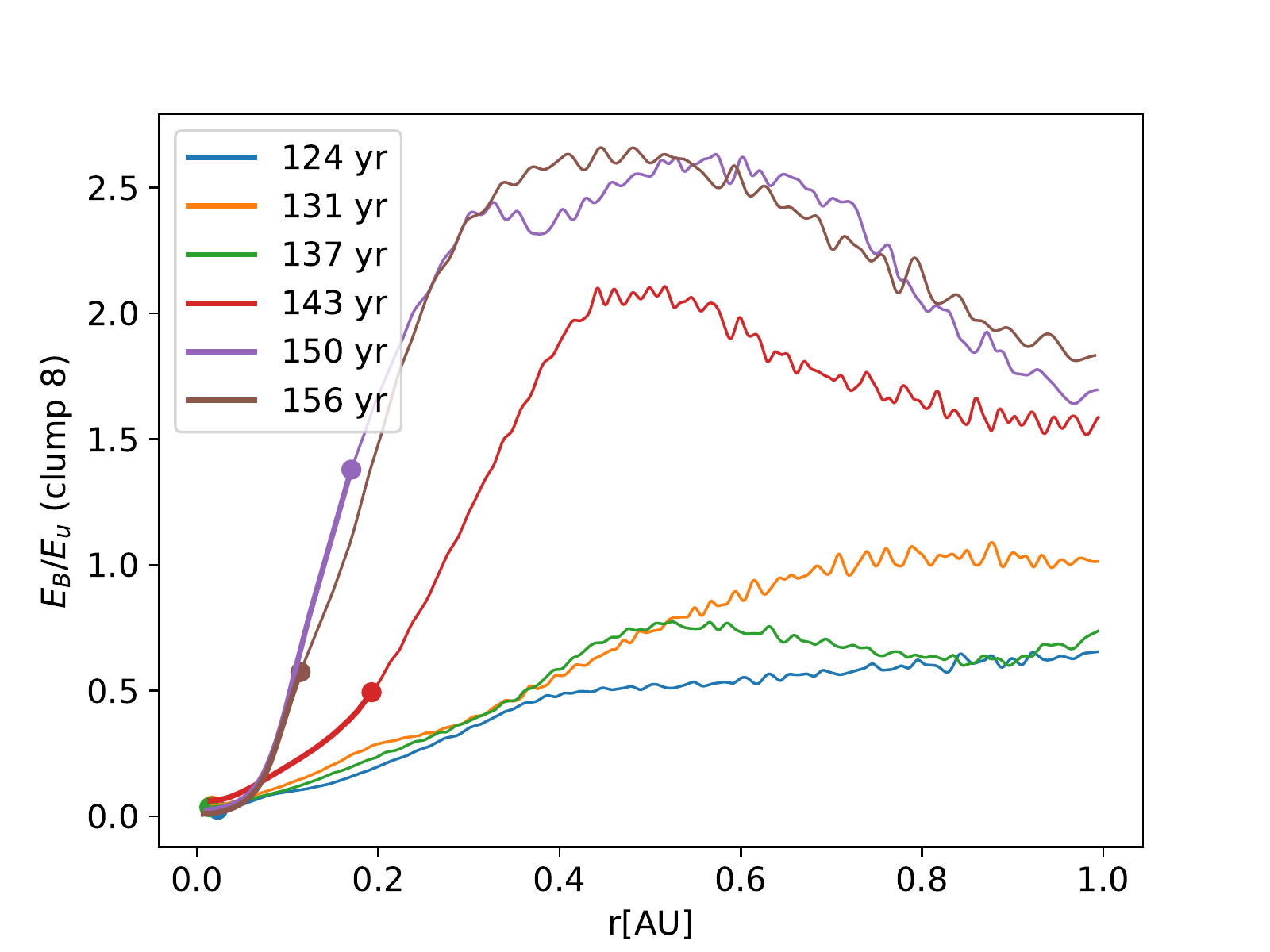}
	\caption{$E_B/E_u$ of clump 8}
\end{subfigure}
\begin{subfigure}{0.33\textwidth}
\centering
		\includegraphics[scale=0.35]{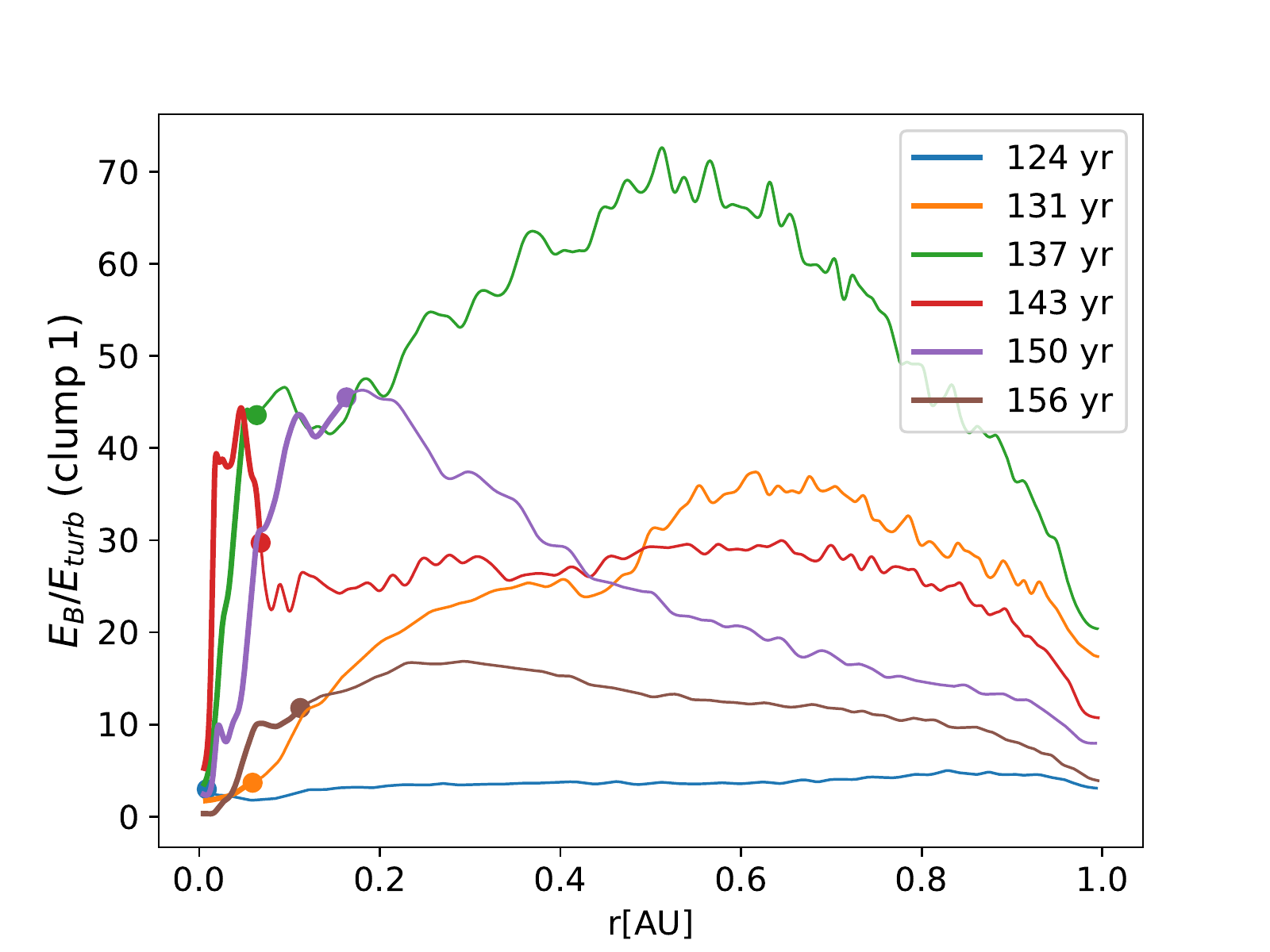}
	\caption{$E_B/E_\text{turb}$ of clump 1}
\end{subfigure}%
\begin{subfigure}{0.33\textwidth}
	\includegraphics[scale=0.35]{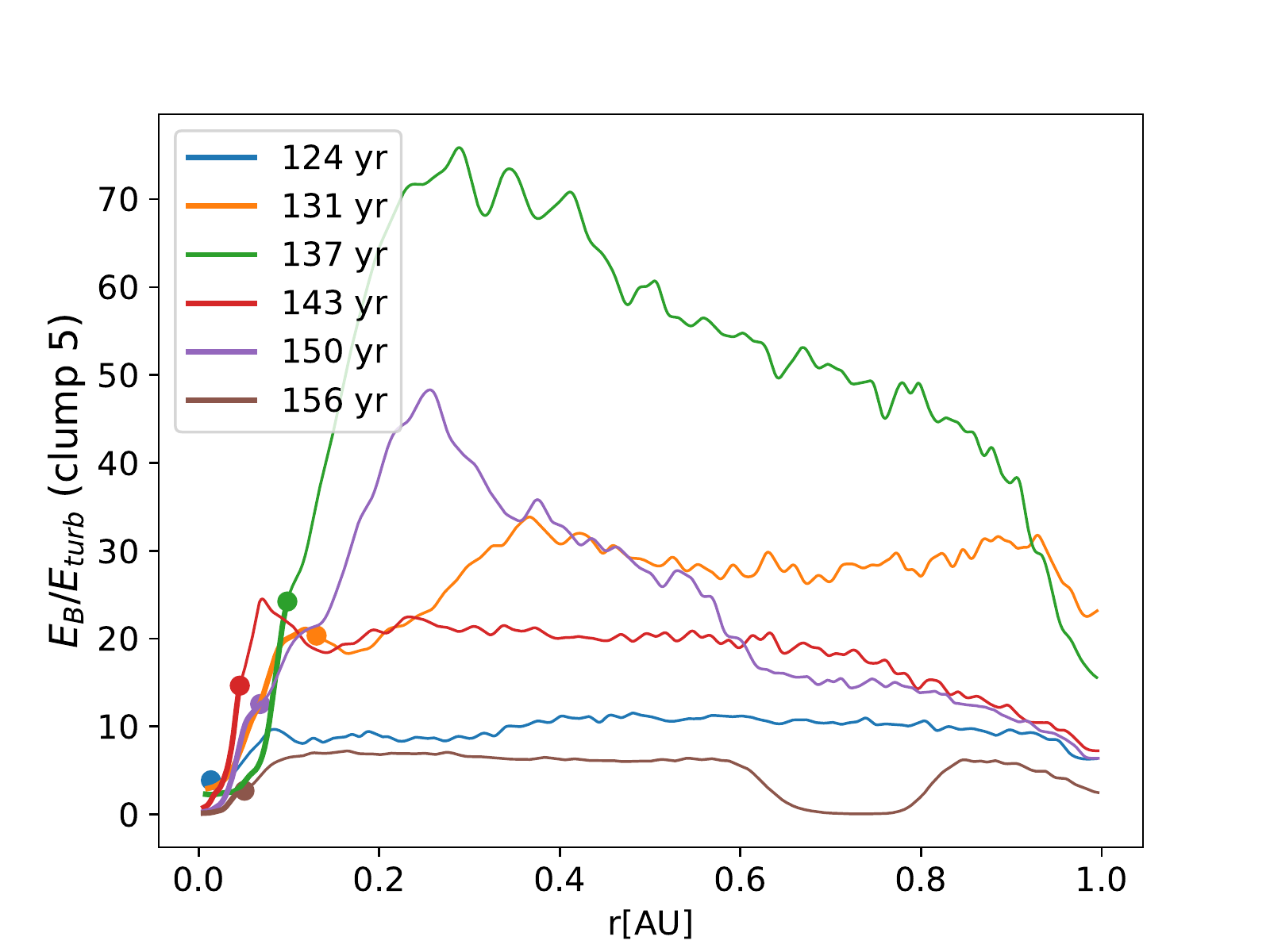}
	\caption{$E_B/E_\text{turb}$ of clump 5}
\end{subfigure}%
\begin{subfigure}{0.33\textwidth}
    \includegraphics[scale=0.35]{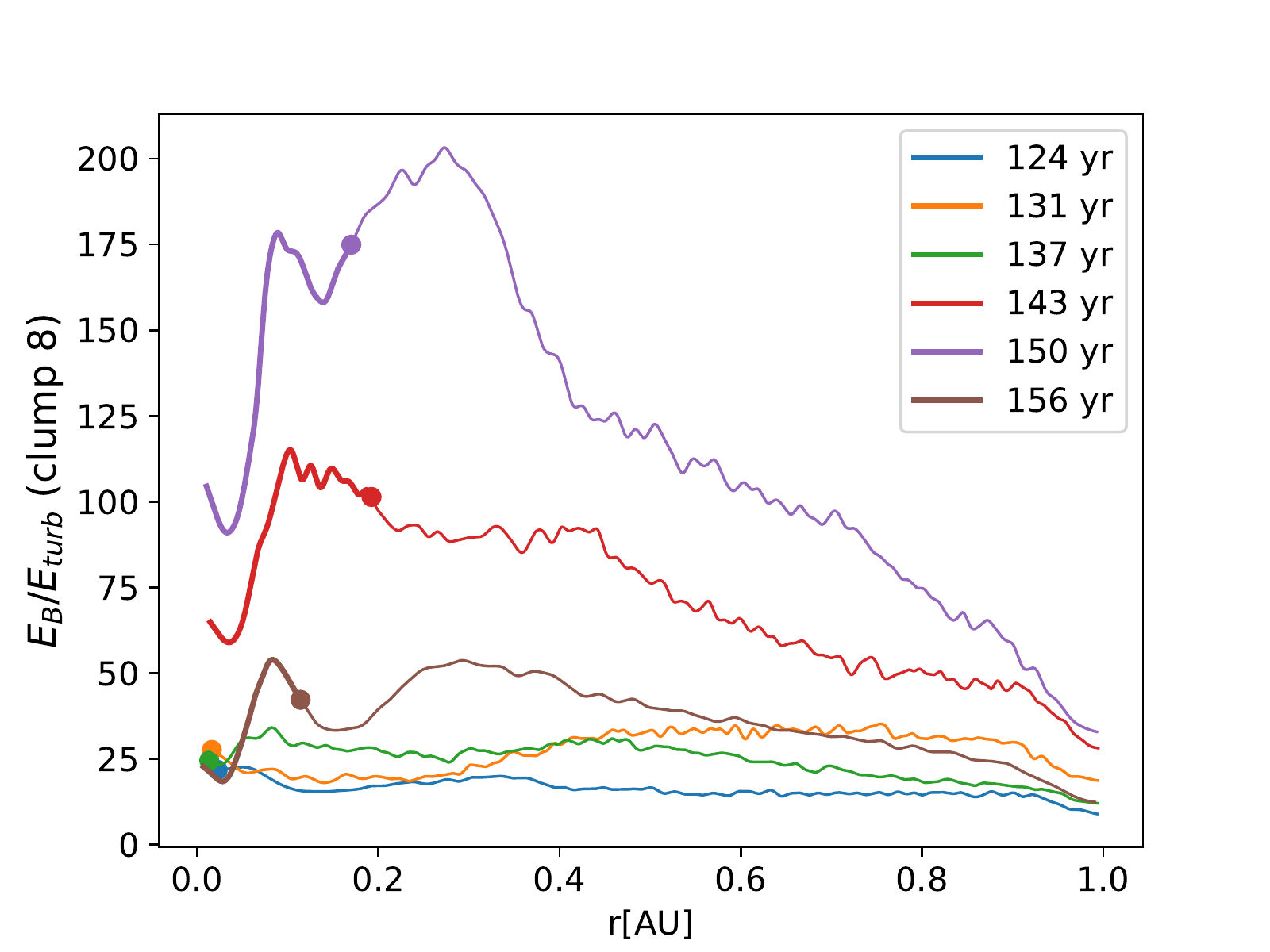}
	\caption{$E_B/E_\text{turb}$ of clump 8}
\end{subfigure}
\begin{subfigure}{0.33\textwidth}
\centering
    \includegraphics[scale=0.35]{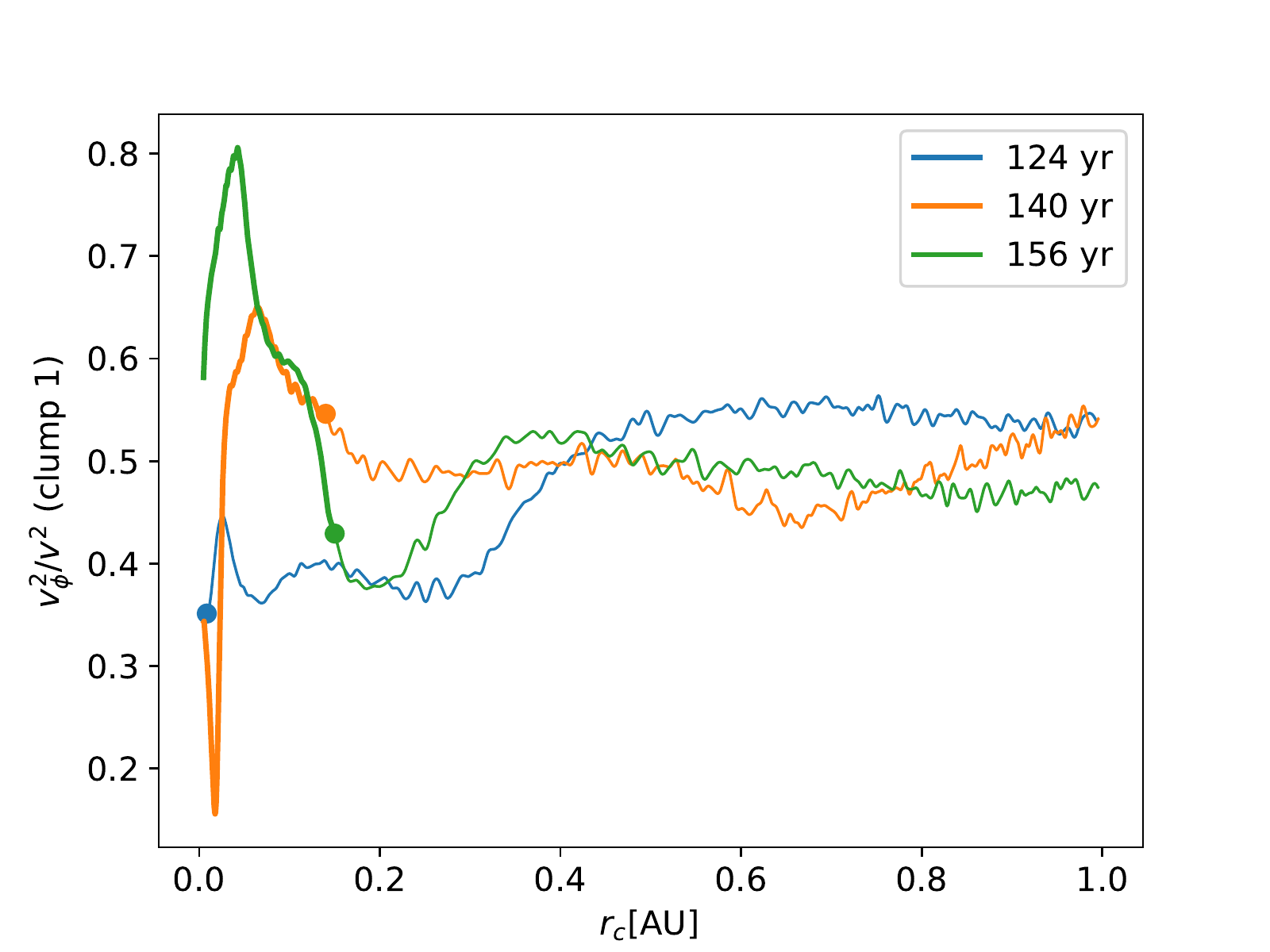}
	\caption{$v_\varphi^2/v^2$ of clump 1}
\end{subfigure}%
\begin{subfigure}{0.33\textwidth}
	\includegraphics[scale=0.35]{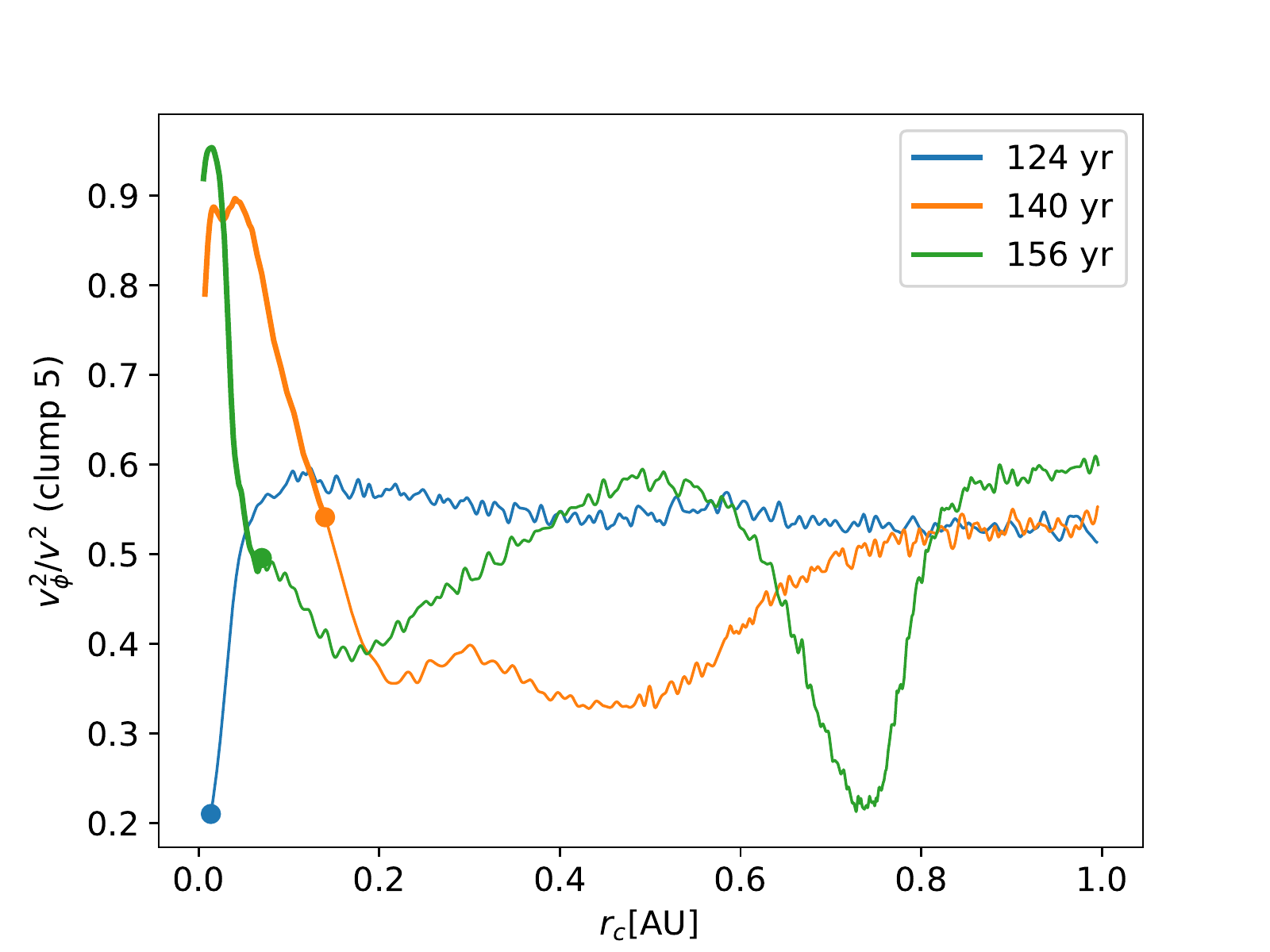}
	\caption{$v_\varphi^2/v^2$ of clump 5}
\end{subfigure}%
\begin{subfigure}{0.33\textwidth}
	\includegraphics[scale=0.35]{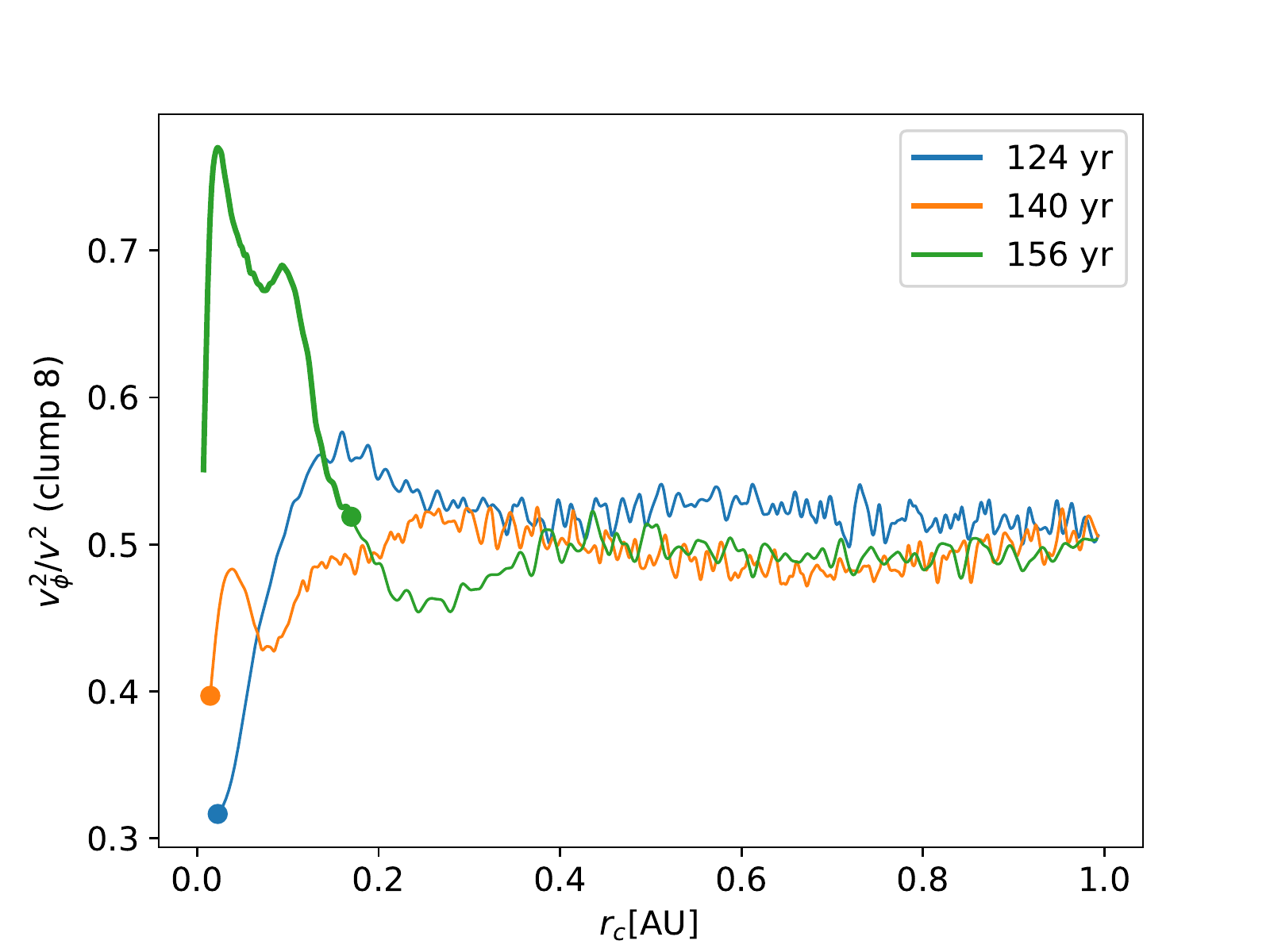}
	\caption{$v_\varphi^2/v^2$ of clump 8}
\end{subfigure}
\caption{Top: Magnetic energy density relative to internal energy density ($\beta_\text{plasma}^{-1}$).
In the vicinity of the clumps, the magnetic energy remains important 
when comparing to the internal energy.
Middle: Magnetic energy density relative to turbulent kinetic energy density. The magnetic energy is generally of a larger magnitude
than the turbulent kinetic energy, especially
at the later stages of the clump's evolution.
Often, this relation increases after the clumps have become bound.
Bottom: Relative rotational energy $v_\phi^2/v^2$.
The plots show radial profiles of the mentioned quantities at different times
and for three different clumps.
Inside the bound radius, the motion is mostly rotation-dominated.}
\label{fig-overview}
\end{figure*}

%Vergleiche mit kinetischer Turbulenz und innerer Energie.
%Komme auf nächstes Kapitel zu (Kräftegleichgewicht).
The middle row of fig. \ref{fig-overview} shows radial profiles of 
the magnetic energy relative to the turbulent kinetic energy:
Again, the different curves show the situation at different stages
in the clump's evolution.
We show the profiles up to a distance of $1 \text{AU}$ since
we are interested in the region close but clearly outside the clump.
This value corresponds to a distance a bit larger than twice as much 
as the maximum bound radius of a clump that we observe in the simulations.

It can be seen that generally the magnetic energy is more important
than the turbulent kinetic energy since it is of a larger magnitude.
This is especially true for the later stages after the clump has
become bound.
The great increase at \hbox{$137$ yr} of clump 5 is because the clump is in an
environment of low density and temperature.
Often an increase in the relation between the magnetic field and the
turbulent kinetic energy can be observed after the clumps have become bound.
For example, in the middle-right plot of fig. \ref{fig-overview},
which shows
the same relation for clump 8,
it is even more clear that after the clump becomes bound,
the magnetic field becomes significantly more important in its surrounding.

It is also interesting to compare the magnetic energy to the internal
energy which is responsible for the gas pressure forces.
This relation is shown in the top-center plot of 
fig. \ref{fig-overview} for clump 5 
and in the top-right plot for clump 8.
Here, it is also visible that the magnetic energy is important
compared to the internal energy.
Again, this is more pronounced for clump 8 than for clump 5
where there is a single spike of the relation at 137 yr.
The decrease in energy around $r=0.7\text{AU}$ at $156$ yr for clump 5 
arises because of another clump forming nearby. 
%(see density plot in fig. \ref{fig-clump-close}).

Both relations, internal energy density to magnetic energy density and 
turbulent kinetic energy density to magnetic energy density
tend to become smaller at small radii inside the clump's bound radius.
This is because of the greatly increased density at these regions.
Because of that, the same temperature yields a much higher internal energy density.
Vice versa, the turbulent kinetic energy density is also increased at this region.
In the next section we characterize the influence of various physical quantities
on the evolution of the 
clump and show their relative importance at different
locations in the clump.

\subsection{Rotation and clump dynamics}
\label{ch-rotation}
Until here, we focused on the properties of the flow 
in the very vicinity of the ensuing clumps.
Their evolutionary path depends also on their
internal properties which are investigated in this section.

\begin{figure}
\centering
\begin{subfigure}{0.23\textwidth}
	\centering
	\includegraphics[width=1.0\textwidth]{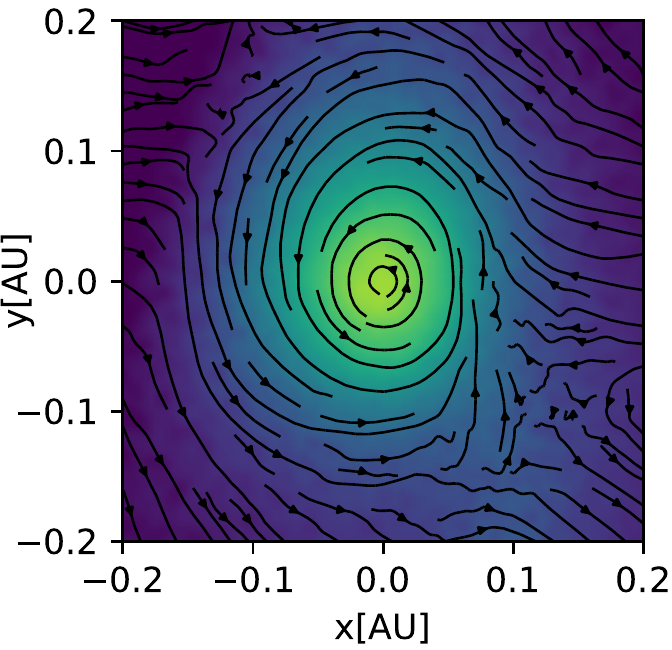}
	\caption{horizontal cut with velocity field lines, clump 5, 141 yr}
\end{subfigure}%
\hfill
\begin{subfigure}{0.23\textwidth}
	\centering
	\includegraphics[width=1.0\textwidth]{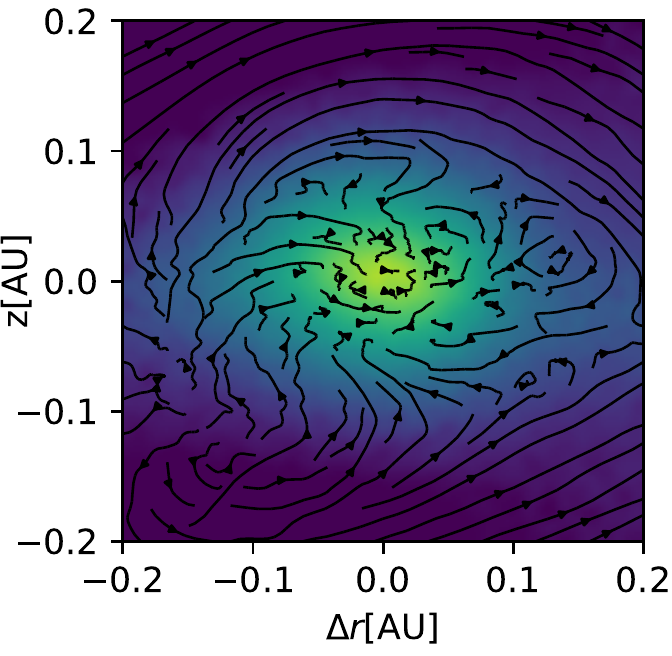}
	\caption{vertical cut with velocity field lines, clump 5, 141 yr}
\end{subfigure}

\vspace{4mm}
\begin{subfigure}{0.23\textwidth}
	\includegraphics[width=1.0\textwidth]{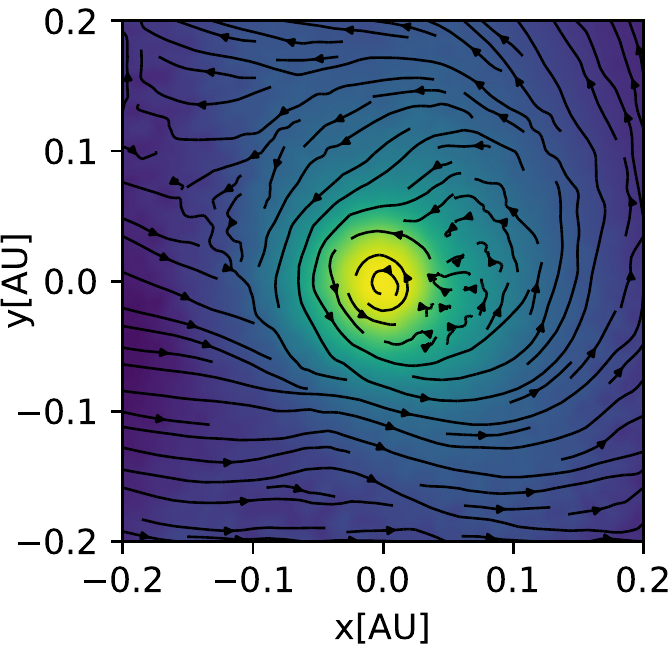}
	\caption{horizontal cut with velocity field lines, clump 5, 158 yr}
\end{subfigure}%
\hfill
\begin{subfigure}{0.23\textwidth}
	\includegraphics[width=1.0\textwidth]{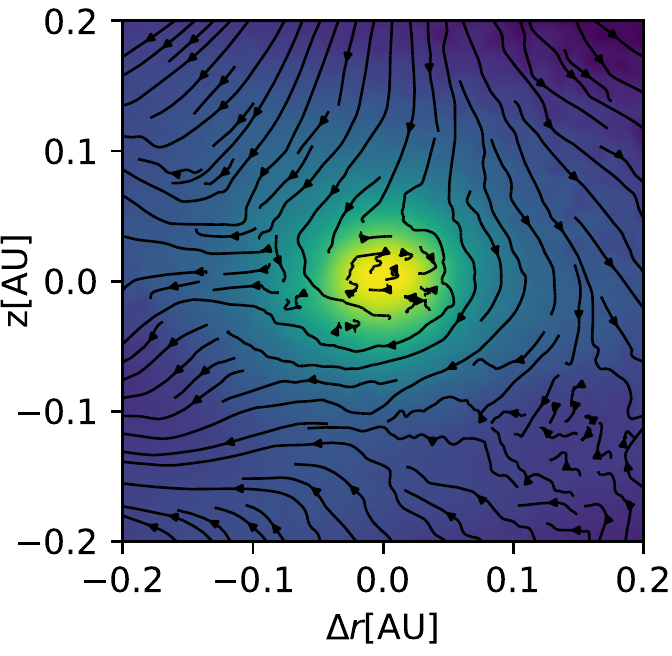}
	\caption{vertical cut with velocity field lines, clump 5, 158 yr}
\end{subfigure}

\vspace{4mm}
\begin{subfigure}{0.46\textwidth}
\centering
	\includegraphics[width=0.8\textwidth]{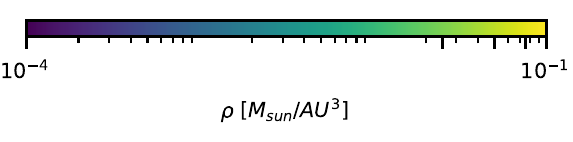}

\end{subfigure}
\caption{Evolution of clump 5: Density at early (top) and later (bottom) stage
(colour) and velocity field (black arrows).
The horizontal cuts on the left are aligned with the disk's plane
and show a contraction of the clump over time.
The vertical cuts on the right show that the clump is first elongated
along the disk's plane, whereas later it is more round.}
\label{fig-evolution-density}
\end{figure}

Rotation has often been reported as dynamically
important in clumps formed via disk instability
\citep{mayer-2004, galvagni, shabram, helled}
Therefore, we investigate its relevance in magnetized 
clumps as well. 
%This is important since rotation has the ability to stabilize
%the system since any long-term surviving clump has to 
%be in an equilibrium between gravity and stabilizing forces.
Additionally, the strength of rotation will also have
implications on
the rotation rate of an eventual planet resulting from further collapse.
Fig. \ref{fig-evolution-density} shows again the configuration
of clump 5 at two different times:
On top at 140 yr, 8 snapshots after the clump became bound
and at the bottom at 158 yr at the end of the analyzed simulations.
The plots show the density with the velocity vector field for
horizontal cuts aligned with the disk's plane (left) and for
vertical cuts perpendicular to the disk (right).

It can be seen at the top left, that the clump at this earlier
time has a wide-spread, almost elliptical region where rotation
around the clump's centre dominates the velocity field.
At this stage the clump is probably sustained %(charged, filled)
with rotation by the differential rotation 
of the surrounding gas of the disk.

In the vertical cut on the top right it can be seen that 
the clump at this stage has an almond-like shape that is
elongated along the mid-plane.
The surrounding material flows around this structure
from smaller to larger disk's radii.
In the interior the flow seems to be erratic without 
showing a preferred pattern.
The elongated shape could be a hint for the rotation to be
important for stabilization at this stage since it only 
exerts a force in the rotating plane. 
%Wie sieht es aus in einem grösseren Abschnitt (0,4 oder 0,6 AU)?

When looking at the later stages at the bottom, 
one sees that the shape of the clump has changed.
From the horizontal cut at the bottom left it
is visible that the clump has become denser than before
and also seems to be concentrated in a smaller region.
The velocity field only shows a clear rotating behaviour
in the inner parts with \hbox{radii $< 0.05 \text{AU}$}.
Further outside the velocity field still
suggests some rotation although the flow seems 
to be in a more undetermined state. 

The vertical cut at the bottom left shows that the clump
has become much rounder than before and is no longer
embedded in this almond-shaped high-density region.
The flow at this stage mostly comes from the upper and lower
end to the centre.

%Wichtig ist Gravitation und gas pressure. Zeige internal energy
%We now focus on the clump's internal properties.
Now the clump's rotation should be determined quantitatively.
As in section \ref{ch-magnet} we show one-dimensional radial
profiles of the clump.
Here it seems more appropriate to use a cylindrical coordinate system
since rotation is defined along an axis.
We start by determining the main rotation axis of the clump.
For that, we consider all particles inside the bound radius
and determine their total angular momentum relative to the
center of mass.
The normalized angular momentum vector gives us the z-component
of a cylindrical coordinate system.
This vector points in the same direction as the orientation 
of the protoplanetary disk.

In that coordinate system we determine 
the azimuthal velocity $v_\phi$ of the particles.
We then consider the squared relation to the total velocity
$v_\phi^2/v^2$
thereby comparing the rotational to the total kinetic energy.
Thus we can determine more quantitatively if the clump
exhibits rotating behaviour.
A radial profile of this relation is shown in the 
middle-bottom plot of fig. \ref{fig-overview}
for clump 5.
A value of $1$ means that all kinetic energy is purely in rotation,
if the energy is equipartitioned we would expect a 
value of  $1/3$.

At the time of the blue curve, before the 
cuts of fig. \ref{fig-evolution-density} the curve is flat over
a large region and sharpy decreasing in the interior.
This decrease seems to come from vertical infall of material
to the centre.
At this stage, the clump is not yet bound.

%First, the profile is flat (blue curve) meaning that no rotation is 
%present at this point.
At a later stage at $140$ yr (yellow curve) which represents
the time of the top plots in fig. \ref{fig-evolution-density} 
the curve begins to increase when approaching the centre.
Here, the clump is bound up until a radius of $\approx 0.15 \text{AU}$.
While at radii further out than $\gtrsim 0.2 \text{AU}$ the curve
is flat, inside it reaches values of $\approx 0.9$.
Inside the bound radius, the clump exhibits strong rotation.
At the even later stage at $156$ yr (green curve) which corresponds
to the time of the bottom plots in fig. \ref{fig-evolution-density} 
the spike is even narrower and higher.
Consistently, the bound radius is also smaller, somewhat below
$0.1 \text{AU}$, up to where the rotation is no longer dominating. 
This seems to indicate that the clump shrunk in radius
during this time.
This also confirms our method of measuring the bound radius 
as described in section \ref{ch-identification}.
In all observed clumps we measure the behaviour of increased
rotation near the centre after they are bound.
Often however, the bound radius is more extended than 
what one would expect from simply estimating where the rotation
curve becomes flat.
On the other hand, if in an inner region the rotation is 
significantly enhanced compared to outside, the clumps seem
to always be bound at least in that part.

\begin{figure}
\subfloat[][MHD] {%
	\includegraphics[scale=0.5]{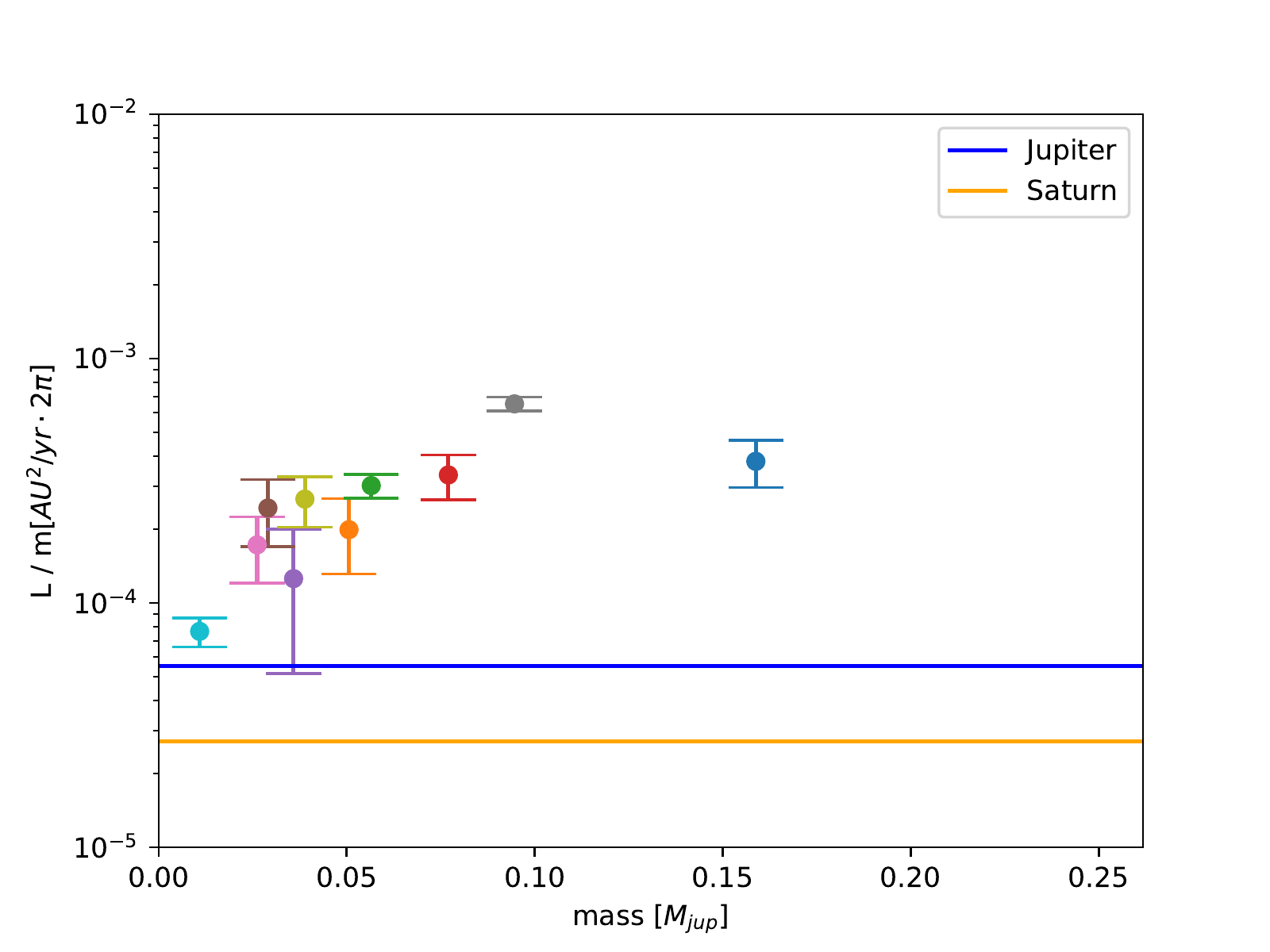}
}
\vspace{0.1mm}
\subfloat[][HD] {%
	\includegraphics[scale=0.5]{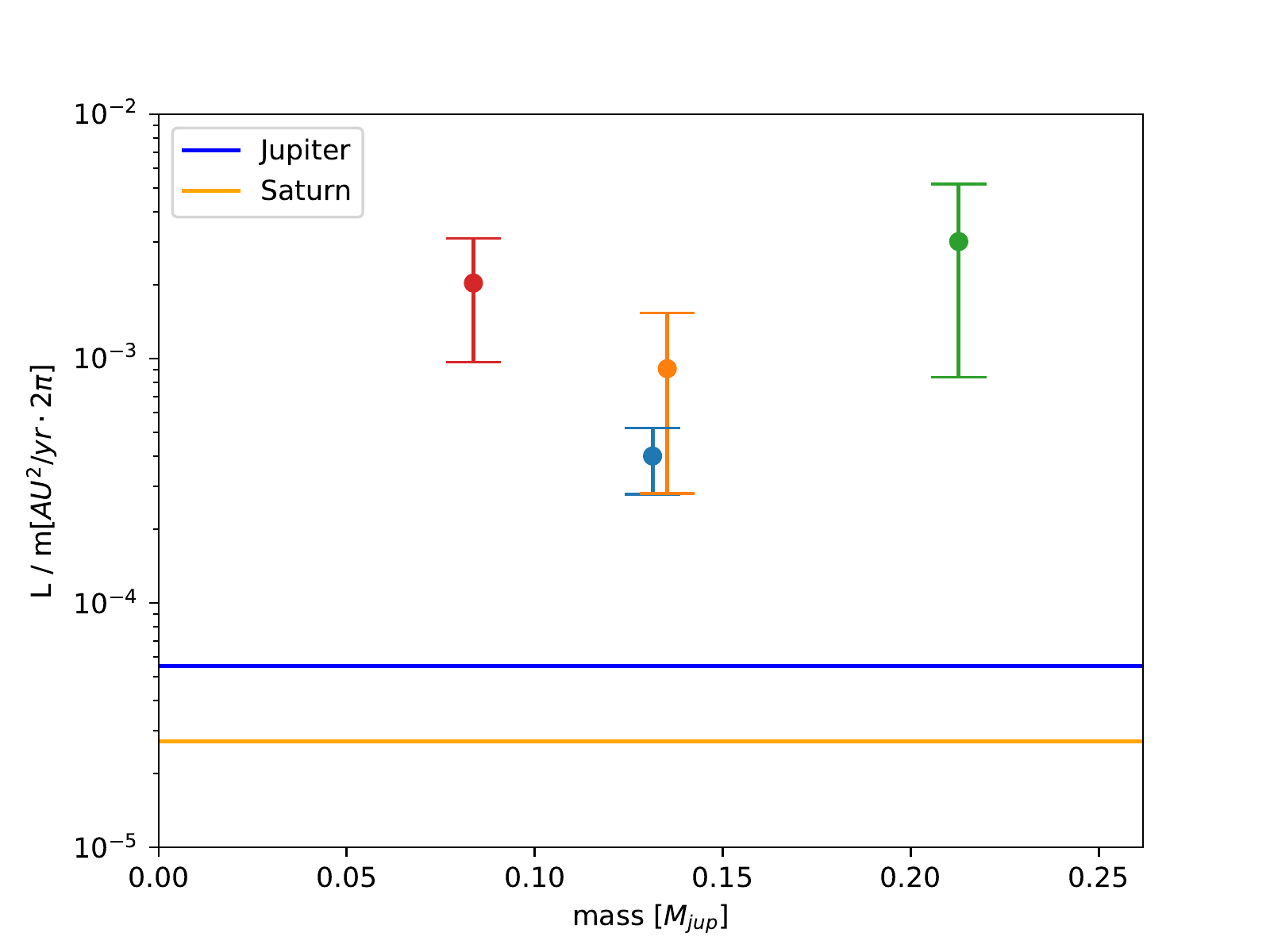}
}
\caption{Specific angular momenta of the clumps arising in the MHD and the HD simulation.
They are significantly lower in the MHD case–
probably because of the dissipation of the magnetic field.
As such, they are closer to those of Jupiter and Saturn
which are also shown for comparison.}
\label{fig-angmom}
\end{figure}

From rotation we can also determine
the total specific angular momentum $L/m$. 
This is plotted for each clump in fig. \ref{fig-angmom}.
It is calculated over the lifetime of each clump respectively where
the dot indicates the mean and the bars are the
standard deviation of the range of observed values.
For comparison, the HD case is also plotted below.
It can be seen that the specific angular momentum measured in 
the MHD case is significantly smaller than in the HD case.
Here, it should be noted that it has already been found in \citet{mayer-2004}
that the angular momentum of protoplanetary clumps
observed in simulations of fragmenting
disks is an order of magnitude too high when comparing with
those of the gas giants in our solar system.
The specific angular momenta of the gas giants \citep{helled-momentum} are also
shown in the plots – the protoplanets in the MHD case are much
closer to them than they are in the HD case.
%Why? Resistivity?
%Why is the angular momentum lower in the MHD case?
The reason for this difference could be the resistivity.
If the magnetic fields are enhanced by the clump's rotation
the resistivity could remove the magnetic energy over time,
preventing a possible saturation of the magnetic field and
thereby leading to a continuous depletion of rotational energy.
This would eventually bring the specific
angular momentum more closely to that of Jupiter.
%Saturation in iMHD? Was würde man erwarten für B?

\begin{figure}
\begin{subfigure}{\columnwidth}
\includegraphics[width=\textwidth]{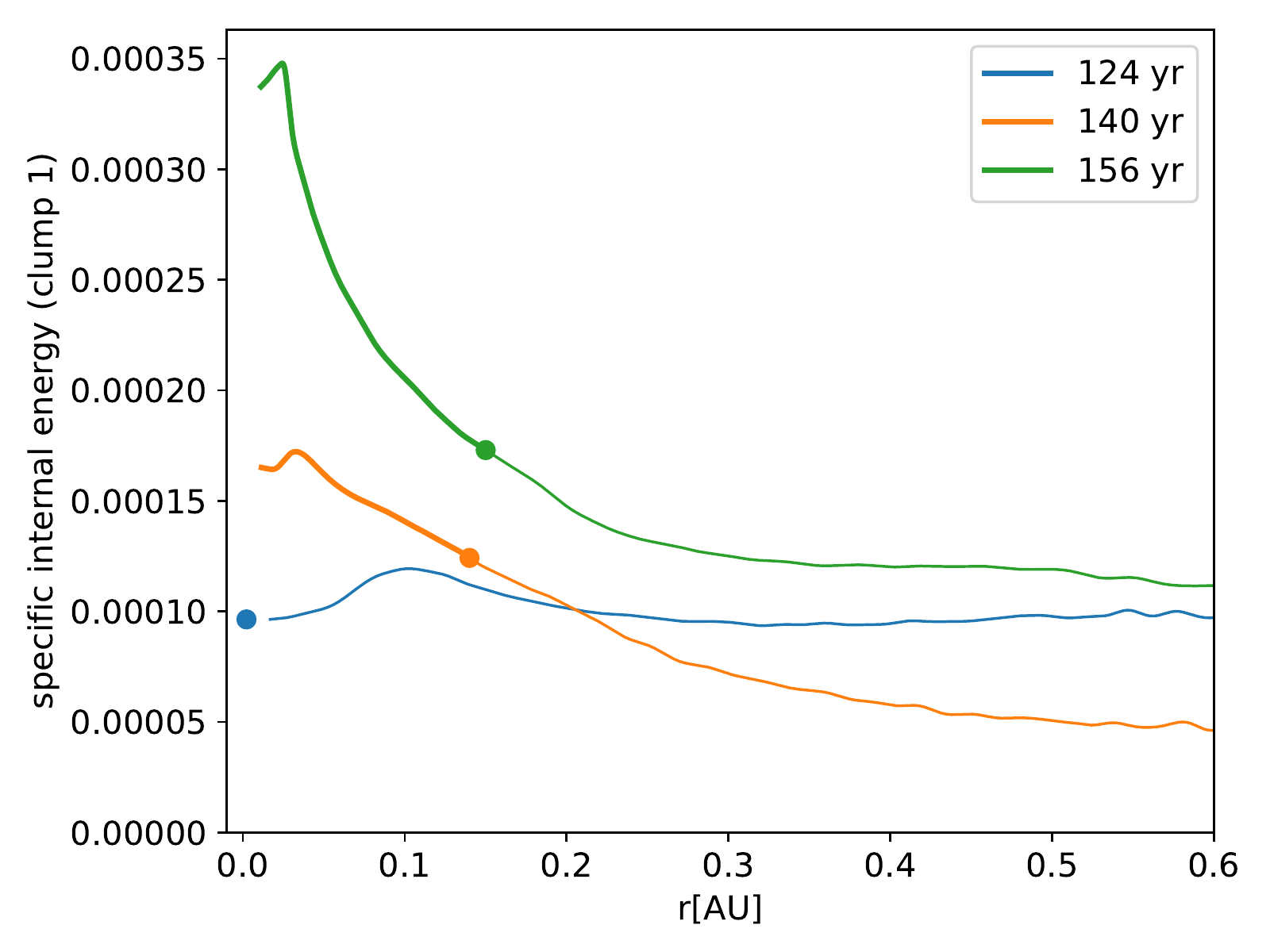}
\caption{clump 1}
\label{fig-int1}
\end{subfigure}

\vspace{0.5cm}
\begin{subfigure}{\columnwidth}
\includegraphics[width=\textwidth]{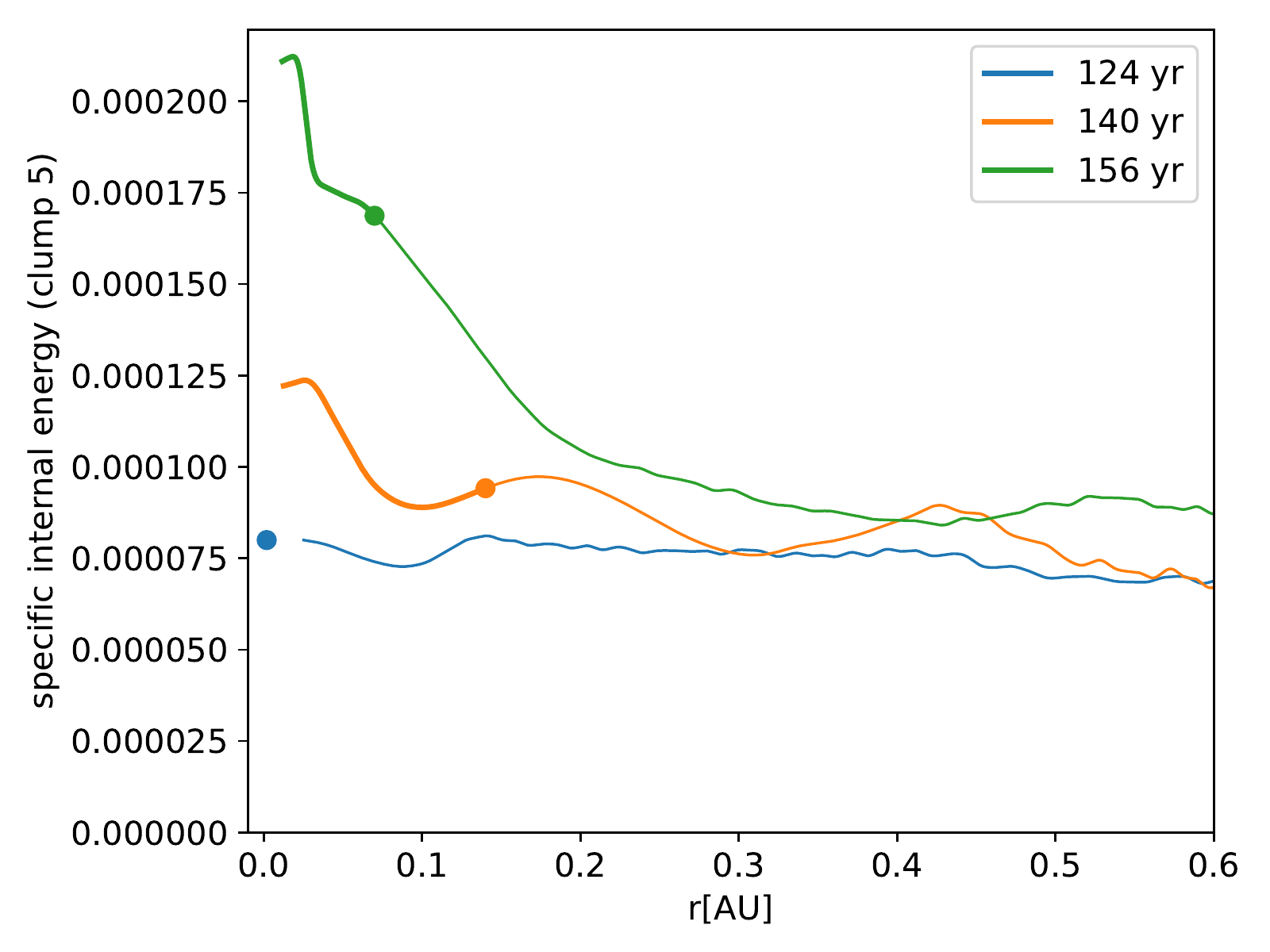}
\caption{clump 5}
\label{fig-int5}
\end{subfigure}
\vspace{0.5cm}

\caption{Radial profiles of the 
specific internal energy of clump 1 and 5.
After the clumps are bound, they heat up in the centre.
The gas pressure, which is determined through the
specific internal energy, is important for stabilizing
the clumps.}
\end{figure}

An important effect of rotation could be the 
stabilization of the clump
against collapse because of the gravitational force.
Another stabilizing force inside the clump
is the gas pressure.
Fig. \ref{fig-int5} shows the evolution of 1d profiles
of the specific internal energy.
This quantity is directly proportional to the gas temperature
and thus also determines the gas pressure.
It can be seen that before the bound stage (blue curve),
the internal energy profile is flat meaning that the
center of the clump forming region has the same temperature
as its surroundings.
At $140$ yr, corresponding to the top plots in fig. 
\ref{fig-evolution-density} the internal energy is slightly 
enhanced in the region inside the bound radius.
However this enhancement is only weak indicating the early
stage in the clump's evolution where it has not reached its 
final density.
At this stage, the relatively cold temperature could 
mean that rotation is more important
leading to the elongated shape of clump 5 at this time
which was described before.
At later times, at $156$ yr (green curve), the internal
energy clearly increases in the centre indicating the
evolved state of the clump. 

Since it was already shown in section \ref{ch-magnet}
that there are significant magnetic fields present inside
the clumps it remains to characterize them and discuss
their effects inside the clump.
The magnetic fields could act in both ways on the clump,
stabilizing or compressing.

%Force balance
We estimate their importance relative to the other forces
(gravity, gas pressure and rotation) by resorting to a 
one-dimensional model of the clump.
For that, we calculate radial profiles of the quantities
thereby ignoring angular features.

The force on a volume element $\delta A\delta x$ consists of several force terms:

\begin{equation}
\delta F = (g\rho \delta x + \Delta P_g + \Delta P_B + \cos (\theta) \rho \delta x v_\varphi^2/r_c)\delta A
\label{eq-force-balance}
\end{equation}
The gravitational acceleration is
\begin{equation}
g = G\frac{M_\text{encl}}{r^2}
\end{equation}
with $M_\text{encl}$ the enclosed mass in a sphere of radius $r$.
The pressure difference between two sides of the volume element is 
defined via the internal energy:
\begin{equation}
\Delta P_g = (\gamma - 1)\rho \Delta u
\end{equation}
The magnetic pressure term is derived from the magnetic 
energy density 
\begin{equation}
\Delta P_g = \frac{\Delta (B^2)}{8\pi}.
\end{equation}
In equation \ref{eq-force-balance} we also subtract a 
centrifugal force term representing the stabilizing effect
of rotation.
The rotation is assumed to happen around a rotation axis.
We define this force in terms of the cylindrical radius
$r_c$ (the distance to the rotation axis),
the angular part of the velocity (defined in the cylindrical
coordinate system)
$v_\phi$ 
and the angle $\theta$ for the angle between the spherical radial
direction and the cylindrical radial direction.

\begin{figure}
\begin{subfigure}{\columnwidth}
\includegraphics[width=\textwidth]{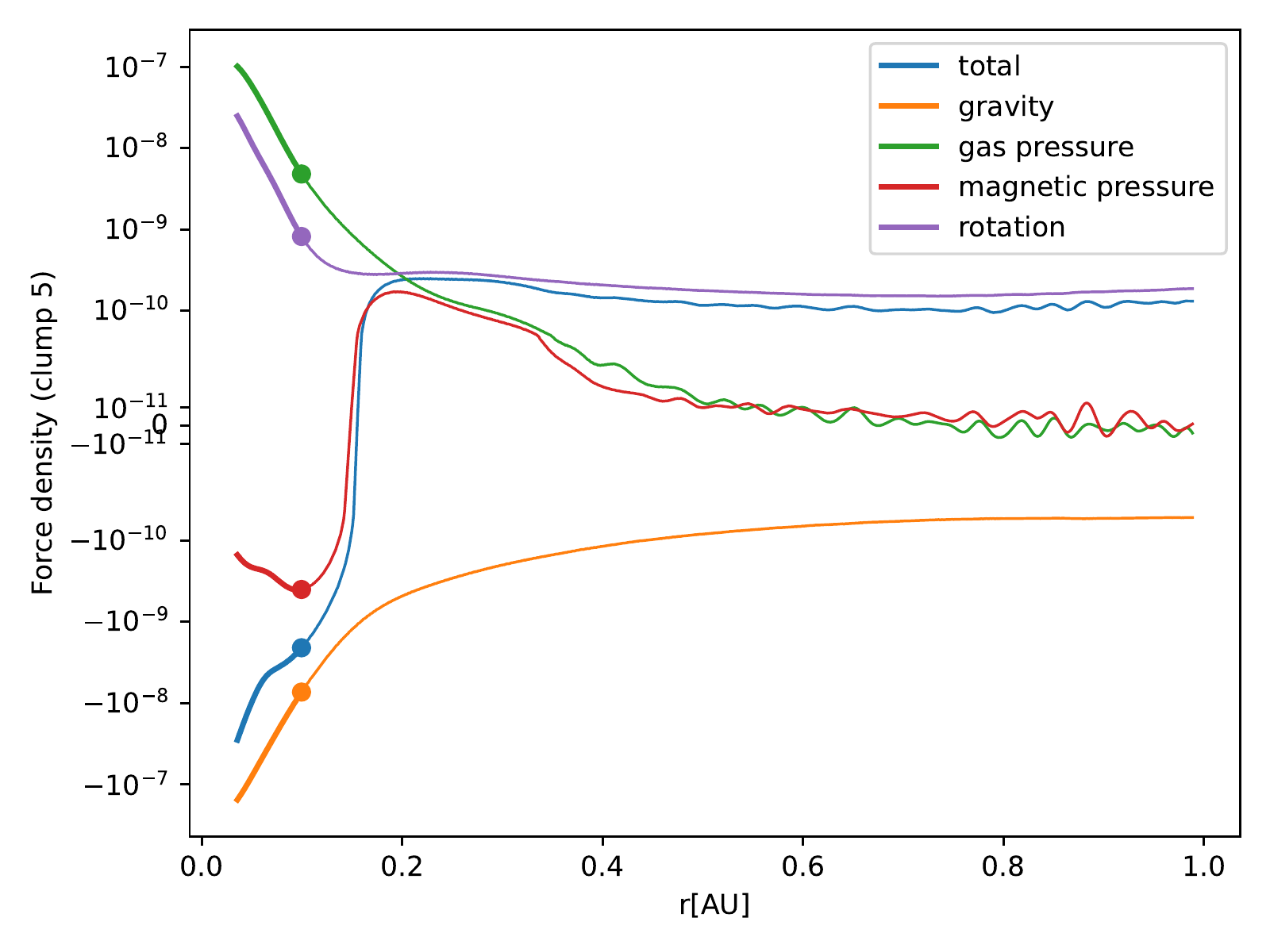}
\caption{Clump 5}
\label{fig-force5}
\end{subfigure}
\vspace{0.3cm}

\begin{subfigure}{\columnwidth}
\includegraphics[width=\textwidth]{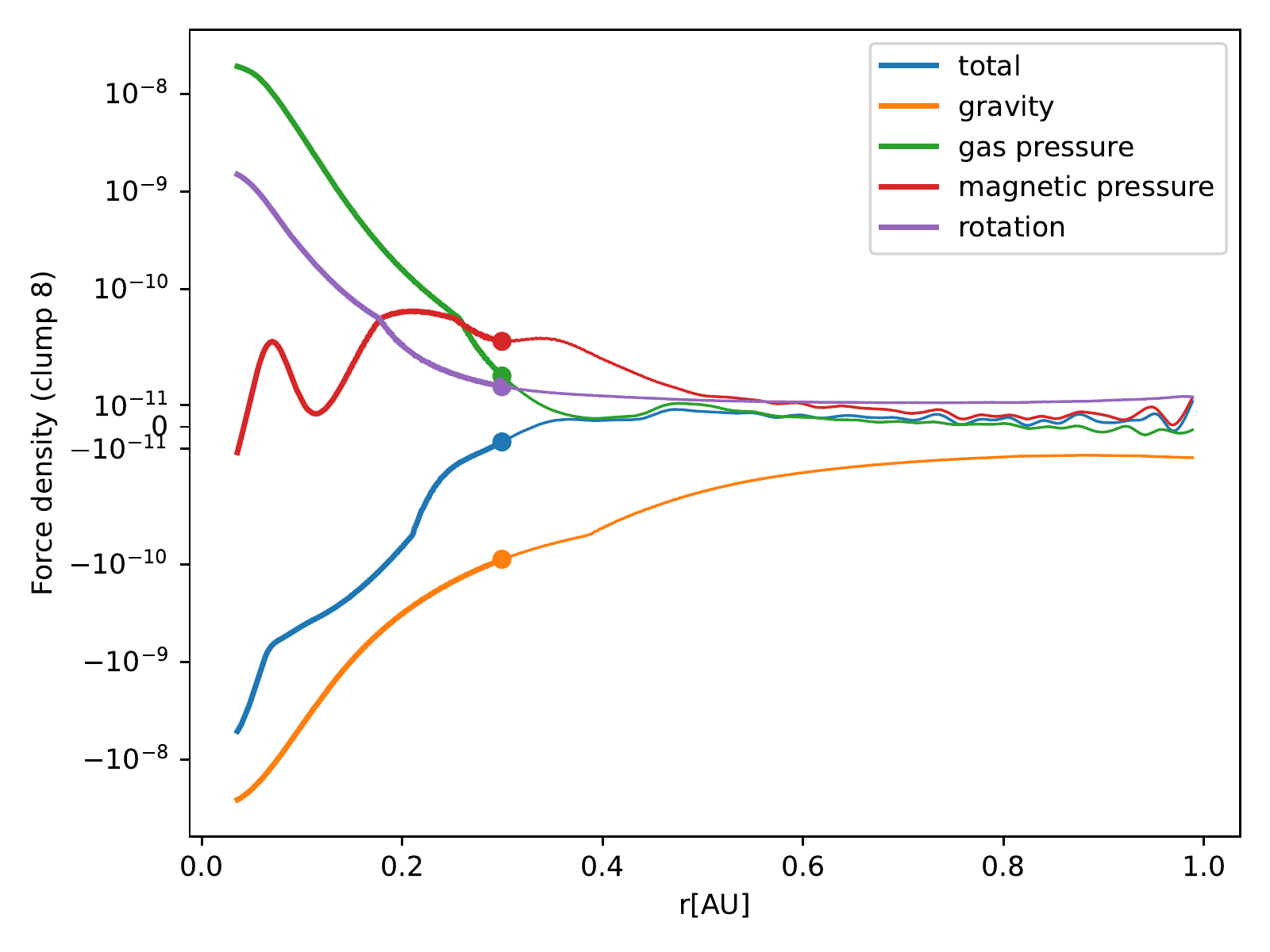}
\caption{Clump 8}
\label{fig-force8}
\end{subfigure}
%\vspace{0.4cm}
\caption{Resulting force density $\delta F/\delta A \delta x$ for two different clumps. \textbf{Top:} Clump 5.
Inside the bound radius (marked with a dot), the gas pressure is the dominating
stabilizing force although rotation also plays a role which could
explain the observed flatness in fig. \ref{fig-evolution-density}.
The magnetic field acts compressing up to a certain radius
until it pushes matter outwards corresponding to the magnetic
shield visible in fig. \ref{fig-evolution-shield}.
\textbf{Bottom:} Clump 8.
While at small radii inside the clump, gas pressure and rotation
dominate over the magnetic field (possibly due to the higher density),
the magnetic field becomes
more important further out, around the clump's bound radius (marked with a dot).}
\end{figure}

Fig. \ref{fig-force5} shows the contribution of the 
various forces at $142$ yr of clump 5.
%at the same time
%as in Fig. \ref{fig-acc5}.
It can be seen that inside the clump's radius
the dominating stabilizing force is the gas pressure, being
larger than the rotational force. 
This is despite the clump showing significant flatness 
(see Fig. \ref{fig-evolution-density} top right).
Inside this clump, the magnetic field exerts a compressing force.
Somewhat outside the clump's radius the magnetic field becomes stronger
and its pressure force points outward being of a similar order
as the gas pressure and the rotational force.
Fig. \ref{fig-force8} shows the same for clump 8 at 
$151$ yr. %(as in \ref{fig-acc8}).
There, it can again be seen that the gas pressure dominates
over the other stabilizing forces inside the clump.
This is observed in all of the clumps from which we can make
the conclusion that the clumps at this stage are pressure-supported
instead of rotation-supported.
For clump 8, it can also be seen that the magnetic field is even
stronger than the other stabilizing forces 
in a region outside the bound radius.
That the magnetic field has the highest contribution
compared to the other forces around and somewhat outside
the bound radius is a general feature we observe
in the clumps.

The resulting radial acceleration from these force contributions shows a characteristic difference 
between taking the magnetic field into account and neglecting it.
As expected, the magnetic field has the greatest effect around the bound radius.
Somewhat outside the bound radius there is for most clumps a region where
the radial acceleration is higher
meaning that material is prevented from accreting on the clump.
At the bound radius the situation is sometimes reversed (e.g. clump 5)
and the magnetic field acts compressing.
Further inside and far outside the effect is small.
This behaviour can be explained by looking at fig. \ref{fig-bpressure}.
At this time, the magnetic field has a sharp peak just outside
the bound radius of the clump.
Therefore the magnetic pressure force points inwards when going closer
to the centre and outwards when going in the other direction.
The first effect can be seen for most of the clumps:
When including the magnetic field, the force balance is shifted 
to the outward direction outside the bound radius.

For clump 8 %(fig. \ref{fig-acc8})
this effect is even more pronounced.
While here, in the situation without the magnetic field, the system
would be collapsing up until a radius of $\approx 0.5\text{AU}$,
if the magnetic field is included, only a region of $\approx 0.3\text{AU}$
has a clear negative force.
%uch für clump 8
%Erklären mit Erstarken des magnetischen Felds.
The other effect of a compressing force at the bound radius 
can however not be observed for the other clumps possibly
because for them the magnetic field is 
dominated by the other forces at this radius.

This observation of an outward pointing force arising from
the magnetic field is consistent with the findings in 
\citet{deng}. 
There the simulations were continued without the magnetic field
after the clump formation and it was found that 
the further evolution of the clumps changed compared to simulations
that continued to include the magnetic field.
Namely, the clumps were disrupted if no magnetic field was present
due to the missing of the shielding effect.

%Konsistent mit Beobachtung von Hongping 2021.
%Kräfte-Plot. Zeige, dass Magnetfeld um Bound radius wichtig wird.
%Ausserdem sind die Clumps durch den Gasdruck dominiert.
%Auch für flachen Plot (vorher dargestellt)?

The analysis presented in this section was carried out at a
time when the clumps have already formed and are gravitationally bound.
It remains however to find reasons for the smaller initial clump
mass at the very onset of fragmentation in the MHD simulations compared
to the HD ones. This will be the focus of the next section.

%\begin{figure}[h]
%\includegraphics[scale=0.5]{./images/interior/cl5_u.pdf}
%\caption{Radial profiles of the 
%specific internal energy of clump 5.
%After the clump is bound, it heats up in the centre.}
%\label{fig-int5}
%\end{figure}
%\begin{figure}[h]
%\includegraphics[scale=0.5]{./images/rot_an_force_density.pdf}
%\caption{}
%\label{}
%\end{figure}

%\begin{figure}[h]
%\includegraphics[scale=0.5]{./images/rot_an_acc.pdf}
%\caption{}
%\label{}
%\end{figure}

\section{A physical description of gravitational instability in magnetized disks}

\label{section-instability}
\subsection{A linear perturbation theory approach}
%\subsection{Magnetic field destabilization}
\label{destabilization}

Here we will try to address how different  the initial
development of clumps is in a magnetized flow as opposed to
an unmagnetized one. This is important since, as we 
reported, the masses of clumps in magnetized disks
are significantly lower than those in unmagnetized
ones since the beginning (since they become bound),
which suggests the effect of magnetic pressure in
stifling gas accretion, suggested in \citep{deng},
can not be the only reason behind the low masses
of clumps (see fig. \ref{fig-masses}).

To this aim, we investigate how the presence of the magnetic 
field could change the fragmentation.
Let us now turn back to Elmegreen's analysis
on fragmentation in magnetized galactic disks \citep{elmegreen}.
Starting with the magneto-hydrodynamical equations he
assumed first-order perturbations.
Then the equations  were evolved numerically and
the response of the system to a perturbation was studied.
We note that the results
presented in this paper include
resistivity.
However, similar results have been
observed for ideal MHD simulations \citep{deng1} where we expect even more prominent differences
since the magnetic field is not restrained.
For simplicity, we consider ideal MHD in this section.
The ideal magneto-hydrodynamical equations describe the gas motion by considering 
gas pressure, self-gravity and magnetic fields:
\begin{equation}
\frac{\partial \rho}{\partial t} + \nabla (\rho v) = 0
\end{equation}
\begin{equation}
\frac{\partial \vec{v}}{\partial t} + \vec{v} \cdot \grad  \vec{v} = -\frac{1}{\rho} c_s^2\grad \rho - \grad \Phi\
+ \frac{1}{\mu_0 \rho} (\vec{\nabla} \times \vec{B}) \times \vec{B}
\end{equation}
\begin{equation}
\Delta \Phi = 4 \pi G \rho
\end{equation}
\begin{equation}
\frac{\partial B}{\partial t} = v\times (v\times B)
\end{equation}

\begin{figure}
\begin{subfigure}{\columnwidth}
\includegraphics[scale=0.5]{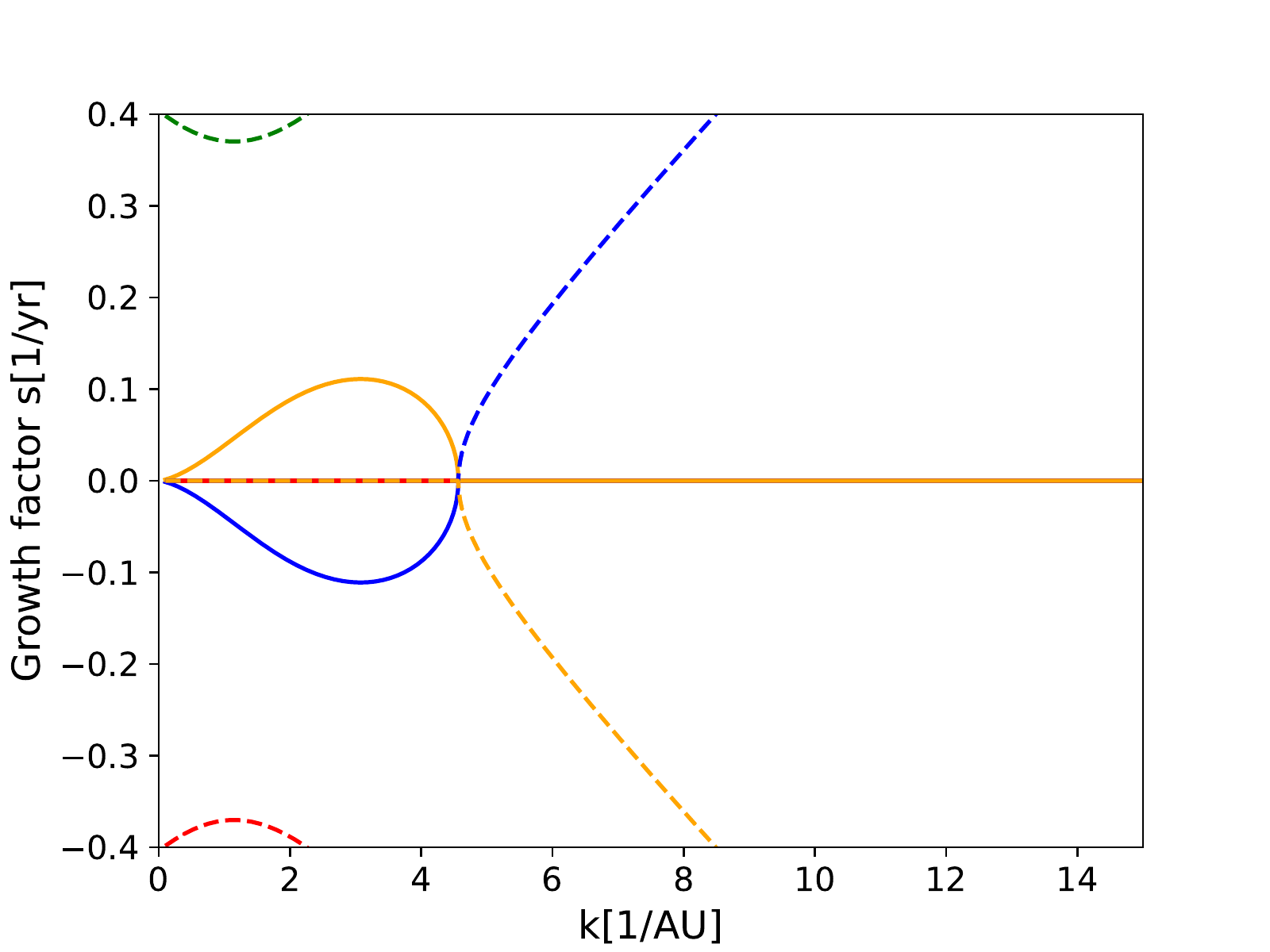}
\caption{Solutions of dispersion relation at t = 0}
\label{disp_solutions_t0}
\end{subfigure}

\begin{subfigure}{\columnwidth}
\includegraphics[scale=0.5]{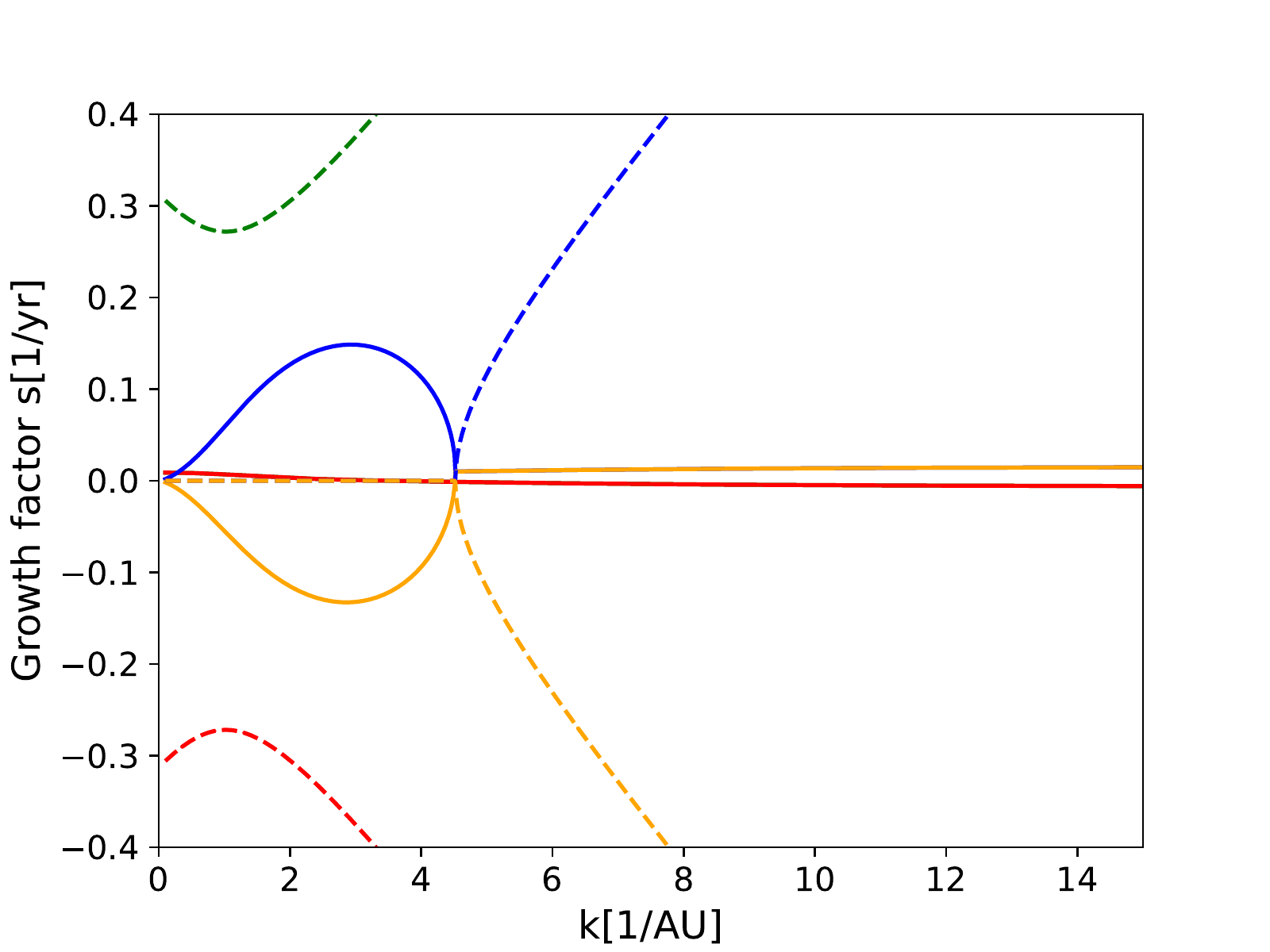}
\caption{Solutions of dispersion relation at t = 0.5 / $\Omega$}
\label{disp-solutions-t05}
\end{subfigure}

\caption{Solutions of the dispersion relation for 
shearing perturbations at different times (different
values of $k_x$), plotted over the wavenumber.
The imaginary parts of the solution are solid
and the real parts are dashed.
If there exists a solution with a non-zero imaginary 
part, perturbations can grow in the corresponding
regime.
At large wavenumbers, there exists only a real (oscillating) solution.}
\end{figure}

In a localized cartesian coordinate system in the disk
where the x points in the radial and y in the transversal direction,
the local background
flow can be approximated as 
\begin{equation}
\begin{pmatrix}
        v_x\\v_y
\end{pmatrix}\
\simeq\
\begin{pmatrix}
        -\Omega(r_0)y\\
        2 \mathcal{A} x + \Omega(r_0)x
\end{pmatrix}
\label{eq-local-cooordinates}
\end{equation}
with $\mathcal{A}$ being one of the Oort's constant \citep{binney}
that represents the shear arising from differential rotation.
By defining a dimensionless shear-parameter $\alpha \coloneqq \mathcal{A} /\Omega$
the shear of the flow can be more conveniently quantified
which makes it radius-independent if the angular velocity
follows a power law in terms of the radius (e.g. in a Keplerian orbit).

Elmegreen considered linear perturbations 
to an equilibrium solution
that are proportional to 
\hbox{$\exp (i k_y (y - 2 \mathcal{A} xt))$}
which means they start azimuthally oriented and are then 
sheared out over time.
He then integrated the perturbative solution numerically over time 
for various parameters of the magnetic field and the shear rate.
Elmegreen found that the effect of the magnetic field depended
hugely on the value of the shear parameter. 
In a strong shear case which corresponds to $\alpha = -0.5$
as for a flat rotation curve of a galactic disk,
the magnetic field stabilized the disk.
Here, an increase in the magnetic field resulted in a stronger damping 
of the perturbations similar to what was found in the last section. 
On the other hand in a weak-shear case which corresponds
to $\alpha = -0.05$ the result however was very different.
Here, it was found that the magnetic field severely destabilized the disk.
While the response of the system to the perturbation was stable without a
magnetic field, even a small value of the magnetic field led to a 
huge amplification of the perturbation.
The stronger the magnetic field, the more unstable the system became.

\citet{elmegreen} %Elmegreen
explained the destabilizing effect of the magnetic 
field intuitively by looking at the magneto-hydrodynamic equations. 
If a region without a magnetic field collapses, the collapse is stabilized by the
Coriolis force.
However, the magnetic field is assumed to be toroidal and thus introduces an asymmetry.
Therefore the magnetic field dampens the radial part (the x-direction) of the 
perturbed velocity and no stabilizing
Coriolis force can arise which gives rise to a huge growth \citep{elmegreen}.

While these are numerical results, 
Gammie used an analytical approach to derive a stabilizing effect of 
the magnetic field for axisymmetric perturbations \citep{gammie}.
From the linearized MHD equations and after solving for
one axisymmetrical mode, he derived a dispersion relation for MHD perturbations.
We now want to examine if a destabilizing effect as found by Elmegreen
can also be seen in Gammie's analytical framework.

\subsection{Dispersion relation}
\label{disp-rel}

In the following paragraphs we look at this dispersion relation but for 
the general case without the restriction to axisymmetric perturbations.
We find two effects:
First, the magnetic field could destabilize the system.
This can happen either in a region of weak shear 
meaning that the
%where if the 
shear parameter is $\alpha \gtrsim -0.15$ 
whereby regions can be destabilized
that are otherwise stable.
Or, if the shear parameter is $\alpha \gtrsim -0.4$ the magnetic field
can at least increase the growth of instabilities that would otherwise
also be present.
The second effect concerns the wavelength of the most unstable perturbation:
Considering a situation where a magnetic field is present,
in regions of weaker shear the wavelength is significantly smaller
potentially leading to smaller-sized objects.

We begin by linearizing
the MHD equations and solving for one mode.
As in \citet{elmegreen}, the perturbations
are now non-axisymmetric but shearing with the flow,
with wavenumber
\begin{equation}
\begin{pmatrix}k_x\\k_y\end{pmatrix}
=\begin{pmatrix}\tilde{k}_x-2\mathcal{A}x t\\k_y\end{pmatrix}.
\label{eq-perturbations}
\end{equation}
%\begin{equation}
%\begin{smallmatrix}(k_x&k_y)\end{smallmatrix}
%=\begin{smallmatrix}(\tilde{k}_x-2\mathcal{A}x t&k_y)\end{smallmatrix}.
%\label{eq-perturbations}
%\end{equation}
Then equations are solved for one angular frequency $\omega$
(so the perturbations are proportional to $e^{i\vec{k}\cdot\vec{x}} e^{i\omega t}$). 
Without any further simplifications this yields a dispersion relation:
%without any further simplifications
%and angular frequency $\omega$
%This yields a dispersion relation

\begin{equation}
\begin{split}
        &\omega^4 + i\omega^3 \left( 4\mathcal{A} \frac{k_y k_x}{k^2} \right)\\
        -&\omega^2 \left( \kappa^2 - 2\pi G\Sigma\lvert k\rvert + c_s^2 \vec{k}^2 + k_x^2\vec{V}_a^2 + k_y^2 (\vec{V}_{a,y}^2 - \vec{V}_{a, x}^2)\right)\\
        -&i\omega 2\mathcal{A} \left(\vec{k}^2 V_{a,x} V_{a,y} + k_x k_y \vec{V}_a^2\right)\\
        +&(k_x^2 V_{a,x}^2 + k_y^2 V_{a,y}^2) \left(-2\pi G \lvert \vec{k}\rvert \Sigma + c_s^2 \vec{k}^2\right)=0
\end{split}
        \label{disprel-full}
\end{equation}

which is a fourth-order polynomial in $\omega$.

\begin{figure}
\begin{subfigure}{\columnwidth}
\includegraphics[width=\textwidth]{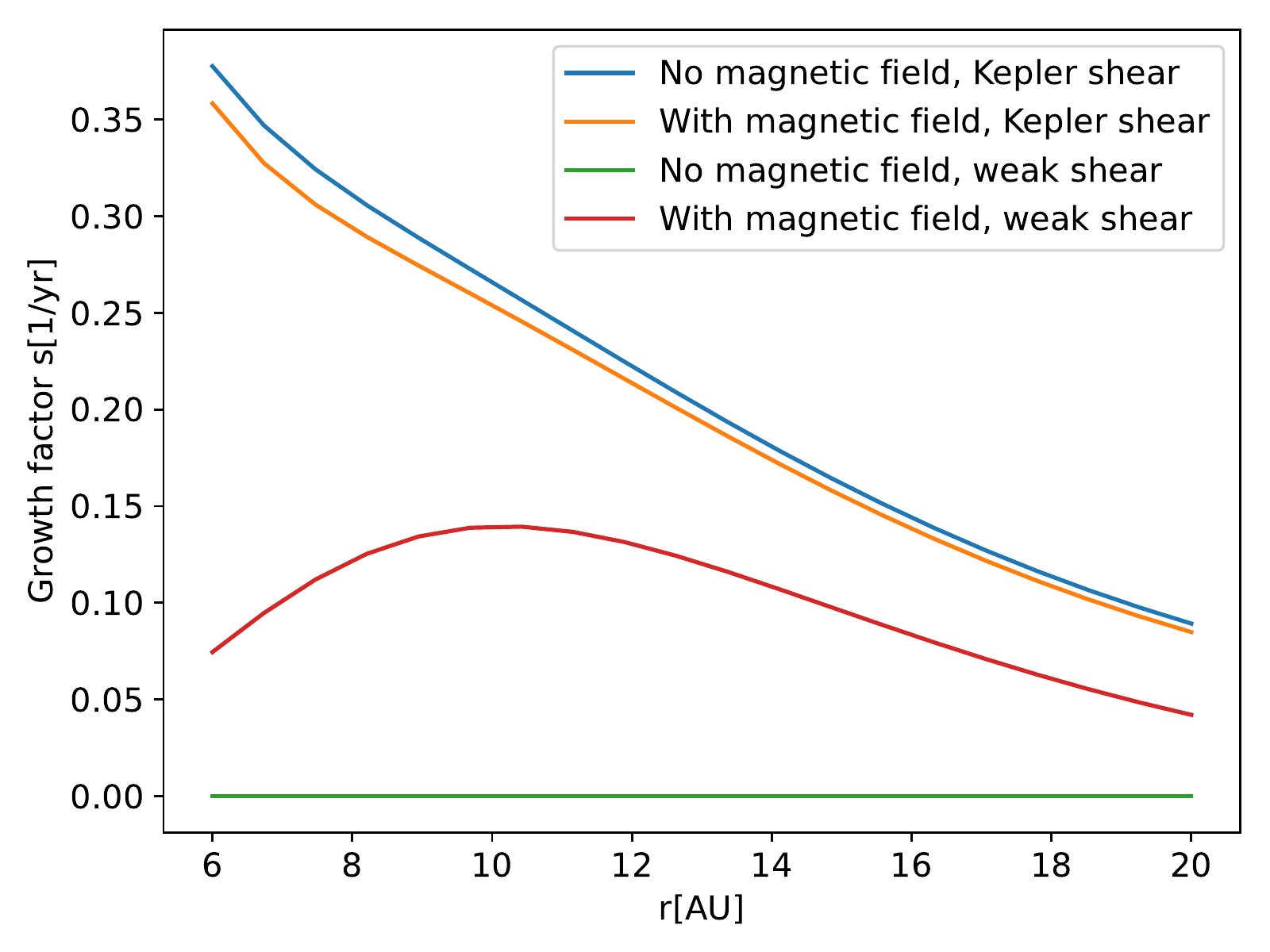}
\caption{Radial profile of the growth factor $s$ for
shearing perturbations obtained from linear perturbation
theory for different configurations.
Whereas in regions with Keplerian shear,
the magnetic field slightly suppresses the perturbations,
in the weak-shear case, the magnetic field
can destabilize perturbations that
would otherwise be stable thus making
otherwise stable regions unstable.}
\label{fig-growthfactor}
\end{subfigure}

\begin{subfigure}{\columnwidth}
\includegraphics[width=\textwidth]{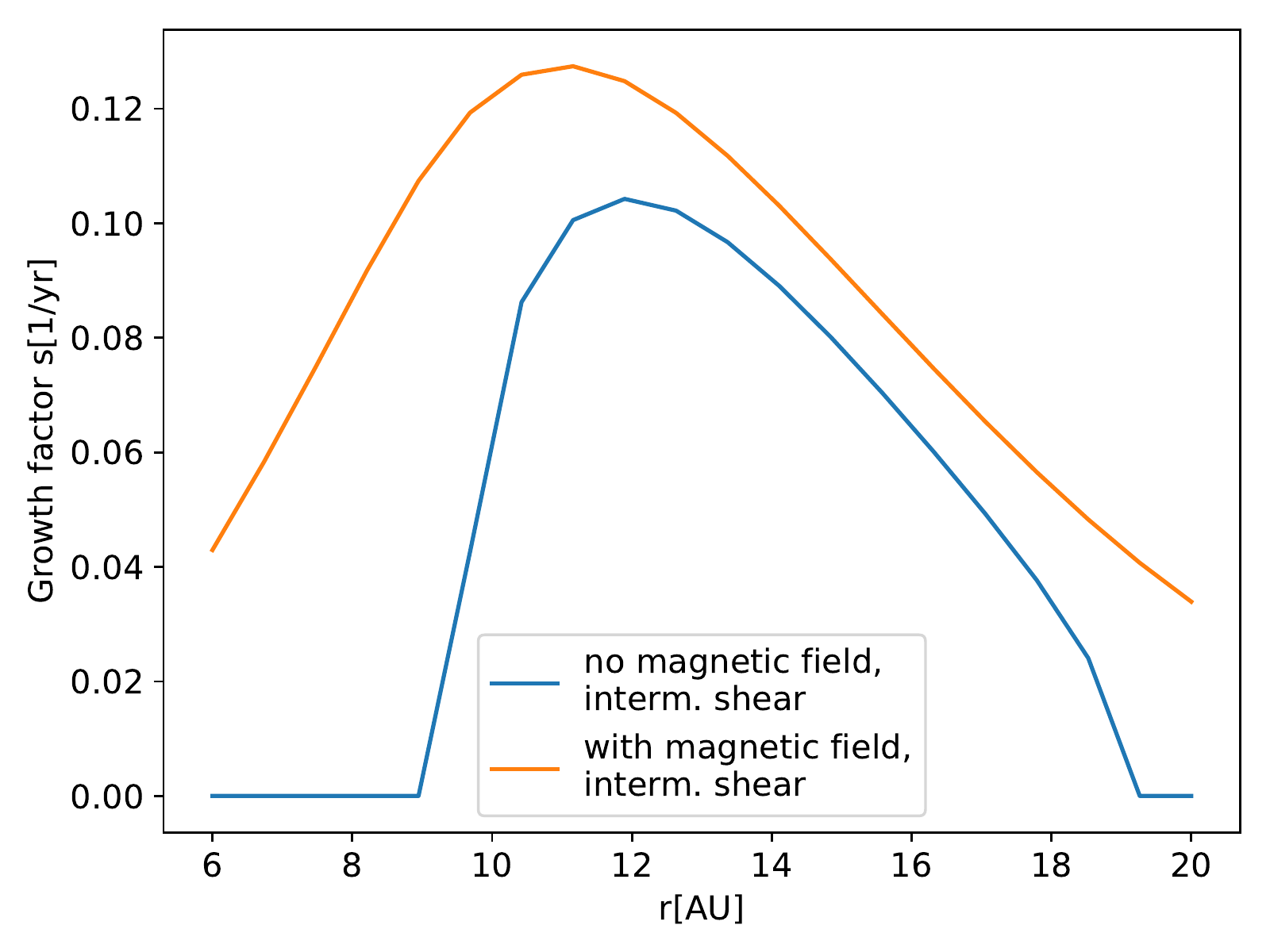}
\caption{Radial profile of the growth factor $s$ for
shearing perturbations obtained from linear perturbation
theory for an intermediate shear rate of $\alpha = -0.4$.
Such regions appear more frequently in the simulation 
(see section \ref{ch-preconditions}).
In them, the magnetic field tends to increase the growth
rate.}
\label{fig-growthfactor-interm}
\end{subfigure}
\caption{Radial profiles of growth factor $s$ obtained 
from linear perturbation theory.}
\end{figure}

In the next step we want to examine the roots
of this polynomial $\{\omega_0\}$ numerically.
%The roots of this polynomial $\{\omega_0\}$ are now
%examined.
In general we expect 4 solutions to the equation.
If they are all real, the solution is an oscillating
wave.
If one of the solutions has an imaginary part, 
perturbations can grow.
Fig. \ref{disp_solutions_t0} in the appendix presents as an example
the situation for
realistic values at approximately $10\text{AU}$ in 
a protoplanetary disk.
%The roots are shown in fig. \ref{disp_solutions}
%which presents the situation in the model from fig.
%\ref{disk_model}
%(using 
%$M_{disk} = 0.1 M_{sun}$)
%at a radius of 
%$10 \text{AU}$.
The 4 solutions depending on the wavenumber 
$k_y$
are drawn in different colours where the imaginary
part is drawn solid and the real part is dashed.
In this case, in the region 
$k_y<5/\text{AU}$
there exists a solution with a non-zero
imaginary part which means that perturbations can grow in this regime.
At large wavenumbers there is only a real (oscillating) solution.

The next step is now to look if this formalism also shows a
destabilizing effect of the magnetic field.
For that, a toy model of a protoplanetary disk is introduced,
using realistic values comparable to what was used to initialize
the simulations from \citet{deng}.
We now calculate a radial profile of the solution.
At each radius, the solutions 
$\omega_0$ 
of the dispersion relation
(\ref{disprel-full})
are calculated for a range of wavenumbers 
$\vec{k}$
with 
$\tilde{k}_x=0$.
At each wavenumber
$k_y$, 
the one of the 4 solutions which is growing the fastest is selected.
%The growth factor is now defined as
%$s\coloneqq \mathfrak{Im}(\omega)$.
The wavenumber that leads to the largest imaginary
part of $\omega$ (which grows the fastest) is then chosen.
Its growth factor, defined as 
$s\coloneqq \mathfrak{Im}(\omega)$,
is then taken as the growth factor at this radius (assuming that the fastest growing mode dominates).

Fig. \ref{fig-growthfactor} shows the growth factor $s$ 
for different configurations over the radius of the 
toy model.
It shows two situations: One where the shear value
is set to the Keplerian shear and another with
weak shear 
where $\alpha = -0.1$.
Both situations are shown with and without a magnetic field.
In the Keplerian shear situation,
the presence of the magnetic field lowers the growth factor a bit.
Here, the system is unstable in both cases since the growth factor is non-zero.
On the other hand, in the weak shear situation ($\alpha = -0.1$), the magnetic field
seems to be required for the system to be unstable.
Without the magnetic field (the green line), the growth factor
is zero everywhere meaning that the system is stable.
When introducing a magnetic field however, the growth-factor is non-zero
and thus the system unstable.
This is even true when introducing a much weaker magnetic field (10 times smaller)
but then the growth factor is also a bit lower.
If the magnetic field is further increased in strength,
the growth factor seems to saturate.
This clearly destabilizing effect seems to be
present for shear rates 
$\alpha \gtrsim -0.15$.
In an intermediate regime until 
($\alpha \sim -0.4$)
(see fig. 
\ref{fig-growthfactor-interm})
the magnetic field increases the growth rate
but does mostly not destabilize regions that would otherwise
be stable.

It is now examined why the magnetic field makes such
a difference in the growth factor at low shear.
Starting from the dispersion relation (eq. \ref{disprel-full})
and now assuming zero-shear $\mathcal{A}=0$
one arrives at a quadratic equation in $\nu \coloneqq \omega^2$:

\begin{equation}
        \begin{split}
        \nu^2\
                -\nu \overbrace{\left(\kappa^2 \
                - 2\pi G\Sigma\lvert k\rvert + c_s^2 \vec{k}^2 + k_x^2\vec{V}_a^2 + k_y^2 (\vec{V}_{a,y}^2 - \vec{V}_{a, x}^2)\right)}^{\eqqcolon q}\\
                +\underbrace{(k_x^2 V_{a,x}^2 + k_y^2 V_{a,y}^2) \left(-2\pi G \lvert \vec{k}\rvert \Sigma + c_s^2 \vec{k}^2\right)}_{\eqqcolon w}=0
        \end{split}
\end{equation}

If there is no magnetic field,
then $w = 0$ and a negative solution
in $\nu$ exists only if $q < 0$ which is
just Toomre's criterion for instability. % (eq. \ref{toomre}).
If $\nu < 0$ then there is a solution
with $\omega$ purely imaginary which means exponential
growth of the perturbation.
Note that however now there is $\kappa = 2\Omega$ 
(since $\mathcal{A}=0$)
which makes the model stable at all radii.
If the model is stable then $Q>1 \iff q>0$ 
(still without the magnetic field).
Now imagine introducing even a small magnetic field.
It can be seen that there exists always a
$\lvert k\rvert$ such that $w<0$.
But if that is the case, then the discriminant is positive
$\mathcal{D}=q^2 - 4w > 0$
meaning that the solutions are real in $\nu$.
The solutions are $\nu = \frac{q\pm\sqrt{\mathcal{D}}}{2}$.
From $w<0$ it follows that $\sqrt{\mathcal{D}} > q$.
Therefore one of the solutions is negative ($\nu>0$).
This again means that $\omega$ is purely imaginary and thus the system
becomes unstable.
In the case of strong shear, where the system is already unstable,
the damping effect of the magnetic field 
can be attributed to the term in $q$
proportional to the Alfvén velocity which acts like a gas pressure.

\begin{figure}
\includegraphics[scale=0.5]{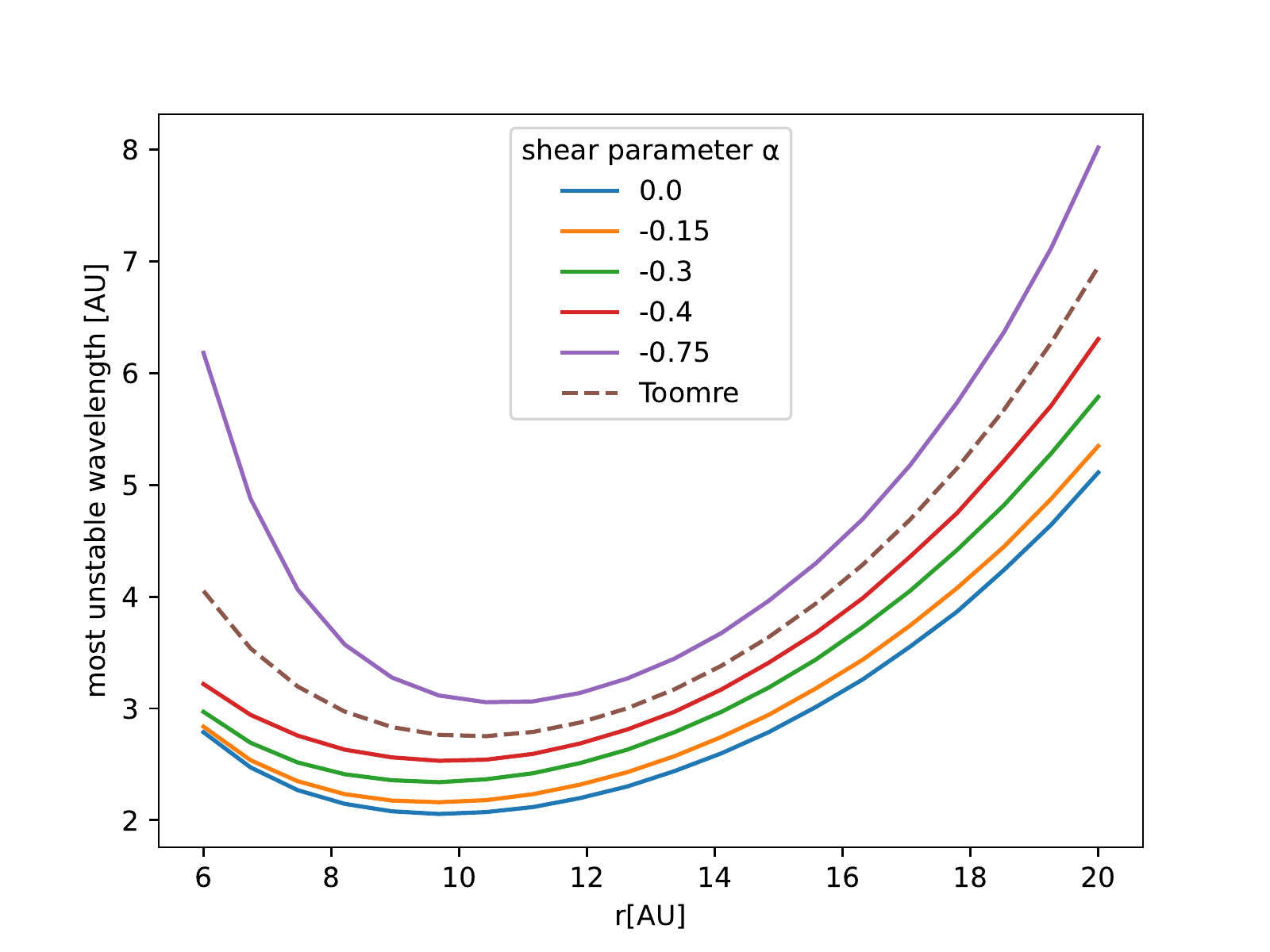}
\caption{Radial profile of the most unstable wavelength for non-axisymmetric
perturbations in a toy model disk with different values
of the shear parameter $\alpha$ and including a magnetic field.
The Toomre most unstable wavelength is shown as a reference.
At weak shear, where the magnetic field may act destabilizing
(see fig. \ref{fig-growthfactor}), the wavelength of the
most unstable perturbation is smaller than Toomre's prediction
potentially contributing to the significantly lower size of fragments
observed in the MHD simulations.}
\label{fig-wavelength}
\end{figure}

%Also lower, compared  to Toomre
%How much smaller, how much smaller the masses
%Does it suffice to explain the smaller masses
What are the expected scales of fragmentation?
The fastest growing mode for certain values of 
the shear parameter is plotted in fig.
\ref{fig-wavelength}.
This includes the presence of a magnetic field.
It can be seen that the scale is lower for weaker shear
meaning that smaller objects may be produced.
Compared to the Toomre most unstable wavelength, the
scale is somewhat reduced to $2/3$ so it would lead 
to masses of $1/3$ the size.

When comparing the masses of the clumps in the MHD simulations
to the ones from the HD simulation (see fig. \ref{fig-masses}) one
can see that this could explain a large part of the difference
between the two cases. 
Still, the predicted mass is much too large when compared
to the clumps actually observed in the simulation.
However, when we phenomenologically combine this magnetic
destabilization effect with the predicted mass according
to \citep{boley-disruption} (see fig. \ref{fig-toomre-mass})
the prediction lies actually in the range of the small clumps
from the simulations.

%Still, the effect could not explain the difference in mass
%between the Toomre prediction (see fig. \ref{fig-toomre-mass}) and what
%is actually observed in the simulations, even
%for the HD case.
%What does it mean? Is it enough?

%Validity
We conclude our discussion
of perturbation theory results with a few comments on the validity of the
approximations made.
With eq. \ref{eq-local-cooordinates} the approximation of 
a local coordinate system was made. 
This is only valid if we consider regions that are much smaller
than the system's length scale. 
This implies that the wavelengths of the perturbation need to be much smaller
than the radial distance from the star $\lambda \ll r$.
This is certainly fulfilled for the weak-shear cases (see fig. \ref{fig-wavelength}).
For the strong-shear cases the analysis could become invalid at small radii.
Further, the WKB approximation was made where the analysis concentrates 
on one mode and does not take
into account mixing. 
However, since for non-axisymmetric perturbations the wavenumber
$\vec{k}(t)$ is time-dependent this can only be justified as long as the change
of $\vec{k}$ is small on the considered time-scale.
This means that the growth factor should be large compared
to Oort's parameter, $s \gg k_x'/\left| \vec{k}\right| = 2 \left| \mathcal{A}\right|\cdot k_y/\left| k\right| \lesssim 2\left|\mathcal{A}\right|$.
For the weak-shear case ($\alpha \gtrsim -0.15$)
this is also fulfilled over the whole region considered in fig. \ref{fig-growthfactor}
since e.g. at $10\text{AU}$ the angular frequency is $\approx 2\pi/30\text{yr}$ 
leading to an Oort's parameter of $\mathcal{A}\approx -0.02/\text{yr}$.
However the solutions for non-axisymmetric perturbations at Keplerian
shear are probably not valid since then the Oort's shear parameter
is comparable to the growth rate.
But still the magnetic field could slightly enhance growth
in the regime of intermediate shear $\alpha \gtrsim -0.4$.

From fig. \ref{disp-solutions-t05} one can see that even if we look
at the behaviour at later times (here $t=0.5/\Omega$) when 
the shape of the perturbations has changed (see eq. \ref{eq-perturbations})
the solutions to the dispersion relation don't change much.
This means that the time-evolved perturbation is still unstable and can grow further.

%The solutions don't change much over time, see fig. \ref{disp_solutions_t05}.
%As can be seen in fig. \ref{fig-growthfactor-interm}
%the magnetic field still increases instabilities
%at later times in the intermediate-shear regime.

After these theoretical considerations it remains to check 
if the conditions used in this section are met in the simulations.
This is done in the next section.

\subsection{Preconditions}
\label{ch-preconditions}

%\begin{figure}
%\includegraphics[scale=0.4]{./images/preconditions/azimuthality.pdf}
%\caption{}
%\label{fig-azimuthality}
%\end{figure}

%\begin{figure}
%\includegraphics[scale=0.75]{./images/preconditions/shear_histogram_mhd.pdf}
%\caption{Histogram of values of 
%the shear parameter $\alpha$ at the beginning of the simulation.}
%\label{fig-shear-hist}
%\end{figure}

\begin{figure}
\centering
\begin{subfigure}{0.5\columnwidth}
	\includegraphics[width=\textwidth]{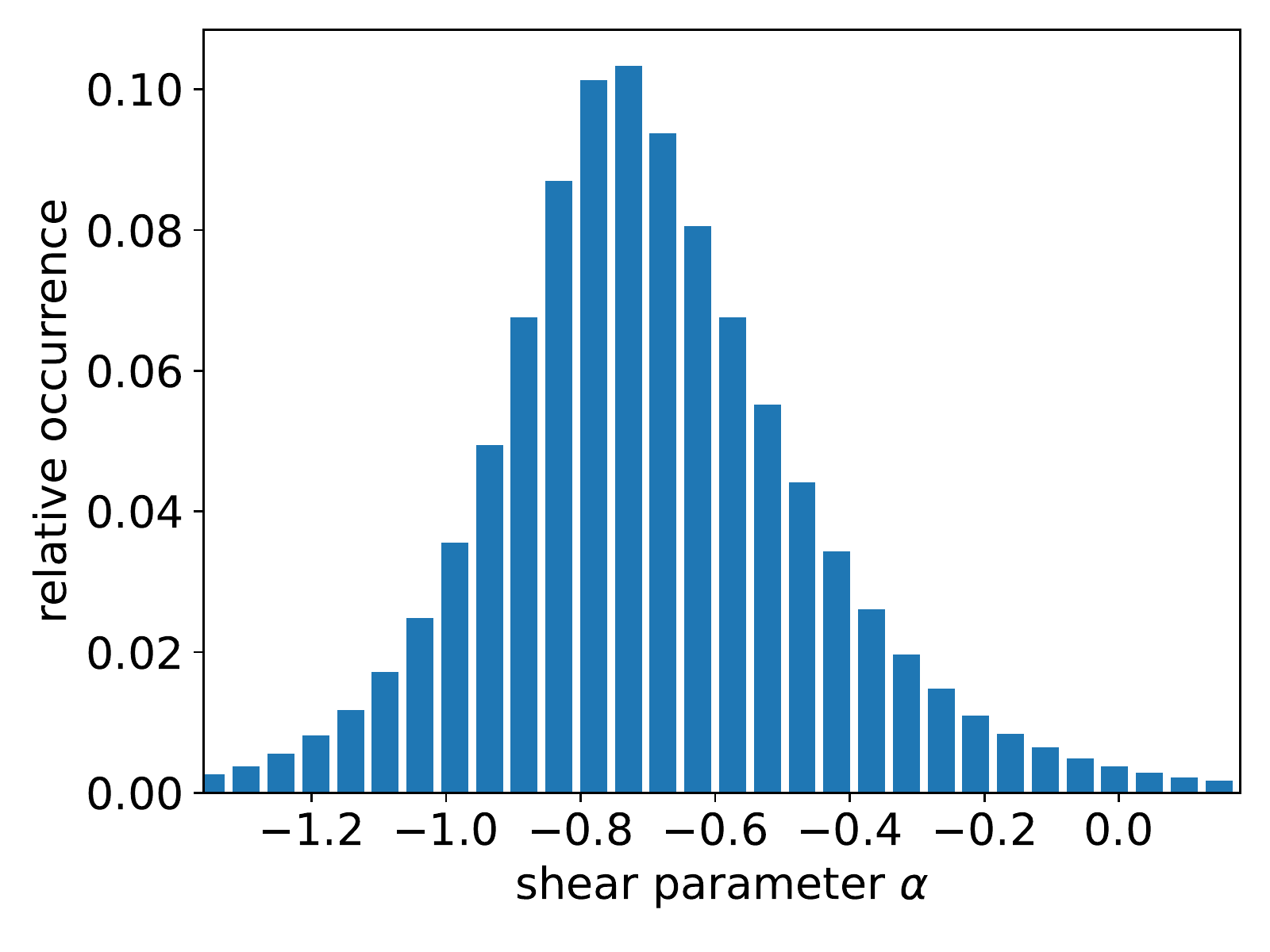}
	\caption{Shear histogram at $t=0$}
\end{subfigure}%
\begin{subfigure}{0.5\columnwidth}
	\includegraphics[width=\textwidth]{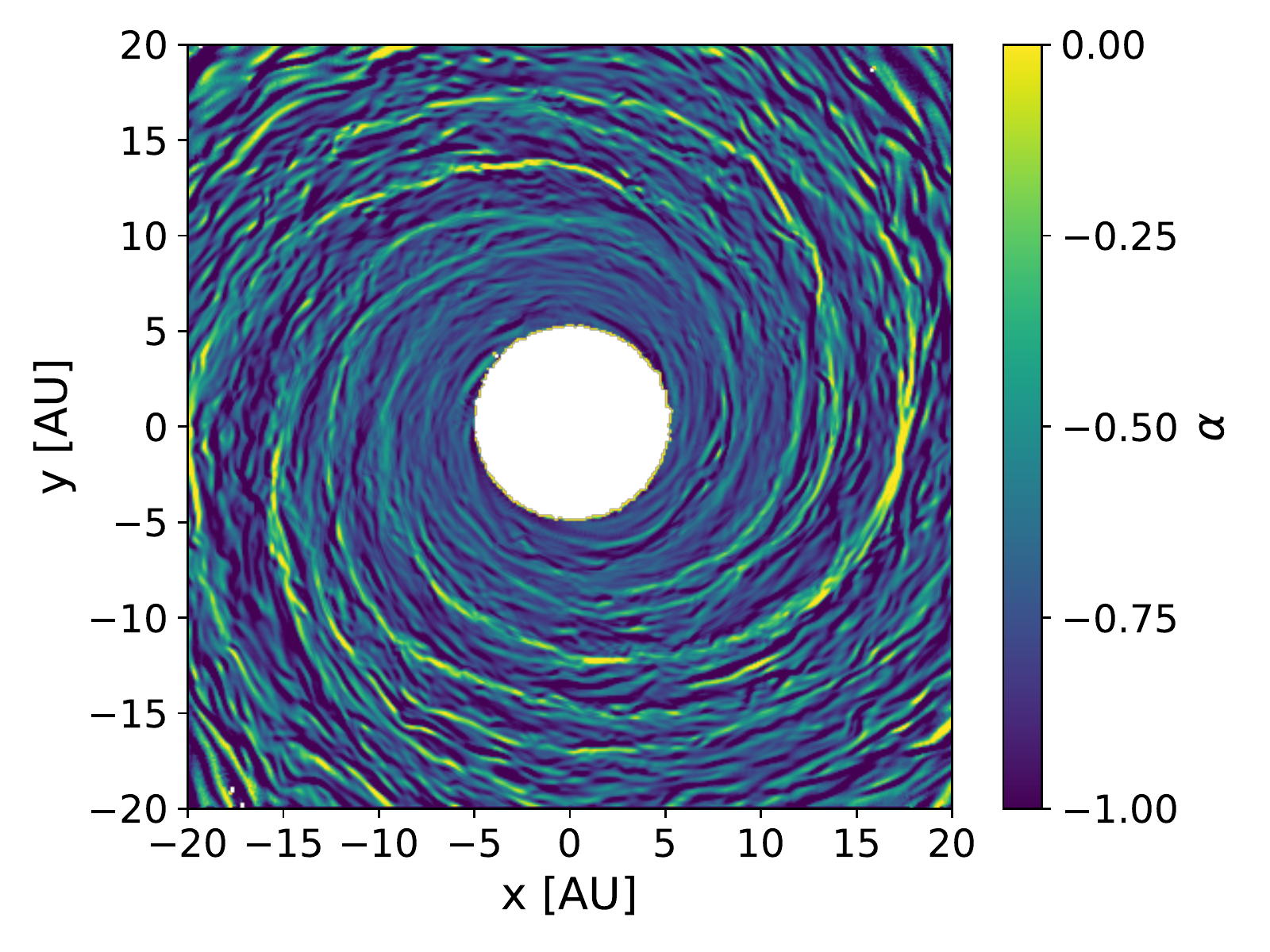}
	\caption{Shear plot at $t = 0$}
\end{subfigure}

\vspace{0.5cm}
\begin{subfigure}{0.5\columnwidth}
	\includegraphics[width=\textwidth]{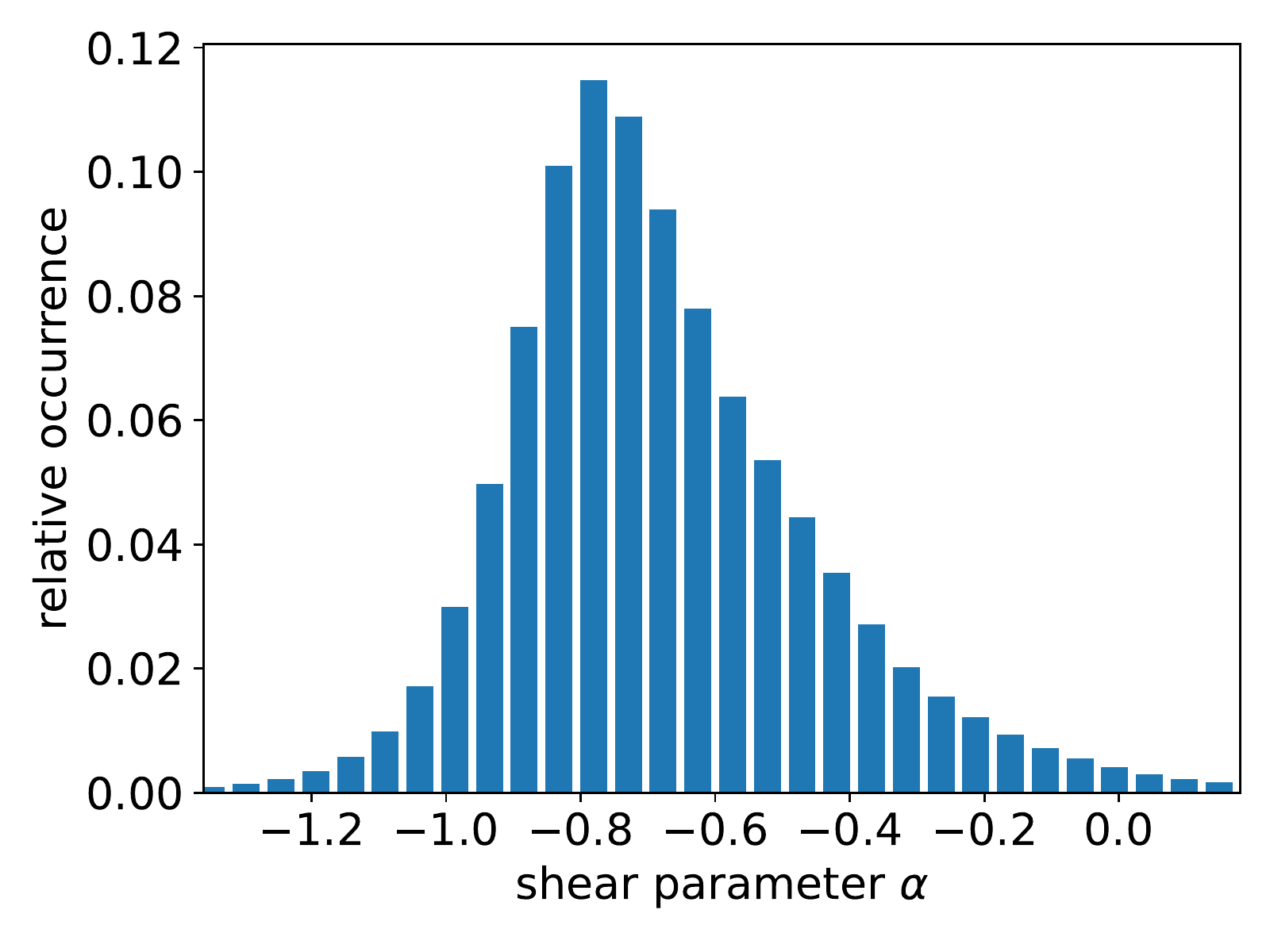}
	\caption{Shear histogram at $t=95\text{yr}$}
\end{subfigure}%
\begin{subfigure}{0.5\columnwidth}
	\includegraphics[width=\textwidth]{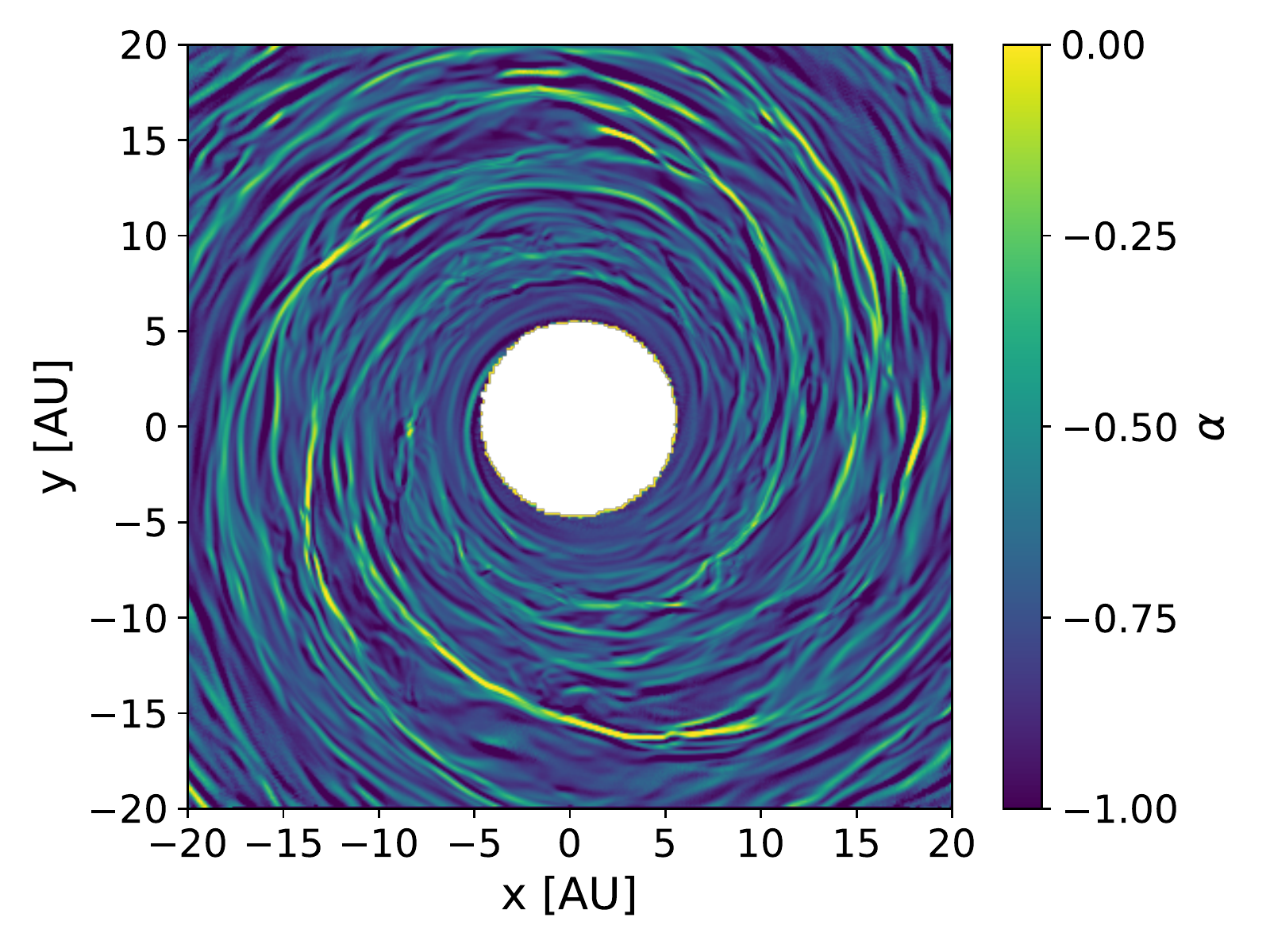}
	\caption{Shear plot at $t = 95\text{yr}$}
\end{subfigure}

\vspace{0.5cm}
\begin{subfigure}{0.5\columnwidth}
	\includegraphics[width=\textwidth]{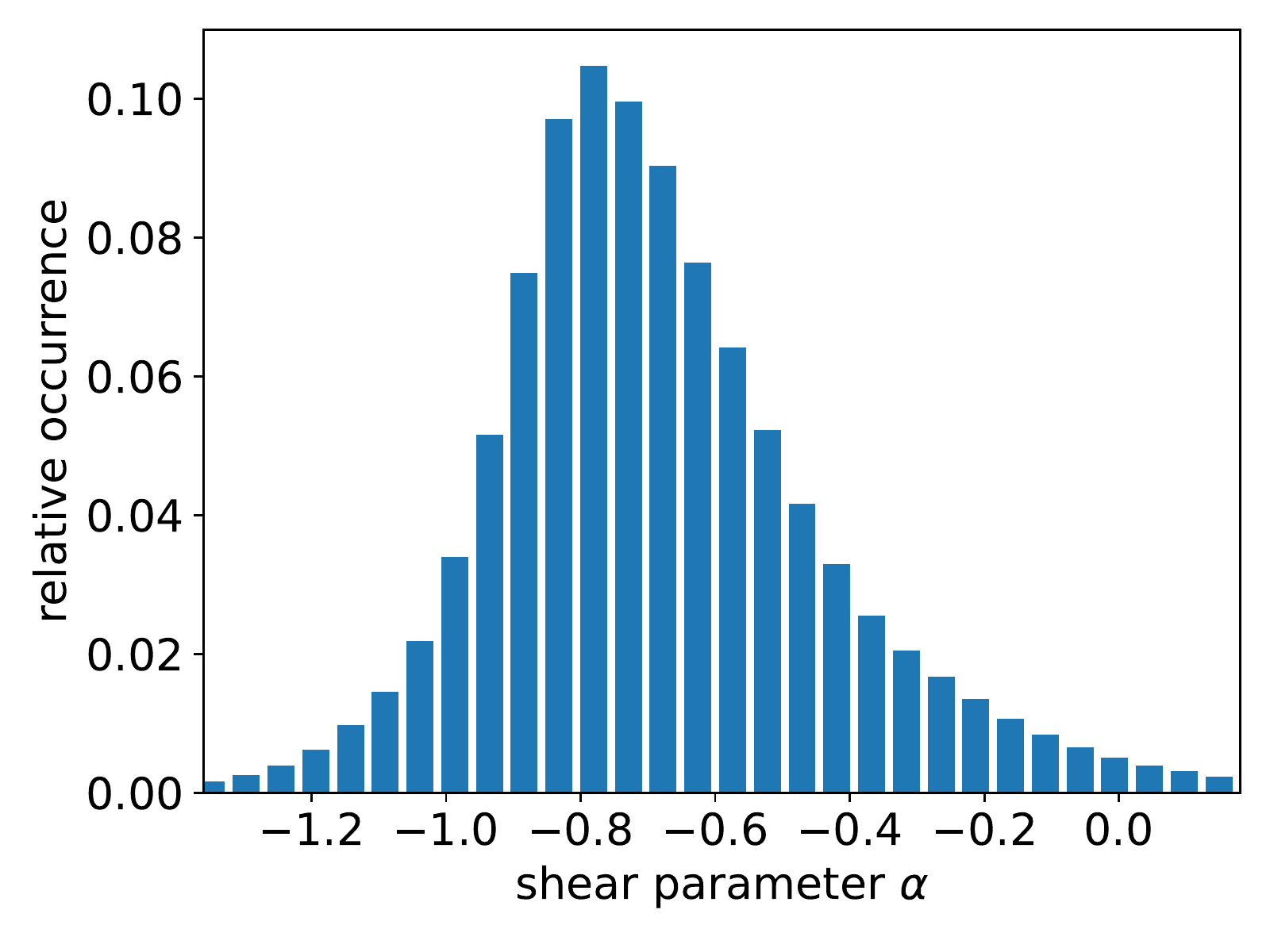}
	\caption{Shear histogram at $t=127\text{yr}$}
\end{subfigure}%
\begin{subfigure}{0.5\columnwidth}
	\includegraphics[width=\textwidth]{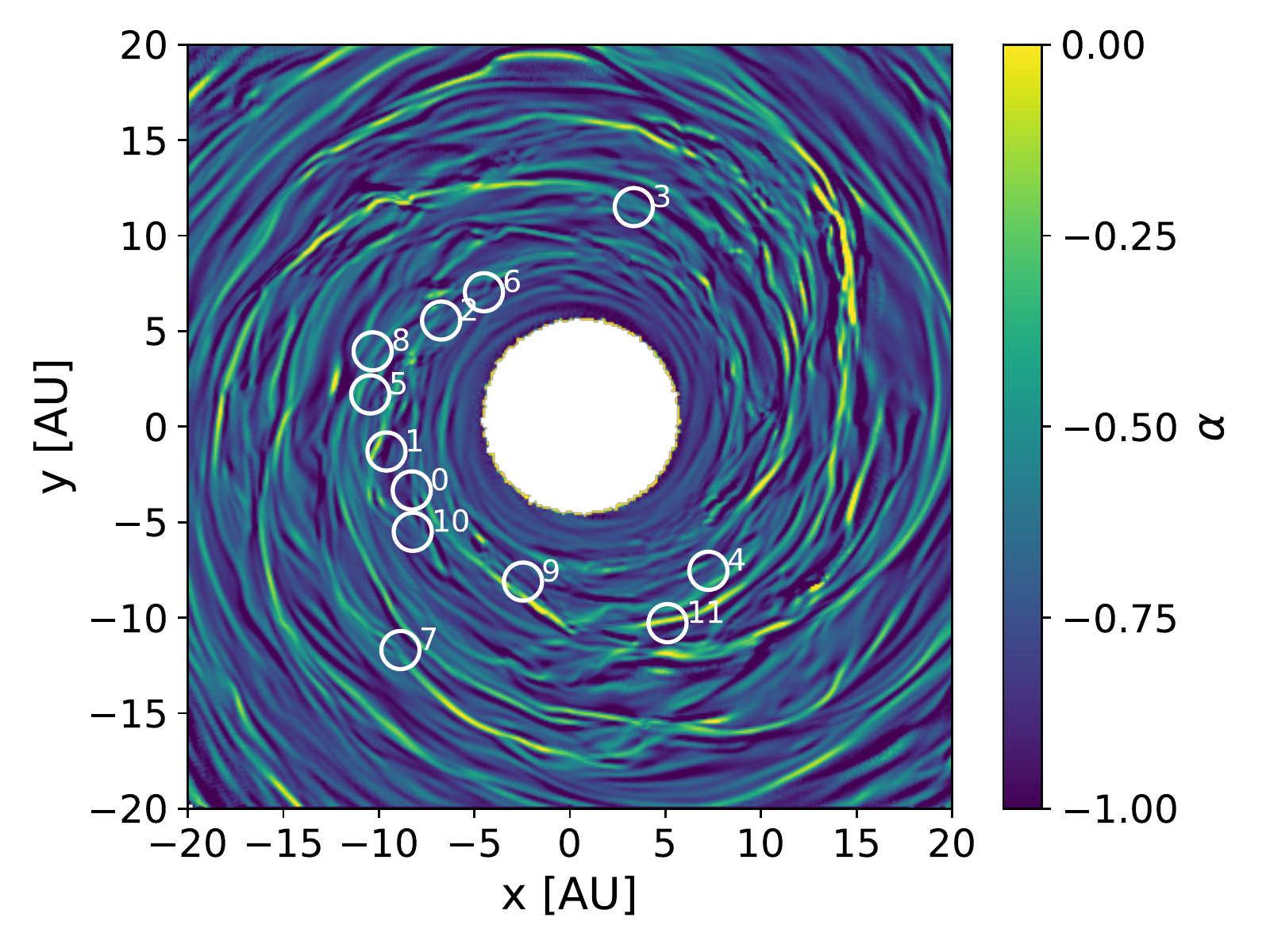}
	\caption{Shear plot at $t =127\text{yr}$}
\end{subfigure}
\caption{Histogram of values of 
the shear parameter $\alpha$ at the beginning of the simulation (left column) and  color-coded intensity
map of the shear parameter in the disk at different times,
at the beginning of the simulation (top),
right before fragmentation (middle),
and after fragmentation (after most clumps
have formed (bottom).
In the last plot, 
the locations of the clumps at this time are shown.
It is clear that low shear values occur primarily
along dense spirals, namely at the typical sites of clump formation.}
\label{fig-shear}
\end{figure}

\begin{figure}
\includegraphics[scale=0.5]{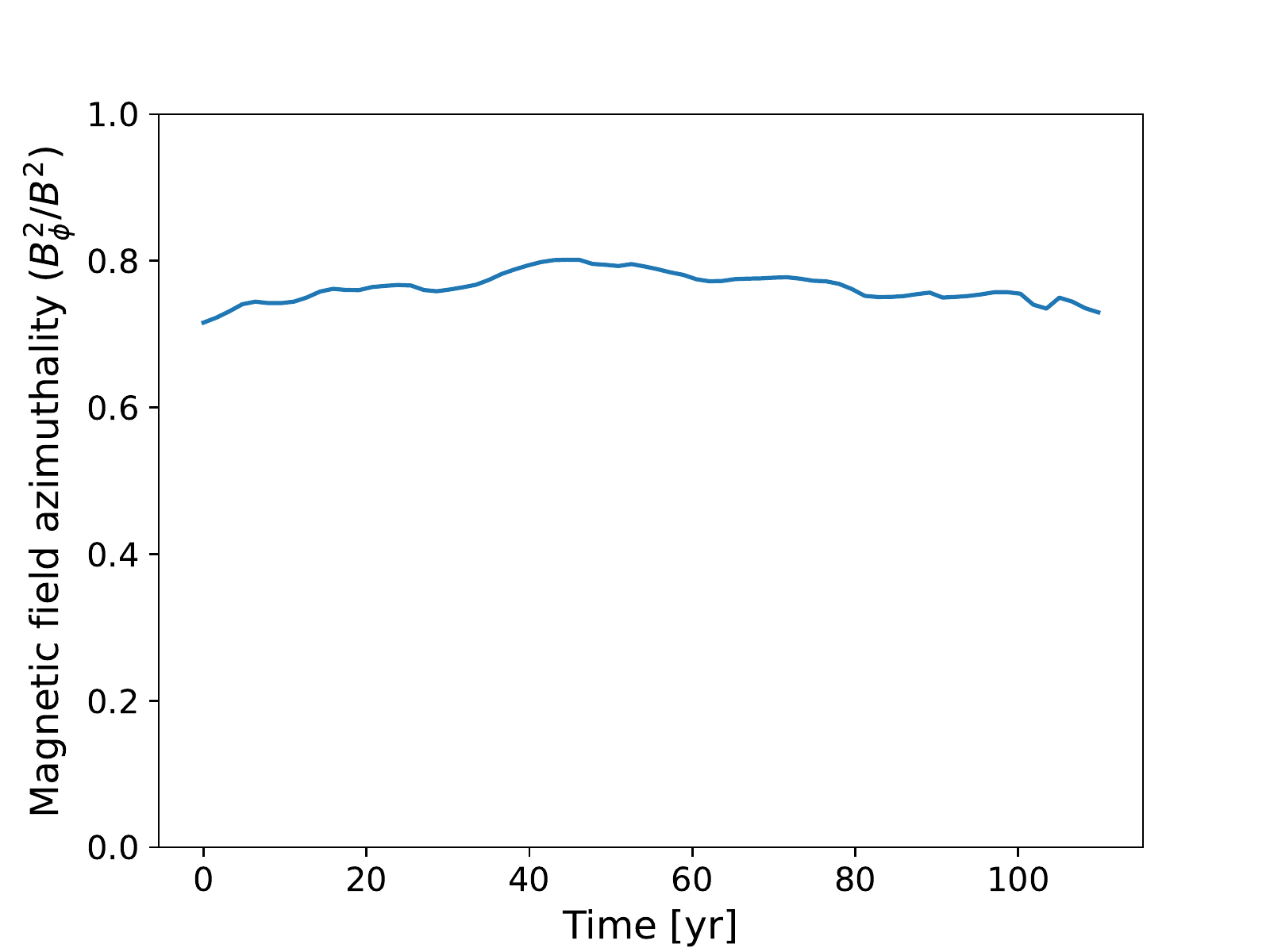}
\caption{Azimuthality of the magnetic field over the simulation time
defined as 
$B_\phi^2 / B^2$ where $B_\phi$ is the toroidal part of the
magnetic field (relative to the disk) and $B$ the total magnetic field.
It is measured taking into account the particles that will later form 
the clumps and thus traces the collapsing regions.}
\label{fig-azimuthality}
\end{figure}

%	\includegraphics[scale=0.55] {./images/shear_map_000.pdf}
%\end{subfigure} 
%\begin{subfigure}{0.5\textwidth} 
%	\includegraphics[scale=0.55] {./images/shear_map_060.pdf}
%\end{subfigure} 
%\begin{subfigure}{0.5\textwidth} 
%	\includegraphics[scale=0.55] {./images/shear_map_080.pdf}

In the next step we want to find out if the process
described in the last section could contribute to the fragmentation results.
To this end, we trace the particles of the clumps back in time
to compute the physical properties of the fragmenting regions
at early stages.
The fragmentation process described by \citet{elmegreen} relies on two
conditions: First, the magnetic field was assumed to be toroidal such
that locally an asymmetry arises between the radial and the azimuthal direction.
Second, the effect requires low-shear regions because only then the
magnetic field amplifies perturbations and also the size of the perturbations
seems to be lower.

Fig. \ref{fig-shear} shows the measured shear parameters $\alpha$
in the simulation at various times.
%Only the regions that will later fragment are taken into account and the
%histogram is created over all snapshots until the first fragments arise.
It can be seen that most values are around the expected Keplerian 
shear $\alpha = -0.75$ but there is a wide dispersion.
The shear values may deviate from the Keplerian value through gas pressure, turbulence
or magnetic field effects.
Regions of low shear $\alpha \gtrsim -0.15$ exist, although they
are not common ($2 \%$).
%These regions would be required to destabilize regions that were otherwise
%stable inducing new instabilities to the system.
However, regions of intermediate shear $\alpha \gtrsim -0.4$ appear more
frequent ($10 \%$).
In these regions the magnetic field enhances perturbations that would
already be unstable without it.
Nevertheless, perturbations in regions of these shear values could lead to smaller
fragmented objects.

In fig. \ref{fig-azimuthality} the relation $B_\phi^2/B^2$ is measured, with 
$B_\phi$ being the azimuthal component of the magnetic field and $B$ the 
total magnetic field.
The relation measures the fraction of the magnetic energy that is in the 
toroidal component.
It is shown over a range of snapshots from the beginning of the simulations
until the first fragments appear; only the regions that contain particles
that will later fragment are taken into account.
It can be seen that the magnetic field is predominantly toroidal
but a significant fraction of the energy is also in the radial and the z-component.
We assume that this is still compatible with the effect described by Elmegreen
because the magnetic field would still be coupled to the contraction
much more in the azimuthal direction than in the other directions and could
thus still suppress the Coriolis force (see section \ref{destabilization}).
When querying the solutions of the dispersion relation (section \ref{disp-rel})
we arrive at very similar solutions if we don't use a perfectly axisymmetric 
magnetic field ($B_x = 0$).

%Still works when checking growthfactor2.py

%of the magnetic field is me
%When looking at the magnetic field (fig. \ref{fig-azimuthality}) it can be seen that
%it is predominantly toroidal in the later fragmenting regions.

\section{Summary and concluding discussion}

As in conventional disk instability, clumps in magnetized self-gravitating disks 
formed from fragmentation sites inside spiral structure. The 
flow state in such a disk has been shown to be more turbulent
relative to non-magnetized disks, due to a combination
of Maxwell and gravitational stresses \citep{deng1}, which
also leads to more flocculent spiral structure.
The initial properties and structure of the clumps in the fragmenting
sites are thus determined by a combination of the gas flows kinematics, the magnetic field, and the thermodynamical state of the medium.
We analyzed both the pre-collapse and post-collapse properties of the
fluid that ends up generating the clumps, which led to numerous
findings on the origin, dynamics and development of magnetized clumps:

\begin{itemize}

\item Clumps forming in magnetized disks have gravitationally bound masses from one to almost two orders of magnitude lower than clumps in 
unmagnetized disks, being typically in the range of Super-Earths
and Neptune-sized bodies.

\item When comparing the energy scales at the time right before %
the formation of the clumps it is found that %
the magnetic energy is smaller than the internal energy but 
dominates over the kinetic turbulence energy.
Since the energy stored in the magnetic field
is much greater than  that in the turbulent motion of particles 
its role in determining directly the properties and dynamics
of clumps is most important.
\item After the collapse, the magnetic field is amplified around and inside
the clumps. The peak may be just outside the bound radius in which case 
the magnetic field acts compressing inwards and pushes the surrounding
flow outwards, or it can be at the centre of the clump in which case
it could just isolate the clump from the outside.
In general, we confirm that this "magnetic shield" stifles gas
accretion, suppressing further clump growth.
\item While the magnetic field may have its maximum field strength
inside the clump, relative to the other energy components, it is dominant only at the periphery of the clumps.

\item Thermal gas pressure plays an important role in determining
the clump energetics. It is higher than both rotational energy and the magnetic energy near the centre of the clumps.
The importance of the magnetic field generally increases further outside,
around the bound radius of the clumps.

\item{After clump formation, rotational energy becomes the dominant
form of kinetic energy inside clumps.
In their outermost regions clumps are rotationally supported in both MHD and HD simulations, but rotation is significantly higher in
the HD clumps than in MHD clumps. As a result the MHD clumps have a lower specific angular momentum than the HD clumps, which brings
their spin in better agreement with the spin of gas and ice giant planets
in the Solar System (excessively high spins are a known problem
for conventional HD fragmentation simulations \citep{mayer-2004}).}

\item Beside influencing the evolution of the clumps, the magnetic field
also has an influence on the fragmentation process itself being
responsible for 
significantly smaller initial masses of the clumps. Adapting previous
results of linear perturbation theory for non-axisymmetric perturbations
of a magnetized rotating sheet by \citet{elmegreen} lends evidence
for a destabilizing effect of the magnetic field in low-shear regions
which results in a smaller characteristic scale of fragments.

\end{itemize}

We discussed the fragmentation and early evolution of intermediate-mass protoplanets in the MHD disk.
It remains however, to investigate their long-term evolution to establish that such protoplanets really contribute to the 
observed intermediate-mass planet population.
First, such protoplanets need to survive for a sufficiently long time, therefore
improving the understanding of migration of such clumps will be crucial to determine their further outcome.
Inward-migration may eventually lead to tidal disruption by the host star \citep{boley-disruption}.
To form gas planets,
they have to avoid tidal disruption until they cool enough to undergo their second dynamical collapse due to the dissociation of molecular hydrogen \citep{helled}.
To form a solid core, it is crucial for them to accrete dust and form a core sufficiently fast.
This would be crucial to explain terrestrial intermediate-mass planets.
Even after such processes, the protoplanet can still fall into the star \citep{helled}.
\citet{deng} noted that the protoplanets experience migration both in- and outward, hence they will eventually be distributed over a broad radial range. However, 
the disks were evolved for only $\approx 10$ orbits 
so the question of migration on longer time scales remains to
be investigated. The significantly smaller masses of the clumps
in magnetized disks are also expected to have an overall
impact on the strength of migration. No runaway migration
is expected in the mass range of typical clumps, in contrast
with clumps in conventional disk instability simulations
\citep{baruteau, malik}, which should
significantly increase the chances of clump survival.
Furthermore, the different nature of the background flow
could have an impact on the nature of the migration process
itself.
In \citet{nelson}, who simulated low mass planets in MHD non-self-gravitating disks, it was found that
a planet of 3 $M_\text{earth}$ 
would experience random walk migration instead of a monotonic drift because of the high turbulence in the disk.
Since the disk in the simulations analysed here \citep{deng} is 10 times more massive, the same mass ratio between disk and planet
would correspond to a mass of $\approx 0.05 M_\text{jup}$, namely
compatible with the typical clump mass in our simulations. 
Therefore, in addition to the low mass, this is another way
clumps in magnetized disks would avoid fast migration and survive.

Further improvements in the understanding of the evolution of such protoplanets could be achieved by implementing additional and more accurate physics. Published hydrodynamical simulations of disk instability provide plenty of hints of what physics should be important.
As an example, in \citet{stamatellos} it was found that the inclusion of radiative feedback in the simulations changed the outcome of migration for giant gas planets, namely the outward-migration was prevented and inward-migration also came to a halt because of a gap at the orbit of the planet that arose from the heating of the material that was accreted on to the planet. Likewise, \citet{rowther} 
showed how heating of the inner disk can stifle migration of massive clumps.
Furthermore, in \citet{nayakshin} it was shown that radiative feedback may have the effect of slowing the accretion of matter on to the planet therefore reducing its growth; but again this result is for giant planets. Also, in the very few simulations that have
included even simple approximations to radiative transfer, such as flux-limited diffusion, it has been shown how clump mass growth
is slowed down as their thermal pressure support increases beyond what is predicted by Beta cooling or other simple cooling recipes
\citep{szulagyi}. 
This suggests that also radiative transfer, as well as radiative feedback, should be included in future MHD simulations
of self-gravitating disks.

Additionally, the effect of ambipolar diffusion and the Hall effect should be studied.
The non-ideal effects should be more important near the mid-plane of the disk as in the outer layers the gas is expected to be ionized by the stellar radiation \citep{perez-becker}. While Ohmic dissipation, which we have included, is usually dominant in the highest density
medium as that inside the clumps, ambipolar diffusion could affect magnetic field dissipation in lower density regions, such as
at the periphery of the clumps. For example, it should be investigated if ambipolar diffusion should  affect the "magnetic shield"
developed around MHD clumps, which, as we have seen, plays an important role in their overall mass growth, or if it could have
an influence on the initial stage of fragmentation (although order of magnitude estimates by \citet{deng} suggest that the
dissipation rate should be too low to be dynamically relevant over the short timescales probed by our simulations).
%For example, in \cite{gressel} it was found that in MHD simulations of inner disks that include ambipolar diffusion the
%accretion rate onto the star changed 
%because of the emergence of  magnetocentrifugal winds.
Moreover, one could use the simulations presented here as a starting point for additional
high-resolution simulations of isolated clumps in order to verify their internal structure and study the collapse of MHD clumps similarly to  what has been done for hydrodynamical clumps in \citet{galvagni-precollapse}
%Migration
%kleine: weniger Migration

While the MHD simulation used $\approx 30$ million particles, the
companion HD simulations that we used as a comparison 
used only $\approx 3$ million.
New HD simulations that used $\approx 30$ million particles
have also been conducted starting from the same initial conditions using the procedure described in section \ref{ch-simulations}.
A quick check revealed not much difference from the lower-resolution
HD simulations used in this paper,
neither in the number of resulting clumps nor in the 
angular momentum result (see fig. \ref{fig-angmom})
but we did not analyze them any further.

For the perturbation analysis, we note
that since we considered non-axisymmetric modes,
their wavenumbers change over time leading
to mode mixing.
A more accurate treatment would have to take this effect 
into account as it would impact the fragmentation
process.
%However, the difference of the MHD situation compared to the HD one %described in this paper is already telling.
More in general, one could question the use of linear perturbation
theory as done in this paper, beginning with the fact that we considered
perturbations on a smooth axisymmetric background. In fact, it
is well established in the literature (eg. \citet{durisen}) that fragmentation occurs inside spiral arms,
namely in a non-axisymmetric, already nonlinear flow.
Furthermore, spiral structure typically develops after a transient stage
in which ring-like global perturbations arise in the disk
(eg. \citet{deng-17}). With this in mind,
\citet{deng-ogilvie22} instead of a smooth disk, considered
an already nonlinear ring-like structure as the background state, described
by solitary waves. Then, they studied
the growth of non-axisymmetric perturbations to the solitary modes, identifying
fast growth, which would result in the development of a
spiral structure.  Note that this
is different from the conventional swing amplification mechanism,
which assumes that non-axisymmetric waves are already present
and can increase their amplitude exponentially when they switch
from leading to trailing \citep{goldreich}.
Fragmenting sites would thus correspond to self-gravitating patches
in the growing non-axisymmetric pattern, a calculation that should
be attempted in the future as it could lend a new, more
realistic prediction of the fragmentation scale. Subsequently,
such an approach should be extended to include the effect of the
magnetic field on the mode growth.
%This ring would provide an environment
%where the traditional eigenmodes obtained from
%linear perturbation theory can grow.}

\section*{Acknowledgements}
This work is supported by the Swiss Platform for Advanced Scientific Computing (PASC) project SPH-EXA2.

\section*{Data Availability}
The data files that support our analysis will be made available upon reasonable request.

\bibliography{quellen}

\end{document}